\documentclass[final,3p,times]{elsarticle}

\usepackage{amssymb}
\usepackage{amsmath}
\usepackage{graphicx}
\usepackage{url}
\usepackage{multirow}
\usepackage[table,xcdraw]{xcolor}
\usepackage{color}  % \usepackage{color}
\usepackage{soul}
\journal{Anonymous journal} % Artificial Intelligence % Chaos, Solitons and Fractals % Information Sciences

\begin{document}

\begin{frontmatter}



\title{A Comprehensive Survey on Multi-Agent Cooperative Decision-Making Applications: Scenarios, Approaches, Challenges and Perspectives} % Comprehensive Survey


% \author{} %% Author name

% \affiliation{organization={},%Department and Organization
\author{Weiqiang Jin\fnref{1}}
\ead{weiqiangjin@stu.xjtu.edu.cn}

\author{Hongyang Du\fnref{2}}
\ead{duhy@hku.hk}

\author{Shixiang Tang\fnref{1}}
\ead{shixiangtang@stu.xjtu.edu.cn}

\author{Biao Zhao\corref{cor1}\fnref{1}}
\ead{biaozhao@xjtu.edu.cn}

% \author{Xingwu Tian\fnref{1}}
% \ead{txw\_xjtu@163.com}

% \author{Bohang Shi\fnref{1}}
% \ead{Bh\_567@stu.xjtu.edu.cn}

\author{Guang Yang\corref{cor1}\fnref{3,4,5,6}}
\ead{g.yang@imperial.ac.uk}

\affiliation[1]{organization={School of Information and Communications Engineering, Xi`an Jiaotong University},
            addressline={Innovation Harbour}, 
            city={Xi`an},
            postcode={710049}, 
            state={Shaanxi},
            country={China}}

\affiliation[2]{organization={Department of 
Electrical and Electronic Engineering, The University of Hong Kong (HKU)},
            city={Hong Kong},
            postcode={Hong Kong},
            country={China}}

\affiliation[3]{organization={Bioengineering Department and Imperial-X, Imperial College London},
    city={London},
    postcode={W12 7SL},
    country={UK}}

\affiliation[4]{organization={National Heart and Lung Institute, Imperial College London},
    city={London},
    postcode={SW7 2AZ},
    country={UK}}

\affiliation[5]{organization={Cardiovascular Research Centre, Royal Brompton Hospital},
    city={London},
    postcode={SW3 6NP},
    country={UK}}

\affiliation[6]{organization={School of Biomedical Engineering \& Imaging Sciences,  King's College London},
    city={London},
    postcode={WC2R 2LS},
    country={UK}}

\cortext[cor1]{Corresponding authors: Biao Zhao and Guang Yang.}
% \cortext[cor2]{Corresponding authors: Biao Zhao and Guang Yang.}  % at: National Heart and Lung Institute, Imperial College London, U.K

\begin{abstract}
% Abstract text.
With the rapid development of artificial intelligence, intelligent decision-making techniques have gradually surpassed human levels in various human-machine competitions, especially in complex multi-agent cooperative task scenarios. Multi-agent cooperative decision-making involves multiple agents working together to complete established tasks and achieve specific objectives. These techniques are widely applicable in real-world scenarios such as autonomous driving, drone navigation, disaster rescue, and simulated military confrontations. This paper begins with a comprehensive survey of the leading simulation environments and platforms used for multi-agent cooperative decision-making. Specifically, we provide an in-depth analysis for these simulation environments from various perspectives, including task formats, reward allocation, and the underlying technologies employed. Subsequently, we provide a comprehensive overview of the mainstream intelligent decision-making approaches, algorithms and models for multi-agent systems (MAS). These approaches can be broadly categorized into five types: rule-based (primarily fuzzy logic), game theory-based, evolutionary algorithms-based, deep multi-agent reinforcement learning (MARL)-based, and large language models (LLMs) reasoning-based. Given the significant advantages of MARL and LLMs-based decision-making methods over the traditional rule, game theory, and evolutionary algorithms, this paper focuses on these multi-agent methods utilizing MARL and LLMs-based techniques. We provide an in-depth discussion of these approaches, highlighting their methodology taxonomies, advantages, and drawbacks. Further, several prominent research directions in the future and potential challenges of multi-agent cooperative decision-making are also detailed.
\end{abstract}



\begin{highlights}
\item Provides a comprehensive survey of multi-agent decision-making methods.
\item Analyzes key simulation environments for multi-agent reinforcement learning.
\item Investigate decision-making approaches, including MARL and large language models.
\item Identifies challenges and future research directions in multi-agent collaboration.
\item Reviews real-world applications in transportation, aerial systems, and automation.
\end{highlights}

%% Keywords

\end{frontmatter}


% \noindent\rule{\textwidth}{0.5pt}  % 无缩进的横线

% Insert Table of Contents here
\tableofcontents
% \newpage

\noindent\rule{\textwidth}{0.5pt}  % 无缩进的横线

\section{Introduction}
\label{intro}
% for IEEE journal papers produced under \LaTeX\ using
% IEEEtran.cls version 1.8b and later.
% % You must have at least 2 lines in the paragraph with the drop letter
% % (should never be an issue)
% I wish you the best of success.

 

% \subsection{Subsection Heading Here}
% Subsection text here.

% % needed in second column of first page if using \IEEEpubid
% %\IEEEpubidadjcol

% \subsubsection{Subsubsection Heading Here}
% Subsubsection text here.
%  首先介绍multi-agent cooperative decision-making基本概念
\subsection{Research Backgrounds of Multi-Agent Decision-Making}
With the continuous advancement of science and technology, intelligent decision-making technology has made rapid progress. These technologies have gradually surpassed human capabilities in various human-machine game competitions, even exceeding the top human levels. 
Over the past few decades, especially following the successful application of Deep Q-Networks (DQN) \cite{mnih2013DQNv1,Mnih2015DQNv2} in the Arita game and the victories of  AlphaGo and AlphaZero \cite{Silver2016Go,Silver2017Zero} over top human opponents, these landmark achievements have significantly propelled the advancement of intelligent decision-making research. 
% The field of intelligent decision-making has entered an unprecedented period of prosperity and produced a plethora of results. 

% 用一张图形象的展示single-agent -> multi-agent.
\begin{figure*}[!t]
\centering
\includegraphics[width=\linewidth]{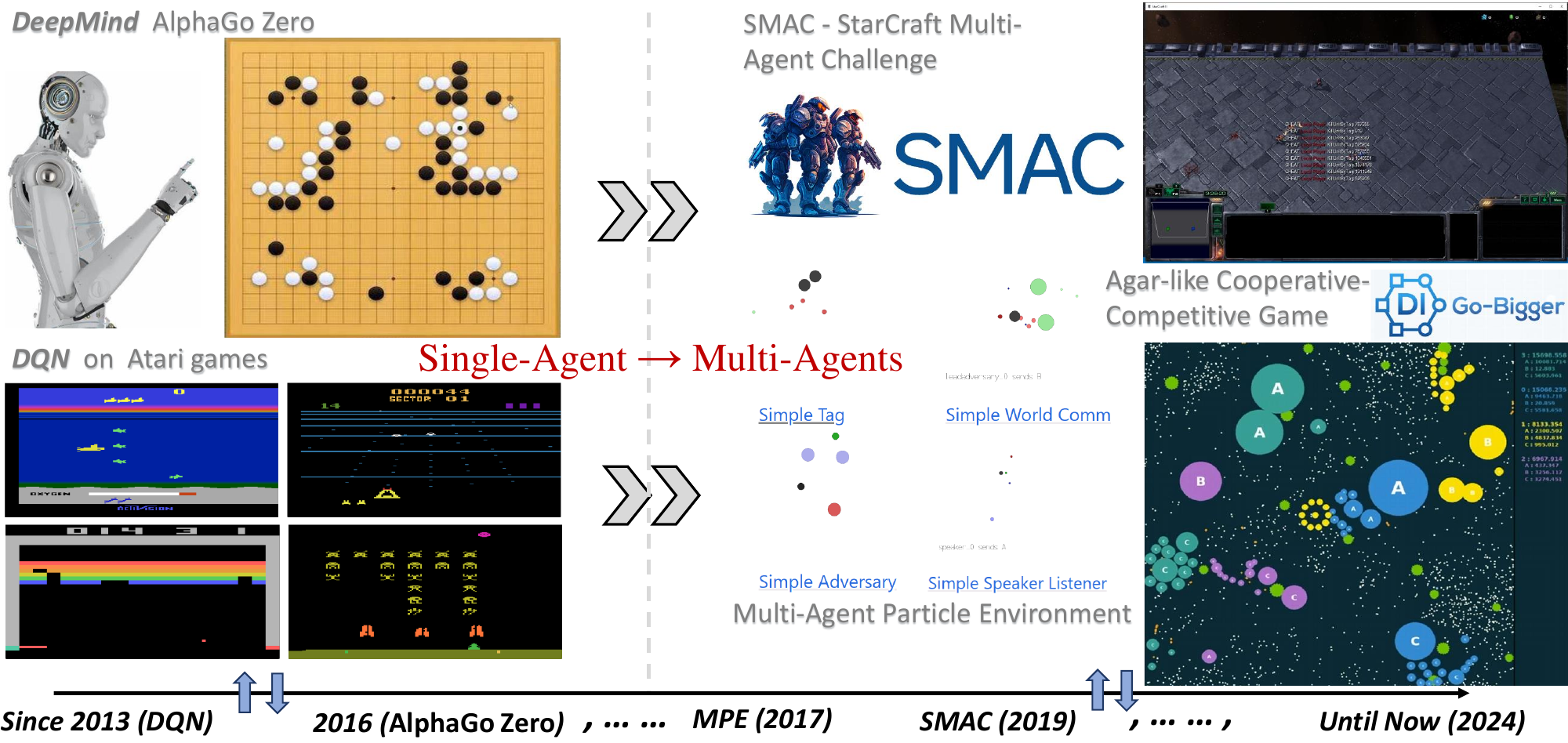}
\caption{An overview of the evolution of scenarios and methods in decision-making from single-agent to multi-agent systems.}
\label{fig_fromsingletomulti}
\end{figure*}

% Multi-Agent Particle Environment (MPE) \cite{lowe2017multi,mordatch2017emergence}.

To meet the growing complexity of real-world applications and the increasing demand for more sophisticated, reliable, and efficient intelligent systems, multi-agent cooperative decision-making has rapidly evolved from simple single-agent scenarios \cite{Li2022Survey,Gronauer2022Survey,Pamul2023survey,James2023Survey}.
Multi-agent cooperative decision-making is a crucial subfield within machine learning (ML) \cite{JIN2024122289} and artificial intelligence (AI) \cite{ZHAOJINPromptMR}. It involves multiple interacting agents working together to complete established tasks across diverse well-designed dynamic simulated environments and various complex real-world systems. 
% An exemplary multi-agent decision-making system tailored for specific scenarios must exhibit precise responsiveness in stochastic scenarios, enhance cooperative decision-making efficiency, and maximize the potential for joint task execution, all while maintaining cost-effectiveness.
%, such as board games and video games, to complex multi-agent cooperative decision-making scenarios. 

As depicted in Figure \ref{fig_fromsingletomulti}, the evolution research progress from single-agent to multi-agent decision-making systems, along with methodological comparisons, highlights that this rapidly advancing field is a crucial step toward achieving human-level AI and the Artificial General Intelligence (AGI) age.
% This field has become a cornerstone in the pursuit of artificial general intelligence (AGI), and its continued development is crucial for tackling the challenges of modern AI.
% Additionally, we will introduce a range of multi-agent cooperative simulation scenario technologies. 
% 简单介绍下各种研发的多智能智能体协同决策的算法模型
Multi-agent cooperative decision-making has a wide range of practical applications and many fundamental theoretical works across various domains.  % practical applications
The service scenarios are extensive, encompassing smart agriculture management \cite{seewald2024multiagentagri,qazzaz2024rescueUAV}, intelligent collaborative robots \cite{collabrobots9431107,collabrobots10039365,collabrobots9107997,MASCENA20134737}, self-driving collaborative obstacle avoidance \cite{JayawardanaMBLautodrive,LuMARLautoDriving,liu2024diversedriving},
% UAV pursuit-evasion \& Multi-UAVs Path Planning \cite{LVMARLonUAVswarm,LUO2024MultiUAVPE,AlexanderUAVPE},
autonomous navigation \cite{UAVNavigation1,UAVNavigation2,UAVNavigation3} as well as joint rescue tasks \cite{qazzaz2024rescueUAV,samad2018multi}.  
% underwater exploration  % intelligent robots collaborative competition
% , intelligent robots collaborative competition and tactical formation,
%, as described in the accompanying Figure \ref{fig_fromsingletomulti}. 
Correspondingly, considering the rapid pace of technological advancement and the multifaceted needs of the real world, in this work, we focus on the comprehensive study of multi-agent cooperative decision-making.

% \begin{CJK*}{UTF8}{gbsn}\end{CJK*}  \textcolor{red}{}
\subsection{Overview of Previous Multi-Agent Surveys}  % of Multi-Agent Decision-Making
% shortcomings deficiencies Weaknesses Inadequacies
Concurrent with the fast-paced advancements in multi-agent cooperative decision-making, there has been a marked increase in systematic literature reviews in this domain \cite{Du2021Survey,Gronauer2022Survey,James2023Survey,NING2024Survey}. These reviews have covered a wide range of topics, from theoretical innovations to practical applications, providing a comprehensive overview of the state-of-the-art. 
% Over the years, researchers have delved into various intelligent algorithms, and application scenarios, summarizing the developments and future directions in multi-agent intelligent decision-making.  % deeply models, systematically latest 

% Concurrently, systematic literature reviews on multi-agent cooperative decision-making have markedly increased, encompassing a wide range of topics from theoretical innovations to practical applications. Researchers have conducted in-depth explorations of various intelligent algorithms, models, and application scenarios, systematically summarizing the latest developments and future directions in intelligent decision-making.

Ning et al. \cite{NING2024Survey} provided a comprehensive overview of the evolution, challenges, and applications of multi-agent reinforcement learning (MARL)-based intelligent agents, including its practical implementation aspects.
% They discuss the potential future directions of RL-based multi-agent system (MAS) and its relevance across various technological fields. 
Gronauer et al. \cite{Gronauer2022Survey} provided an overview of recent developments in multi-agent deep reinforcement learning, focusing on training schemes, emergent agent behaviors, and the unique challenges of the multi-agent domain, while also discussing future research directions. Yang et al. \cite{yang2023utilitytheoryrobot} explored the utility theory application in AI robotics, focusing on how utility AI models can guide decision-making and cooperation in multi-agent/robot systems. 
% It introduces a utility-oriented needs paradigm, reviews relevant literature, and suggests future research directions to address related open challenges. 
Orr et al. \cite{James2023Survey} reviewed recent advancements in MARL, particularly its applications in multi-robot systems, while discussing current challenges and potential future applications. Du et al. \cite{Du2021Survey} provided a systematic overview of multi-agent deep reinforcement learning for MAS, focusing on its challenges, methodologies, and applications. Pamul et al. \cite{Pamul2023survey} provided a comprehensive analysis of the application of MARL in connected and automated vehicles (CAVs), identifying current developments, existing research directions, and challenges. Hernandez-Leal et al. \cite{hernandezleal2019survey} provided a comprehensive overview of approaches to addressing opponent-induced non-stationarity in multi-agent learning, categorizing algorithms into a new framework and offering insights into their effectiveness across different environments. The survey by Zhu et al. \cite{Zhu2024Survey} provided a systematic classification and analysis of MARL systems that incorporate communication, encompassing recent advanced Comm-MARL research and identifying key dimensions that influence the design and development of these multi-agent systems. 
% Li et al. \cite{Li2022Survey} reviewed the applications of MARL in next-generation Internet technologies, focusing on its use in solving challenges existed in these emerging networks like network access, power control, computation offloading, and network security. Nguyen et al. \cite{Nguyen2020survey} discussed the main challenges and solutions in MARL, analyzes the advantages and disadvantages of these methods and their applications in real-world scenarios, and offers insights for the future development of more practical multi-agent systems.

\subsection{Motivations of the Current Survey}
% 叙述下当前其他综述的不足，以及本综述的目的
However, despite the growing body of work in this field, existing surveys often have noticeable limitations \cite{Du2021Survey,Gronauer2022Survey,NING2024Survey,Zhu2024Survey}.
Specifically, our thorough investigation reveals that most current reviews and surveys share several common and significant significant drawbacks and limitations:  
% widespread  % common general universal pervasive catholic  % systematic
\begin{itemize}  % surveys
\item \textbf{Limited Research Scope:} Previous literature reviews \cite{hernandezleal2019survey,Zhu2024Survey} predominantly remain within the primary framework of reinforcement learning and have not broken through theoretical limitations, resulting in a lack of comprehensive coverage. 
\item \textbf{Neglect of Environments:} Previous literature reviews \cite{Nguyen2020survey,Gronauer2022Survey,Wang2024LLMsurvey} have largely concentrated on methodological and algorithmic advancements, frequently overlooking the essential role of simulation environments and platforms in multi-agent intelligent decision-making.
\item \textbf{Under-emphasis of Project Implementation:} Prior surveys \cite{NING2024Survey,Zhu2024Survey,Wang2024LLMsurvey} often focus on theoretical models and overlook detailed implementation aspects, including code-bases and project architectures. This gap limits readers' ability to fully understand and apply the findings.  % literature reviews
\end{itemize}

% TODO ####################################################
% \%  \textcolor{red}{Our motivation of this survey.}
% \textcolor{red}{.}
%%TODO 到时候把上面的介绍的limitations的list的简介放在这里，然后删除1.3章节，貌似1.3不重要且啰嗦
To address the aforementioned limitations and challenges, we recognize the need for more systematic and comprehensive reviews in the multi-agent intelligent decision-making field. 

% The motivation for this survey stems from the limitations observed in existing literature reviews, aiming to advance the field of multi-agent cooperative decision-making. 
Firstly, current reviews overly emphasize deep reinforcement learning and fail to adequately consider other potentially effective intelligent decision-making methods \cite{Du2021Survey,Li2022Survey,Pamul2023survey,Zhu2024Survey}.
% , such as rule-based (primarily fuzzy logic), game theory-based, evolutionary algorithms-based, and other advanced decision-making systems \cite{Du2021Survey,Li2022Survey,Pamul2023survey,Zhu2024Survey}. % intelligent fuzzy logic-based, 
% This narrow focus limits a comprehensive understanding of the diverse and complex challenges in decision-making. 
% Therefore, our survey seeks to broaden the scope by incorporating these alternative approaches, providing a more holistic perspective to aid researchers in exploring a wider range of solutions.
Secondly, with the rapid development of large language models (LLMs), their potential in natural language processing, knowledge representation, and complex decision-making has become increasingly apparent. However, current surveys have largely overlooked their integration. 
% This survey aims to bridge this gap by exploring the application of LLMs in multi-agent systems, examining their potential impact on intelligent decision-making, and proposing future research directions to align with the latest technological advancements in the field.
% 说下多智能体智能协同博弈不可或缺的两个要素：环境 (仿真平台) & 智能体 (AI方法 算法模型)
Additionally, existing reviews often neglect the critical role of simulation environments in the development of multi-agent systems. 
However, simulation environments are not merely auxiliary tools but are an integral part of the MAS development and evaluation process. The agents' learning and decision-making processes are influenced and constrained by these environments, making it equally important to understand and develop these environments as it is to focus on the algorithms themselves.
% These environments are essential for translating theoretical advancements into practical applications. 
% Based on these, our survey focuses on the impact of simulation environments on cooperative decision-making outcomes, addressing this significant gap in the literature.
Finally, the lack of attention to practical implementation details in current reviews has resulted in a disconnect between theory and practice. This survey will delve into the specifics of project implementation, including code structures, system architecture, and the challenges encountered during development, to enhance research reproducibility and facilitate the effective translation of theoretical research into practical applications. 
% This approach aims to drive innovation in multi-agent cooperative decision-making technologies, making them more broadly applicable in complex real-world scenarios.

% In this survey, we conduct a comprehensive investigation and provide a detailed introduction to the current state-of-the-art approaches in multi-agent cooperative decision-making. 
Building on the motivations outlined earlier, this survey extends beyond the scope of previous reviews, which were often limited to specific areas of discussion. We treat multi-agent environments as equally important components, alongside the methods and techniques, and provide a thorough introduction to the most advanced algorithms and simulation environments. Moreover, we categorize various multi-agent cooperative decision-making methods from a more fundamental implementation perspective.
In summary, this survey seeks to provide a more comprehensive and practical framework for research in multi-agent cooperative decision-making, thereby advancing the continuous development of this critical field.

% % ######################################################################
% Good reviews should not be limited to the learning paradigms of reinforcement learning. Instead, they should encompass a broad spectrum of intelligent decision-making theories. This includes, but is not limited to, collaborative decision-making in multi-agent systems, strategy formulation based on game theory, and the application of fuzzy logic in uncertain environments, especially the innovative applications of large language models in complex decision-making. %, especially the recent large language models (LLMs) reasoning-based approaches
% % It is crucial to focus on the integration and cross-application of emerging technologies. 
% Exploring how large language models (LLMs) can be combined with other intelligent decision-making methods to improve decision-making efficiency and accuracy in broader scenarios is essential. Through such systematic and multi-perspective reviews, the latest developments and future directions in the field of intelligent decision-making can be more comprehensively reflected, ultimately driving further breakthroughs and innovations in this multi-agent cooperation research.
% % ######################################################################

\subsection{The Survey Overview / Contents Organization}

% % ################################################################
% MAKE YOUR DREAM 
% % ################################################################
% Specifically, in this survey, we will deeply analyze the basic principles, methodological schemes classification, experimental simulation environments, realistic application scenarios, and possible challenges of this research field, comprehensively elucidating its potential applications and future development directions. 
% potential
% 讲述本综述的内容组织架构，以及展望未来的研究方向

% 在这里画整个survey的框架和讲解图
\begin{figure*}[!t]
\centering
\includegraphics[width=\linewidth]{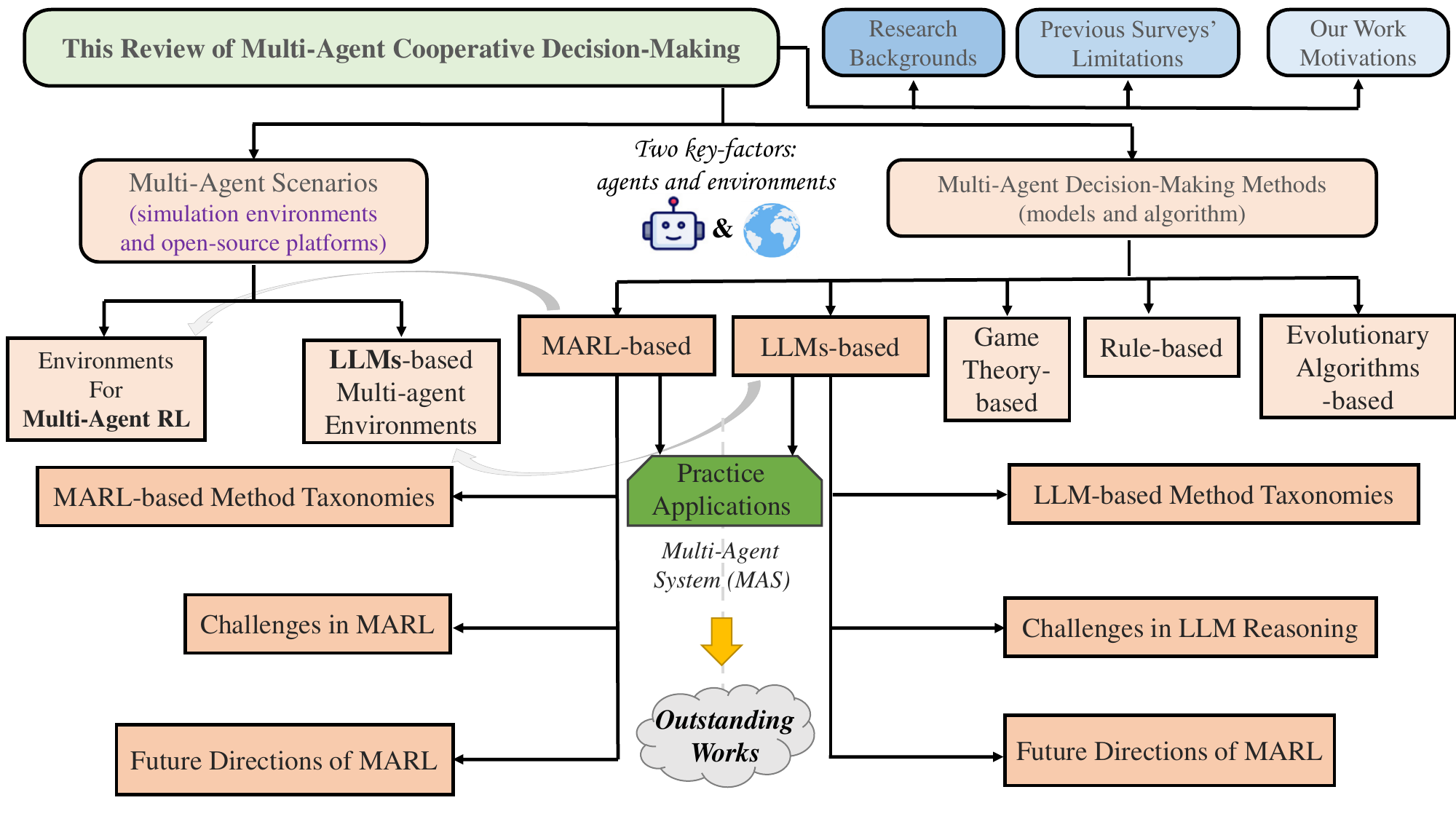}
\caption{Illustration of our systematic review of multi-agent intelligent decision-making research. Compared to previous reviews, we have incorporated comprehensive introduction and analysis, with each segment corresponding to a specific chapter in the survey.}  % extensive and
\label{fig_overall}
\end{figure*}

As depicted in Figure \ref{fig_overall}, we have structured the survey to reflect our research approach, with each main and sub-branch corresponding to a specific part:
First, in Section \ref{intro}, we introduce the research background of multi-agent cooperative decision-making, discuss the drawbacks of previous surveys, and outline the organizational structure of this survey.
%
% Next, in Section \ref{resbackMADK}, we present the preliminaries and essential background knowledge on multi-agent cooperative decision-making. 
% This includes the historical evolution of multi-agent decision-making technologies, the distinctions between multi-agent and single-agent decision-making, and an in-depth discussion of the limitations identified in previous surveys. 
%
% Next, we introduce the current mainstream paradigms in multi-agent cooperative decision-making research in Section \ref{coreTEX}. 
Given that MARL and LLMs-based intelligent decision-making methods demonstrate significant advantages and future potential,
% our research primarily focuses on these paradigms within multi-agent cooperative scenarios. O
our primary attentions are placed on Deep MARL-based and LLMs-based methods due to their superior ability to manage dynamic and uncertain environments.  % Special
In Section \ref{madktax}, we then delve into mainstream intelligent decision-making approaches, algorithms, and models. We categorize these approaches, with a continued focus on MARL-based and LLMs-based methods, discussing their methodologies, advantages, and limitations.
Following this, in Section \ref{simenvs}, we provide an in-depth analysis of the leading simulation environments and platforms for multi-agent cooperative decision-making, again focusing on Deep MARL-based and LLMs-based methods. 
% This section examines these environments from multiple levels, including various technical details and application scenarios.
%
Furthermore, in Section \ref{praappsMADM}, we discuss the practical applications of multi-agent decision-making systems, such as autonomous driving, UAV navigation, and collaborative robotics. 
% This section offers a detailed exploration of how these systems are applied across various fields.
%
Finally, in Sections \ref{issuesMADM} and \ref{futresMADM}, we explore the potential challenges and future research directions of multi-agent cooperative decision-making. 
% By providing this structured overview, we aim to illuminate the landscape of multi-agent learning and guide future research and development in this rapidly evolving field.

\subsection{How to read this survey?}
% 对于XXX感兴趣的读者，可以refer Sec XXX.
This survey caters to a diverse readership, each with varying levels of expertise and interest in different aspects of multi-agent decision-making systems. 
% Below, we provide guidance on navigating the content based on your specific focus areas:
To help readers efficiently find the content that interests them, we offer the following guide, providing direction based on different topics:

\begin{itemize}
\item For those interested in rule (fuzzy logic)-based, game theory-based, and evolutionary algorithm-based decision-making research, please refer to Section \ref{rulexx}, \ref{gamexx}, and \ref{EAxxx}. This section provides a comprehensive analysis of the rule and game-based methods in multi-agent systems, detailing their corresponding technological taxonomies, features, and limitations.
% \item If you are interested in the basic concepts and foundational knowledge of multi-agent decision-making, we recommend starting with Section \ref{resbackMADK}. This section introduces the core principles of multi-agent decision-making, including the evolution from single-agent to multi-agent research, the Technological Comparisons between Single-Agent and Multi-Agent. %  and the key challenges involved

% , which provides an in-depth exploration of various simulation platforms used for multi-agent decision-making, analyzing these environments in terms of task formats, reward allocation, and the underlying technologies that support these systems.
\item For those interested in MARL-based decision-making research, please refer to Section \ref{MARLtaxonomies}. This section provides a comprehensive analysis of the deep MARL-based methods in multi-agent systems, detailing their corresponding technological taxonomies, advantages, and limitations.
\item If you are focused on decision-making based on LLMs, Section \ref{LLMtaxonomies} will offer you an in-depth exploration, with the corresponding technological taxonomies, advantages, and limitations. This part discusses the unique capabilities of LLMs in multi-agent environments and their potential applications, especially in reasoning and decision-making.
\item For readers focused on the well-known simulation environments of MAS, we suggest reading Section \ref{simenvs}, which primarily covers an introduction to MARL-based Simulation Environments (Section \ref{marlsimenvs}) and LLMs Reasoning-based Simulation Environments (Section \ref{llmsimenvs}).
\item If your interest lies in the practical applications of multi-agent decision-making systems, Section \ref{praappsMADM} will be of particular relevance. This section offers a detailed discussion of how these systems are applied across various fields, such as autonomous driving, UAV navigation, and collaborative robotics.
\item If you are interested in the challenges and problems faced by existing multi-agent decision-making methods, Section \ref{issuesMADM} provides an in-depth discussion, exploring the limitations of current approaches and unresolved issues in the field, offering insights into these challenges.  % difficulties
\item Finally, if you wish to learn about future research directions and the prospects for multi-agent decision-making technique, we recommend reading Section \ref{futresMADM}. This section looks ahead to future research trends and potential breakthroughs, exploring key directions that could drive the field forward.  % the development of 
\end{itemize}

\section{Multi-Agent Decision-Making Taxonomies} % 分类讲解 % Method
%然后介绍MARL的分类 
\label{madktax}
% 在这里讲一个总结，Multi-Agent Decision-Making 的很多种分类，然后细化到每个子章节，并且要提及5~7章节。
% This section focuses on the taxonomies of decision-making in multi-agent systems, aiming to provide a systematic framework and technical support for the design and optimization of such systems. 
This section discusses the taxonomies of decision-making in multi-agent systems and their related techniques.
The multi-agent cooperative decision-making methods can be broadly classified into five categories: rule-based (primarily fuzzy logic), game theory-based, evolutionary algorithms-based methods, MARL-based approaches, and LLMs-based methods \cite{ZHAO2023126708}.
Although these rule-based, game theory-based, and evolutionary algorithms-based solutions demonstrate a degree of effectiveness, they typically rely heavily on pre-designed strategies and assumptions. This dependence limits their adaptability to changing and complex environments and ill-suited for handling highly dynamic and uncertain scenarios.
In contrast, DRL-based and LLMs reasoning-based solutions offer more dynamic and flexible approaches, capable of learning and adapting to new strategies on the fly.
Therefore, these methods have significant advantages in dealing with dynamic and uncertain environments.
Thus, special research attentions are placed on DRL-based and LLMs-based methods due to their significant advantages in handling dynamic and uncertain environments.

% The analysis is conducted from multiple perspectives, including agent interaction dynamics, mainstream paradigms of cooperative decision-making, MARL (multi-agent reinforcement learning), and LLM (large language model)-driven multi-agent systems, aiming to provide a systematic framework and technical foundation for the design and optimization of multi-agent decision-making.

Specifically, 
% Section \ref{agentintera} analyzes agent interaction dynamics in MAS, categorizing them into four typical types: fully cooperative, fully competitive, mixed cooperative-competitive, and self-interested, while discussing their impact on overall system behavior. Subsequently, 
Sections \ref{mainparamulti}, \ref{MARLtaxonomies}, and \ref{LLMtaxonomies} introduce mainstream paradigms of cooperative decision-making, MARL-based decision-making methods, and LLMs-based multi-agent systems, respectively. 

\subsection{Mainstream Paradigms of Multi-Agent Cooperative Decision-Making}
\label{mainparamulti}
% Learning
% % game theory: Guo gametheory8781205 / Lanctot gametheoryLanctot / Wang wang2019gametheoretic
In multi-agent cooperative decision-making, several mainstream paradigms exist, each leveraging different techniques to tackle challenges associated with coordination, learning, adaptability, and optimization among autonomous agents. These paradigms utilize diverse approaches, including rule-based  (primarily fuzzy logic) systems \cite{Miki5665369,Yarahmadi10570102,Wu1999,Zhang6762934,Ren4276424,Gu1635080}, game theory-based \cite{Wang9763486,gametheory8781205,Schwung9152119,wang2019gametheoretic,Lin8217366,Wang10143172}, evolutionary algorithms-based \cite{Larry1998evoluation,Liu2010evoluation,Daan2015evoluation,Daan2021conevolution,yuan2024evoagentautoma,Zhang2024Evolutionary}, MARL-based \cite{QMIX,KRAEMER2016CTDEori,FastQMIX,QTRAN,WeightedQMIX,Kurach2020gfootball,ICLRQPLEX}, and LLMs-based \cite{li2024rescueLLM,Wang2024LLMsurvey,xu24pmlrLLMWerewolf,mordatch2017emergence} multi-agent decision-making systems. 
% hybrid intelligent systems % fuzzy logic-based \cite{},
Table \ref{table_maincate} provides an overview of five key methodologies used in multi-agent decision-making systems. 
% These methods contribute unique strengths to the field. 
Each of these methods has distinct strengths and applications, depending on the problem context and the complexity of interactions between agents. RL serves as a foundational approach, optimizing agent strategies through iterative learning. LLMs enhance communication and strategic planning, often integrating with RL for improved decision-making. Rule-based systems ensure transparency and consistency, making them ideal for applications with strict protocol adherence. Game-theoretic models enable the analysis of strategic interactions, improving decision-making in competitive environments. Evolutionary algorithms simulate natural selection, optimizing agent strategies dynamically. The interplay between these methodologies highlights their complementary nature, where hybrid approaches often yield the most effective solutions in complex, real-world applications.
Furthermore, Table \ref{table_mainpara} presents a comprehensive overview of existing methods within these paradigms, offering readers a detailed and structured reference.

\begin{table*}[ht]
\renewcommand{\arraystretch}{1.5}
\caption{Mainstream Multi-Agent Decision-Making Systems’ Categories, and their Characteristics, Features.}
    \label{table_maincate}
    \centering
    % \footnotesize  % \large \scriptsize
    % \setlength{\tabcolsep}{1.25mm}{
        \scalebox{0.85}{
\begin{tabular}{lp{7cm}p{6cm}}
\hline
\textbf{Method Category} & \textbf{Characteristics} & \textbf{Relationships with \newline Other Multi-Agent Systems} \\ 
\hline
Reinforcement Learning (RL) & Optimizes strategies through iterative feedback and learning from the environment; suitable for dynamic and complex scenarios. & A foundational method that provides decision optimization mechanisms for other approaches. \\ 
\hline
Large Language Models (LLMs) & Enhances communication and strategic planning; strong natural language processing capabilities improve predictive analytics. & Often integrated with RL to improve decision-making and interaction while addressing the rigidity of rule-based systems. \\ 
\hline
Rule-Based Systems & Provides structured decision-making frameworks ensuring transparency and consistency, particularly in environments requiring strict protocol adherence. & Can be combined with LLMs to create hybrid systems that balance structure with flexibility. \\ 
\hline
Game Theory-Based Models & Analyzes strategic interactions among agents to optimize decision-making in competitive environments. & Works with RL to improve equilibrium computation and integrates with LLMs to enhance strategic interaction efficiency. \\ 
\hline
Evolutionary Algorithms & Simulates natural selection to optimize strategies, improving adaptability and robustness in dynamic environments. & Frequently integrated with RL and LLMs, especially for decision optimization in dynamic and competitive settings. \\ 
\hline
\end{tabular}
    }
% }
\end{table*}

% Below, each paradigm is detailed, beginning with rule-based approaches, which predominantly utilize fuzzy logic.

\begin{table*}[ht]
\renewcommand{\arraystretch}{1.5}
\caption{Representative Methods in Mainstream Paradigms of Multi-Agent Cooperative Decision-Making.}
    \label{table_mainpara}
    \centering
    % \footnotesize  % \large \scriptsize
    % \setlength{\tabcolsep}{1.25mm}{
        \scalebox{0.85}{
\begin{tabular}{cp{12cm}} % {|c|p{12cm}|}
        \hline
        \textbf{Paradigm} & \textbf{Representative Methods and Key References} \\
        \hline
        Rule-Based (Primarily Fuzzy Logic) & 
        Miki et al. \cite{Miki5665369}, Yarahmadi et al. \cite{Yarahmadi10570102}, Wu et al. \cite{Wu1999}, Zhang et al. \cite{Zhang6762934}, Ren et al. \cite{Ren4276424}, Gu et al. \cite{Gu1635080}, Schwartz et al. \cite{Schwartz9002707}, Harmati et al. \cite{Harmati4777014}, Khuen et al. \cite{Khuen4077713}, Yan et al. \cite{Yan10297417}, Vicerra et al. \cite{Vicerra7372985}, Gu et al. \cite{Gu1626787}, Maruyama et al. \cite{Maruyama9494454}, Peng et al. \cite{Peng4621748}, Yang et al. \cite{Yang4058888} \\
        \hline
        Game Theory-based & 
        Wang et al. \cite{Wang9763486}, Guo et al. \cite{gametheory8781205}, Schwung et al. \cite{Schwung9152119}, Wang et al. \cite{wang2019gametheoretic}, Lin et al. \cite{Lin8217366}, Wang et al. \cite{Wang10143172}, Wang et al. \cite{Wang10126545}, Lanctot et al. \cite{gametheoryLanctot}, Guo et al. \cite{GuoGameTheory}, Zhang et al. \cite{Zhang8764909}, Kong et al. \cite{Kong8484212}, Wang et al. \cite{Wang10295775}, Dong et al. \cite{Dong9257354}, Nguyen et al. \cite{Nguyen6471883}, Schwung et al. \cite{Schwung9152119}, Khan et al. \cite{Khan4228007} \\
        \hline
        Evolutionary Algorithms-based & 
        Liu et al. \cite{Liu2010evoluation}, Xu et al. \cite{Xu2022evoluation}, Daan et al. \cite{Daan2015evoluation}, Franciszek et al. \cite{Franciszek2022volutionary}, Larry et al. \cite{Larry1998evoluation}, Daan et al. \cite{Daan2021conevolution}, Liu et al. \cite{Liu2024evolutionlarge}, Yuan et al. \cite{yuan2024evoagentautoma}, Dong et al. \cite{Dong2024Evolutionary}, Chen et al. \cite{Chen2023evolution}, Zhang et al. \cite{Zhang2024Evolutionary} \\
        \hline
        MARL-based & 
        Wai et al. \cite{wai2018MARL}, Hu et al. \cite{hu2024pdmultiagent}, Son et al. \cite{QTRAN}, Yu et al. \cite{FastQMIX}, Rashid et al. \cite{QMIX}, Rashid et al. \cite{WeightedQMIX}, Sunehag et al. \cite{VDN}, Huang et al. \cite{collabrobots9107997}, Xu et al. \cite{xu24pmlrLLMWerewolf}, Yun et al. \cite{quantumMetaMARL}, Mao et al. \cite{MetaMARL}, Kraemer et al. \cite{KRAEMER2016CTDEori}, Kouzeghar et al. \cite{kouzeghar2023UAVpursuit}, Gao et al. \cite{gaoMaCA}, Liu et al. \cite{liu2024diversedriving}, Qi et al. \cite{qimilitaryMARL}, Vinyals et al. \cite{Vinyals2019sc2}, Lu et al. \cite{LuMARLautoDriving}, Chu et al. \cite{ChuMARLonTrafficSingal}, et al. \cite{UAVNavigation1}, Kurach et al. \cite{Kurach2020gfootball}, Lv et al. \cite{LVMARLonUAVswarm}, Radac et al. \cite{reinforcementDQNIJSS}, Wang et al. \cite{ICLRQPLEX}, Liu et al. \cite{liu2024grounded} \\
        \hline
        LLMs-based & 
        Mordatch et al. \cite{mordatch2017emergence}, Zhang et al. \cite{zhang2024building}, Xu et al. \cite{xu24pmlrLLMWerewolf}, Li et al. \cite{li2024rescueLLM}, Wang et al. \cite{Wang2024LLMsurvey}, Zhao et al. \cite{ZHAOJINPromptMR}, Hou et al. \cite{hou2024coact}, Puig et al. \cite{Puig2018VirtualHomeSH,puig2021watchandhelp}, Gao et al. \cite{gao2024agentscope}, Xiao et al. \cite{xiao2024chainofexperts}, Wang et al. \cite{wang2024mllmtool}, Wu et al. \cite{wu2023autogen}, Wen et al. \cite{MAT}, Chen et al. \cite{chen2024agentverse}, Liu et al. \cite{liu2024dynamic}, Chen et al. \cite{AutoAgents2024ijcai}, Hong et al. \cite{hong2024metagpt,hong2024data}, XAgent Team \cite{xagent2023}, Wang et al. \cite{wang2024LangGraph,wangJ2024LangGraph}, Zheng et al. \cite{zheng2024planagent}, Zhang et al. \cite{zhang2025crewai,duan2024crewaiLangGraph}, Cao et al. \cite{LLMmarlsurvey} \\ %  \textcolor{red}{TODO.} , Hou et al. \cite{hou2024coact}  Zhang et al. \cite{zhang2024building},
        \hline
    \end{tabular}
    }
% }
\end{table*}

% vinyals2017starcraftiiPYSC2   

\subsubsection{Rule-based}  % Fuzzy logic-based
\label{rulexx}
% et al. \cite{Ishibuchi1006642} examines the adaptability of fuzzy rule-based systems to gradual and sudden changes in the environment of a market selection game.
Rule-based Multi-Agent decision-making Systems (MAS) have emerged as a critical paradigm in artificial intelligence, offering structured decision-making frameworks across various domains
% These methods have been widely adopted in multi-agent systems (MAS) due to its ability to handle uncertainty, imprecise data, and dynamic environments 
\cite{Marti5285014,NekhaiFuzzy,Ren4276424,Ramezani4530444}. 
% Fuzzy logic enables agents to make adaptive, human-like decisions by mapping inputs to linguistic rules rather than strict mathematical models.
These systems enhance transparency, reliability, and interpretability, making them essential for high-stakes applications. Table~\ref{tabruleclassification} presents a structured classification of rule-based MAS, detailing their characteristics, advantages, and representative applications.

Despite their advantages, rule-based MAS face challenges in adaptability and scalability, necessitating hybrid approaches that integrate machine learning, fuzzy logic, and expert-driven methodologies. This study categorizes rule-based MAS into five primary types based on their underlying mechanisms and application domains. The classification aims to provide insights into the strengths and limitations of each approach, aiding researchers and practitioners in selecting the most suitable methodology for their respective applications.

\begin{table*}[ht]
\renewcommand{\arraystretch}{1.5}
\caption{Detailed Classification of Rule-Based Multi-Agent Cooperative Decision-Making.}
    \label{tabruleclassification}
    \centering
    % \footnotesize  % \large \scriptsize
    % \setlength{\tabcolsep}{1.25mm}{
        \scalebox{0.85}{
    \begin{tabular}{m{2.5cm}m{6cm}m{9cm}}
        \hline
        \textbf{Category} & \textbf{Characteristics} & \textbf{Representative Applications} \\
        \hline
        
        \textbf{Fuzzy Logic-Based MAS} & 
        - Uses fuzzy logic to handle uncertainty and imprecise data. \newline
        - Enhances decision-making in complex environments. \newline
        - Provides flexible reasoning mechanisms. &
        - \textit{Fuzzy Game-Theoretic Modeling of a Multi-Agent Cybersecurity Management System for an Agricultural Enterprise} \cite{NekhaiFuzzy} \newline
        - \textit{A Fuzzy-Based Approach for Partner Selection in Multi-Agent Systems \cite{Ren4276424}} \newline
        - \textit{Select Reliable Strategy in Multi-Agent Systems Using Fuzzy Logic-Based Fusion} \cite{Ramezani4530444} \\
        
        \hline

        \textbf{Reinforcement Learning-Integrated Rule-Based MAS} & 
        - Combines rule-based decision-making with reinforcement learning. \newline
        - Improves adaptability in dynamic environments. \newline
        - Balances predefined rules with learned policies. &
        - \textit{Improving the Resource Allocation in IoT Systems Based on the Integration of Reinforcement Learning and Rule-Based Approaches in Multi-agent Systems} \cite{Yarahmadi10570102} \newline
        - \textit{Fuzzy Q-Learning-Based Multi-agent System for Intelligent Traffic Control by a Game Theory Approach} \cite{Daeichian2018} \newline
        - \textit{Fuzzy Multi-Agent Cooperative Q-Learning} \cite{Gu1635080} \\
        
        \hline

        \textbf{Expert System-Based MAS} & 
        - Incorporates expert knowledge into MAS. \newline
        - Provides domain-specific reasoning. \newline
        - Ensures high accuracy in specialized fields. &
        - \textit{A Rule-Based Multi-Agent System for Road Traffic Management} \cite{Marti5285014} \newline
        - \textit{A Rule-Based Multi-Agent System for Testing Distributed Applications} \cite{Charaf6320205} \\
        
        \hline

        \textbf{Hybrid Intelligence MAS} & 
        - Integrates symbolic reasoning with machine learning. \newline
        - Enables self-learning and adaptation. \newline
        - Enhances MAS robustness and efficiency. &
        - \textit{A Multiple Level MIMO Fuzzy Logic Based Intelligence for Multiple Agent Cooperative Robot System} \cite{Vicerra7372985} \newline
        - \textit{A Reasoning System for Fuzzy Distributed Knowledge Representation in Multi-Agent Systems} \cite{Maruyama9494454} \newline
        - \textit{A Coordination Model Using Fuzzy Reinforcement Learning for Multi-Agent System} \cite{Peng4621748} \\
        
        \hline

        \textbf{Ensemble Learning-Based MAS} & 
        - Uses ensemble techniques to enhance decision-making accuracy. \newline
        - Integrates multiple models for robustness. \newline
        - Optimizes policy learning for collaborative agents. &
        - \textit{A Contrastive-Enhanced Ensemble Framework for Efficient Multi-Agent Reinforcement Learning} \cite{DU2024123158} \newline
        - \textit{Decentralized Ensemble Learning Based on Sample Exchange among Multiple Agents} \cite{YongDecentralizedEns} \newline
        - \textit{Explainability and Interpretability of an Ensemble Multi-agent System for Supervised Learning} \cite{BlancoEnsemble} \newline
        - \textit{A Multiagent Fuzzy Policy Reinforcement Learning Algorithm with Application to Leader-Follower Robotic Systems} \cite{Yang4058888} \\
        
        \hline
    \end{tabular}
        }
% }
\end{table*}

\begin{figure}[!t]
\centering
\includegraphics[width=0.95\linewidth]{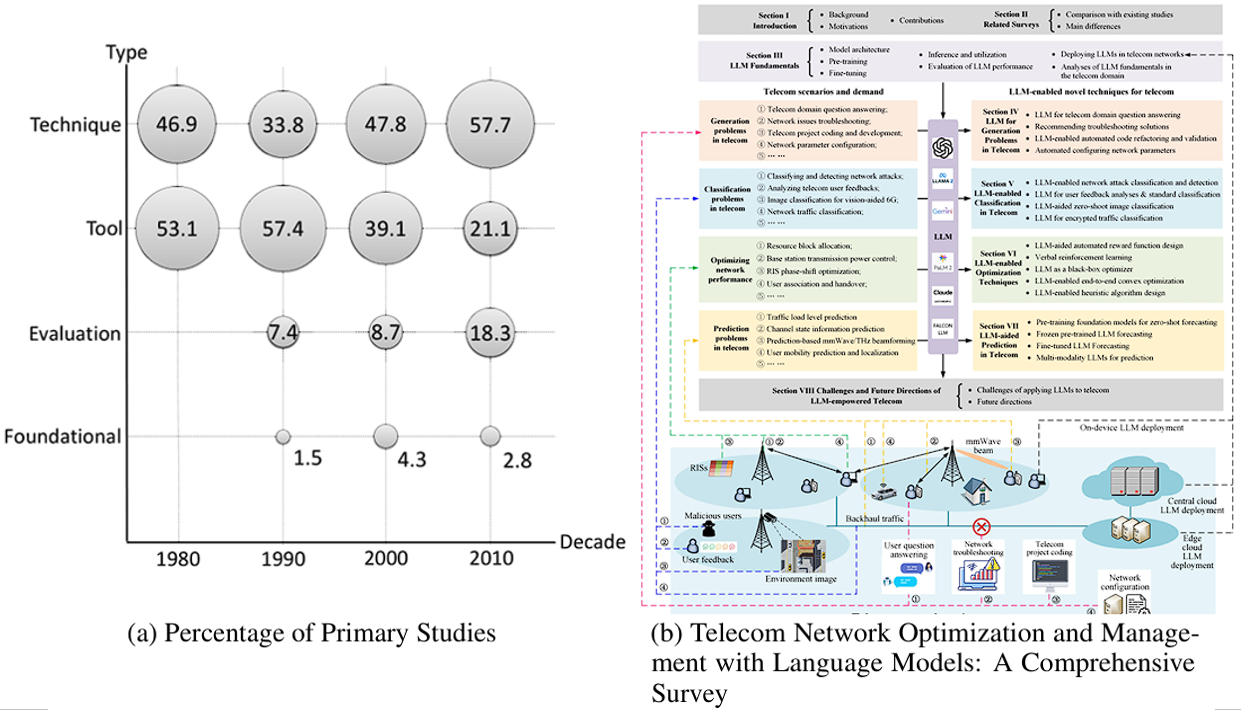}
\caption{Typical Examples of Applications of Rule-Based Systems \cite{Nunes2017,zhou2024modelllm}.}  % extensive and
\label{fig_RuleMAS}
\end{figure}

Rule-based systems are widely applied across domains, providing structured and reliable decision making frameworks. 
As shown in Figure~\ref{fig_RuleMAS}, the application of rule-based systems spans diverse fields. The analysis of "Percentage of Primary Studies" offers insights into evolving research focuses, while the survey on "Telecom Network Optimization and Management with Language Models" highlights the integration of LLMs in telecom operations. These examples underscore the versatility and applicability of rule based systems across domains, from academic research to technological implementations \cite{Nunes2017,zhou2024modelllm}.
In healthcare, systems like IvyGPT improve medical diagnostics by enhancing contextual understanding and reliability \cite{WangRSIvyGPT}. In industrial automation, frameworks such as Agents4PLC automate PLC code generation and verification, enhancing efficiency and safety \cite{hassani2024enhancinglegalc,liu2024agents4plc}.
Miki et al. \cite{Miki5665369} presented a rule-based multi-agent control algorithm that utilizes local information instead of absolute coordinates, making it more practical for real-world applications.
Charaf et al. \cite{Charaf6320205} introduced a rule-based multi-agent system to address coordination challenges, such as controllability and observability, in distributed testing environments.
Yarahmadi et al. \cite{Yarahmadi10570102} reviewed the applications of multi-agent systems in Cyber-Physical Systems (CPS) and the Internet of Things (IoT), proposing a combination of learning and rule-based reasoning to improve decision-making in MAS.
Marti et al. \cite{Marti5285014} presented an expert rule-based system using multi-agent technology to support traffic management during weather-related issues.
Daeichian et al. \cite{Daeichian2018} used fuzzy logic in combination with Q-learning and game theory to control traffic lights autonomously. 
% The fuzzy Q-learning approach allows each agent to make decisions based on both past experiences and neighboring agents' actions, helping to adapt to the dynamic nature of traffic networks.
% Wu et al. \cite{Wu1999} introduced a fuzzy-theoretic game framework that integrates fuzzy logic with game theory to handle uncertainty in utility values during multi-agent decision making. 
% It defines concepts like fuzzy dominant relations, fuzzy Nash equilibrium, and fuzzy strategies, allowing agents to make decisions under conditions of vague or imprecise utilities.
Nekhai et al. \cite{NekhaiFuzzy} devised a cybersecurity management model for agricultural enterprises using a multi-agent system (MAS) based on fuzzy logical reasoning. 
% It integrates Cyber Situational Awareness to model interactions between attack and protection agents, aiming to develop winning strategies rather than the best solution.
% Ramezani et al. \cite{Ramezani4530444} applied fuzzy logic to multi-agent decision-making in soccer robot teams, combining cooperative and non-cooperative game strategies. 
% Fuzzy Integral Operators (FIO) are used to integrate sensory data, allowing flexible adjustments in team strategy based on real-time game dynamics.
Zhang et al. \cite{Zhang6762934} introduced a new online method for optimal coordination control in multi-agent differential games, combining fuzzy logic, and adaptive dynamic programming. 
% The approach uses a Generalized Fuzzy Hyperbolic Model (GFHM) to approximate the value functions of Hamilton–Jacobi equations, simplifying the process with a single-network architecture for each agent.
% Ren et al. \cite{Ren4276424} presented a fuzzy logic-based approach for partner selection in multi-agent systems, emphasizing flexibility and adaptability in dynamic environments. % By combining fuzzy logic with the extended dual concern model, agents can balance their own benefits with those of potential partners, considering attitudes and negotiation outcomes.
Gu et al. \cite{Gu1635080} introduced a cooperative reinforcement learning algorithm for multi-agent systems using a leader-follower framework, modeled as a Stackelberg game. 
% The leader-follower Q-learning algorithm is extended to continuous state spaces with fuzzy logic integration.
Schwartz et al. \cite{Schwartz9002707} introduced a multi-agent fuzzy actor-critic learning algorithm for differential games. 
% It uses an object-oriented approach to model agent relationships and defines the fuzzy inference system as a network structure with agent attributes represented as rule sets.
Harmati et al. \cite{Harmati4777014} proposed a game-theoretic model for coordinating multiple robots in target tracking, using a semi-cooperative Stackelberg equilibrium and a fuzzy inference system for high-level cost tuning.
% Khuen et al. \cite{Khuen4077713} introduced an Adaptive Fuzzy Logic (AFL) approach for multi-agent systems with negotiation capabilities, focusing on resource allocation.
% The AFL model enables agents to learn and adapt to the behavior of other agents during negotiations involving multiple issues like price and time.
Yan et al. \cite{Yan10297417} proposed a graphical game-based adaptive fuzzy optimal bipartite containment control scheme for high-order nonlinear multi-agent systems (MASs). % It simplifies the existing approaches by removing the identifier-actor-critic structure and integrating fuzzy logic with Nash equilibrium-seeking via integral reinforcement learning.
Vicerra et al. \cite{Vicerra7372985} proposed a multi-agent robot system using a pure fuzzy logic-based artificial intelligence model. 
%The model features multiple fuzzy logic levels in parallel and series configurations, with a MIMO fuzzy logic block for full AI control.
% Gu et al. \cite{Gu1626787} presented a fuzzy logic-based policy gradient multi-agent reinforcement learning algorithm for leader-follower systems, where fuzzy logic controllers act as policies.
Maruyama et al. \cite{Maruyama9494454} extended the classical framework for reasoning about distributed knowledge, incorporating fuzzy logic to handle uncertainty and degrees of certainty within multi-agent systems. 
%Building on the work of Fagin, Halpern, Moses, and Vardi, they introduced fuzzy distributed knowledge and prove several important properties of fuzzy modal systems.
Peng et al. \cite{Peng4621748} proposed a two-layer coordination model for multi-agent systems using fuzzy reinforcement learning. 
%In the architecture, agents employ a fuzzy inference system to select optimal local behaviors and consider the intentions and actions of others.
Yang et al. \cite{Yang4058888} presented a multi-agent reinforcement learning algorithm with fuzzy policy to address control challenges in cooperative multi-agent systems, particularly for autonomous robotic formations. 
% The algorithm fine-tunes the parameters of the fuzzy policy using gradient-based reinforcement learning, enhancing the performance of an initial controller.

As depicted from Table~\ref{tabruleclassification}, the categorization of rule-based MAS highlights the diverse methodologies employed in enhancing multi-agent decision-making. Each category presents unique advantages and trade-offs:

\begin{itemize}
    \item \textbf{Fuzzy Logic-Based:} Particularly effective in scenarios requiring nuanced decision-making under uncertainty. These systems excel in cybersecurity, partner selection, and industrial automation.
    \item \textbf{Reinforcement Learning-Integrated Rule-Based MAS:} Provides a balance between structured rule enforcement and adaptive learning, making it suitable for applications such as traffic control and IoT resource allocation.
    \item \textbf{Expert System-Based:} Ensures high reliability and accuracy in domain-specific applications by leveraging expert knowledge. Widely used in fields requiring precise control, such as road traffic management.
    \item \textbf{Hybrid Intelligence:} Merges symbolic reasoning with learning-based models, enhancing adaptability. This category is particularly useful for complex cooperative robotic systems and distributed knowledge representation.
    \item \textbf{Ensemble Learning-Based:} Optimizes multi-agent collaboration through ensemble learning techniques, offering improved policy robustness and decision-making precision.
\end{itemize}

Rule-based systems ensure consistent and predictable agent behavior, enhancing the robustness and reliability of collaborative efforts. This consistency is vital for effective decision-making, reducing risks associated with unpredictability and fostering synergy among agents \cite{Nunes2017,zhou2024modelllm,liu2024agents4plc,hassani2024enhancinglegalc}. Rule-based systems provide a reliable benchmark for evaluating model efficacy, particularly in high baseline performance scenarios, as seen with evaluation frameworks like SciAssess \cite{cai2024sciassess}. Their transparency and interpretability facilitate clear decision-making processes, simplifying auditing and modifications, which is crucial in applications requiring accountability \cite{jing2024translating}.

Moreover, the integration of rule-based systems with methodologies like LLMs and deep reinforcement learning creates hybrid systems that enhance robustness and flexibility, addressing challenges across domains \cite{WangRSIvyGPT}. Despite their advantages, rule-based systems face limitations in incorporating nuanced human expertise and adapting to dynamic environments, necessitating ongoing refinement to maintain effectiveness \cite{scherbakov2024emergence,pezeshkpour2024multiconditional,jing2024translating}.

Overall, rule-based decision-making remains a foundational approach in MAS, offering interpretability and robustness in uncertain environments. 
% Despite their effectiveness, rule-based MAS face limitations in handling dynamic and evolving environments. 
% The integration of large language models (LLMs) and advanced machine learning techniques presents a promising direction for future research. 
By combining structured rule-based approaches with adaptive learning mechanisms, MAS can achieve greater scalability and intelligence, paving the way for more robust and autonomous systems.
In the future, rule-based decision-making systems will be further integrated with LLMs, hierarchical decision architectures, and multi-agent planning, enabling more precise and adaptive decision-making in complex real-world scenarios.

\subsubsection{Game Theory-based}
\label{gamexx}

Game theory provides a structured framework for analyzing strategic interactions in multi-agent systems. It enables agents to make rational decisions in cooperative, competitive, or mixed scenarios through equilibrium-based optimization \cite{Li2022Survey,Nguyen2020survey}. Traditional methods such as Nash equilibrium and Stackelberg games form the foundation, while modern approaches integrate reinforcement learning and Bayesian inference to enhance adaptability in dynamic environments.

Wang et al. \cite{Wang9763486} provided a broad discussion on game-theoretic approaches in multi-agent systems, covering cooperative and non-cooperative scenarios. Guo et al. \cite{gametheory8781205} applied game theory to multi-agent path planning, leveraging Nash equilibrium to optimize navigation and obstacle avoidance. Zhang et al. \cite{Zhang8764909} developed a distributed control algorithm that ensures optimal coverage while maintaining network connectivity.

Beyond fundamental decision-making, game theory has been applied in communication networks and energy systems. Wang et al. \cite{wang2019gametheoretic} utilized game-theoretic learning to enhance resource allocation in wireless networks while countering adversarial actions like jamming. Lin et al. \cite{Lin8217366} introduced potential game theory to optimize distributed energy management in microgrids, where agents autonomously coordinate power distribution. Dong et al. \cite{Dong9257354} further extended this approach using a hierarchical Stackelberg model for energy trading, balancing incentives between microgrids and individual agents.

Incorporating machine learning with game theory has also led to advances in multi-agent optimization. Schwung et al. \cite{Schwung9152119} combined potential game theory with reinforcement learning for adaptive production scheduling, while Wang et al. \cite{Wang10295775} designed a Nash equilibrium-based fault-tolerant control strategy for multi-agent systems. Additionally, game-theoretic methods have been explored for distributed computing, as shown by Khan et al. \cite{Khan4228007}, who developed a replica placement strategy to minimize data access delays in distributed systems.

Overall, game theory remains a cornerstone of multi-agent decision-making, offering well-defined theoretical guarantees while enabling dynamic adaptation through hybrid approaches. Future research will likely focus on integrating game theory with deep learning and large language models to enhance strategic reasoning in high-dimensional, uncertain environments.

\subsubsection{Evolutionary Algorithms-based}
\label{EAxxx}

Evolutionary algorithms (EAs) have become a robust framework for optimizing strategies in multi-agent systems (MAS) by drawing inspiration from natural evolution. It provides a bio-inspired approach to optimization in multi-agent systems by leveraging principles such as natural selection, mutation, and recombination \cite{Franciszek2022volutionary,Daan2021conevolution,Dong2024Evolutionary}.
By allowing agents to evolve their strategies iteratively, EAs are particularly effective for problems requiring continuous learning, large-scale coordination, and self-organized behavior.

Table~\ref{tabEAclassification} presents a structured classification of evolutionary algorithms-based MAS (focusing on the specific dimensions: optimization strategies and learning methodologies), detailing their characteristics, and representative applications. 

The first category, GA-based MAS, employs genetic algorithms (GAs) that simulate population-level evolution through selection, mutation, and crossover \cite{Xu2022Liying,JIN2025100645,6721993,AKOPOV2019103}. These methods \cite{Liu2010evoluation,Xu2022evoluation,Larry1998evoluation} are particularly effective for continuous learning scenarios and large-scale coordination, as they enable agents to explore vast and complex solution spaces adaptively. 

Deep Neuroevolution Methods represent an advanced approach by integrating deep neural networks with evolutionary strategies \cite{ZiyuanMultiagent,10152759,Daan2021conevolution}. These methods \cite{such2018deepneuro,stember2021deepneuro} are designed for high-dimensional optimization problems, where conventional reinforcement learning might struggle. The deep neuroevolution paradigm enhances exploration and accelerates convergence in dynamic environments.

Coevolutionary Algorithms take a more interactive approach, evolving multiple agents simultaneously while considering their interdependencies \cite{ZHOU2024103676,Dorronsoro2013,Chatziagorakis,Daan2015evoluation}. This leverages competitive and cooperative interactions, often integrating game theory or cellular automata to simulate real-world dynamics.

Hybrid Evolutionary Reinforcement Learning Methods combine the strengths of evolutionary algorithms and reinforcement learning \cite{yuan2024evoagentautoma,Liu2024evolutionlarge,yuan2024evoagentautoma}. By partitioning the learning process into stages, these approaches enhance scalability and adaptability, particularly in environments characterized by noise and uncertainty \cite{10637292,Dong2024Adaptive,Lin2024ERL}. 

Lastly, Evolutionary Game Theory-based Approaches utilize game-theoretic models to promote cooperation and competition among agents \cite{Zhang2024Evolutionary,NekhaiFuzzy,PARESCHI2017201}. These methods often incorporate dynamical systems concepts, such as kinetic models based on Boltzmann equations, to balance local and global objectives in policy optimization \cite{Franciszek2022volutionary,Chen2023evolution}.

\begin{table*}[ht]
\renewcommand{\arraystretch}{1.5}
\caption{Detailed Classification of Evolutionary Algorithms-Based Multi-Agent Cooperative Decision-Making.}
    \label{tabEAclassification}
    \centering
        \scalebox{0.85}{
\begin{tabular}{p{2.5cm}p{9.5cm}p{6cm}}
\hline
\textbf{Category} & \textbf{Key Characteristics} & \textbf{Representative Methods and References} \\
\hline
\textbf{Pure GA-based MAS} & Population-based search leveraging natural selection, mutation, and crossover. Suitable for continuous learning, global optimization, and large-scale coordination, enabling adaptive exploration of complex solution spaces \cite{Liu2010evoluation,Xu2022evoluation,Larry1998evoluation}. & - \textit{Multi-Agent Genetic Algorithm (MAGA)} \cite{6721993,AKOPOV2019103}; \newline
- \textit{Hardware-based MAS using nanoclusters for parallel computation} \cite{Xu2022Liying,JIN2025100645}
\\
\hline
\textbf{Deep- \newline Neuroevolution \newline Methods} & Integration of deep neural networks with evolutionary strategies. Emphasizes optimization in high-dimensional tasks and dynamic environments, enhancing exploration and adaptability \cite{Daan2021conevolution}. & 
- \textit{Deep Neuroevolution (DNE)} \cite{such2018deepneuro,stember2021deepneuro}; \newline
- \textit{Multi-Agent Transformer architectures for cooperative decision-making} \cite{ZiyuanMultiagent,10152759} \\
\hline
\textbf{Coevolutionary \newline Algorithms} & Simultaneous evolution of multiple agents considering both competitive and cooperative interactions. Often combined with game theory and cellular automata to simulate complex agent dynamics \cite{Daan2015evoluation}. & - \textit{Extended Coevolutionary Theory frameworks} \cite{KALLIS2010690,MADHOK20061}; \newline
- \textit{Cellular automata integrated with game theory} \cite{ZHOU2024103676,Dorronsoro2013,Chatziagorakis}
% \newline
% - \textit{Three-strategy decision model} \cite{} 
\\
\hline
\textbf{Hybrid \newline Evolutionary \newline Reinforcement Learning Methods} & Combines evolutionary algorithms with reinforcement learning. Uses staged strategy exploration and optimization to enhance scalability and adaptability in noisy and dynamic environments \cite{yuan2024evoagentautoma,Liu2024evolutionlarge}. & - \textit{Evolutionary Reinforcement Learning (ERL)} \cite{10637292,Dong2024Adaptive,Lin2024ERL}; \newline
- \textit{EvoAgent evolution framework integrating LLMs} \cite{yuan2024evoagentautoma} \\
\hline
\textbf{Evolutionary Game Theory-based \newline Approaches} & Application of game theoretic models to drive long-term cooperation and competition. Utilizes statistical physics or dynamical systems models to balance local and global interests in policy optimization \cite{Franciszek2022volutionary,Chen2023evolution,Dong2024Evolutionary,Zhang2024Evolutionary}. & 
- \textit{Kinetic decision-making models based on Boltzmann equations} \cite{PARESCHI2017201,bobylev2023boltzman,ChenOptimal}; \newline 
- \textit{Policy optimization in agricultural water conservation projects} \cite{Zhang2024Evolutionary,NekhaiFuzzy,OlofintoyeEvolutionary,NASERI2021566,Zhang2024Evolutionary}; 
% \newline
% - \textit{Combination with three-strategy models} \cite{} 
\\
\hline
\end{tabular}
                }
\end{table*}

Liu et al. \cite{Liu2010evoluation} introduced the Multi-Agent Genetic Algorithm (MAGA), where agents interact through competition and cooperation to optimize global solutions. Xu et al. \cite{Xu2022evoluation} extended this idea to hardware-based multi-agent systems, using nanoclusters as physical agents to achieve large-scale parallel computation. Daan et al. \cite{Daan2015evoluation} explored the role of evolutionary strategies in dynamic environments such as financial markets, smart grids, and robotics, demonstrating how adaptive algorithms can handle real-world uncertainties.
Franciszek et al. \cite{Franciszek2022volutionary} proposed a self-optimization model integrating cellular automata and game theory, simulating competitive evolutionary interactions among agents. Larry et al. \cite{Larry1998evoluation} analyzed the trade-offs between mutation and recombination, showing that mutation can sometimes outperform traditional recombination strategies in evolutionary computing. To further enhance adaptability, Daan et al. \cite{Daan2021conevolution} introduced Deep Neuroevolution (DNE), applying coevolutionary techniques to complex multi-agent scenarios, including Atari games.

Recent studies have focused on scaling evolutionary learning to larger agent populations. Liu et al. \cite{Liu2024evolutionlarge} developed Evolutionary Reinforcement Learning (ERL), a scalable approach that partitions learning into multiple stages, ensuring better adaptability in multi-agent environments. Yuan et al. \cite{yuan2024evoagentautoma} introduced EvoAgent, a framework that extends LLMs-based autonomous agents into multi-agent systems using evolutionary techniques like mutation and selection.

Evolutionary game theory has also been explored to improve cooperative behavior. Dong et al. \cite{Dong2024Evolutionary} designed a three-strategy decision model, where agents adopt conservative or adaptive strategies based on their interactions with neighbors, fostering long-term cooperation. Chen et al. \cite{Chen2023evolution} proposed a kinetic decision-making model grounded in rarefied gas dynamics, offering a new perspective on agent evolution using the Boltzmann equation. Zhang et al. \cite{Zhang2024Evolutionary} applied evolutionary game theory to policy optimization, analyzing cooperation strategies among governments, enterprises, and farmers in agricultural water conservation projects.

Overall, evolutionary algorithms provide a robust framework for decentralized decision-making, allowing agents to self-improve and adapt in uncertain environments. In the future, evolutionary algorithms will be further integrated with deep learning, hierarchical evolution, and large-scale multi-agent coordination, enabling more adaptive, autonomous MAS.

\begin{figure*}[!t]
\centering
\includegraphics[width=0.985\linewidth]{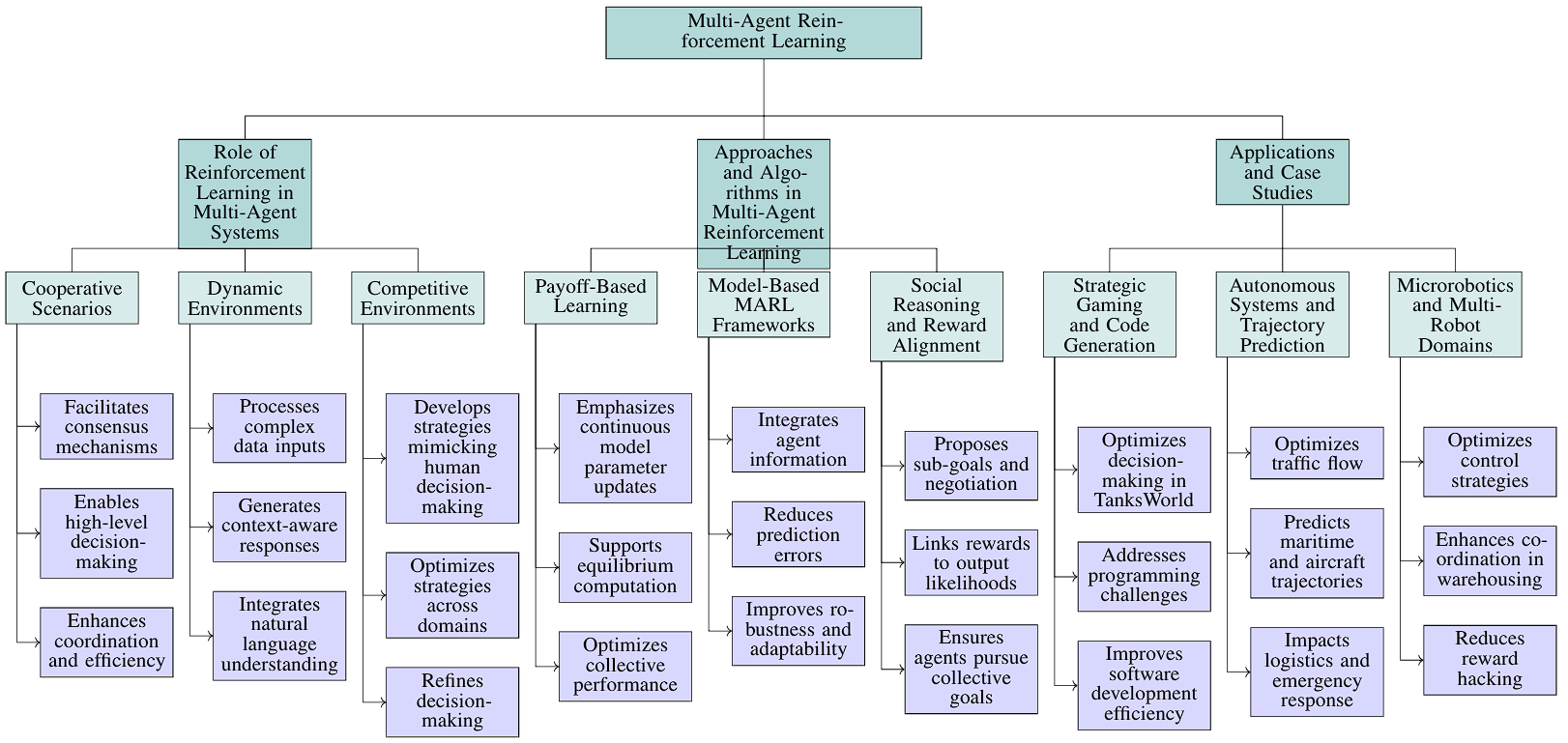}
\caption{The hierarchical structure of MARL-based MAS, detailing its roles, approaches, algorithms, and applications. It highlights MARL's contributions to cooperative, dynamic, and competitive environments, MARL methodologies like payoff-based learning and model integration, and diverse applications in strategic gaming, autonomous systems and microrobotics.}
\label{fig_MARLALL}
\end{figure*}

\subsubsection{MARL-based Multi-Agent Systems} 
The integration of Multi-Agent Reinforcement Learning (MARL) within multi-agent systems establishes a dynamic framework that enhances agents’ decision-making capabilities. In this section, we explore how MARL empowers agents to learn from environmental interactions and adapt their strategies. As illustrated in Figure~\ref{fig_MARLALL}, the hierarchical structure of MARL is depicted detailing its roles, approaches, algorithms, and applications. It highlights MARL's contributions to cooperative, dynamic, and competitive environments, as well as MARL methodologies such as payoff-based learning and model integration. Furthermore, it underscores the diverse applications of MARL in strategic gaming, autonomous systems, and microrobotics.

Before introducing the MARL-based multi-agent systems (MAS), we provide a detailed discussion in \ref{app_MDPPOMDP} on the key technological comparisons and methodological principles of both DRL-based single-agent systems and MARL-based MAS. This helps readers build the necessary background knowledge for better understanding the following discussions.

\textit{Multi-Agent Reinforcement Learning} offers a structured framework to tackle decision-making in MAS, where autonomous agents interact with each other and a shared environment.
The MAS research in MARL is broadly divided into three paradigms: \textit{Centralized Training with Centralized Execution (CTCE)} \cite{amato2024introductionDTE,zhou2023CTEenough}, \textit{Decentralized Training with Decentralized Execution (DTDE)} \cite{amato2024introductionDTE}, and \textit{Centralized Training with Decentralized Execution (CTDE)} \cite{KRAEMER2016CTDEori,amato2024CTDEsurvey}. Each paradigm is designed to address specific challenges such as coordination, scalability, and policy optimization, providing tailored solutions for diverse scenarios.

% et al. \cite{} .
% et al. \cite{} .
% et al. \cite{} .
% \cite{} .
% \cite{} .
% wai2018MARL hu2024pdmultiagent QTRAN FastQMIX WeightedQMIX QMIX VDN collabrobots9107997 KRAEMER2016CTDEori quantumMetaMARL MetaMARL xu24pmlrLLMWerewolf gaoMaCA qimilitaryMARL kouzeghar2023UAVpursuit liu2024diversedriving LuMARLautoDriving ChuMARLonTrafficSingal LVMARLonUAVswarm Kurach2020gfootball UAVNavigation1 reinforcementDQNIJSS vinyals2017starcraftiiPYSC2 Vinyals2019sc2  

\textbf{Centralized Training with Centralized Execution}
The CTCE paradigm \cite{Sharma2021CTE,zhou2023CTEenough} relies on a central controller that governs all agents by aggregating their observations, actions, and rewards to make joint decisions. While this paradigm enables high levels of coordination, its scalability is limited in large-scale systems. Multi-Agent DQN (MADQN) \cite{DQNori,mnih2013DQNv1,Mnih2015DQNv2} is a representative method, employing parameter-sharing mechanisms to handle cooperative tasks effectively. However, its reliance on centralized control restricts its applicability in dynamic environments with numerous agents.

\textbf{Decentralized Training with Decentralized Execution}
The DTDE paradigm \cite{amato2024introductionDTE} emphasizes independent learning and execution, where each agent interacts with the environment individually and updates its policy based solely on local observations and rewards. This paradigm excels in scalability and robustness, especially in scenarios with limited communication. Notable methods include \textit{Independent Q-Learning (IQL)} \cite{claus1998IQL,LauerDistributedQLearning} and \textit{Decentralized REINFORCE} \cite{BOWLING2002DecMARL}, which allow agents to learn autonomously. Despite its advantages, DTDE faces challenges such as learning non-stationarity, where the environment changes as other agents adapt, and difficulty in addressing the credit assignment problem in cooperative settings.

\textbf{Centralized Training with Decentralized Execution}
The CTDE paradigm \cite{KRAEMER2016CTDEori,CTDEMAS,zhou2023CTDEenough} combines the strengths of centralized training and decentralized execution, making it the most prominent paradigm in MARL research. During training, a central controller aggregates information from all agents to optimize their policies, but during execution, each agent operates independently based on its own observations. CTDE addresses key challenges like non-stationarity and scalability, with methods such as \textit{Value Decomposition Networks (VDN)} \cite{VDN} and \textit{QMIX} \cite{QMIX,FastQMIX} for value-based learning, \textit{Multi-Agent Deep Deterministic Policy Gradient (MADDPG)} \cite{MADDPG} for actor-critic frameworks, and \textit{Multi-Agent Proximal Policy Optimization (MAPPO)} \cite{yu2022surprising} for policy gradient optimization. These approaches are widely applied in complex environments like StarCraft II \cite{samvelyan19smac,SMACv2} and the Multi-Agent Particle Environment (MPE) \cite{lowe2017multi,MPE9619081}.

\textbf{Communication-based MARL Algorithms}
Additionally, communication-based MARL algorithms have emerged to enhance coordination by enabling agents to share critical information during training and execution. Examples include \textit{Attentional Communication (ATOC)} \cite{jiang2018learning} and \textit{Targeted Multi-Agent Communication (TarMAC)} \cite{SunaaaiT2MAC}, which use advanced mechanisms to improve the efficiency and effectiveness of inter-agent communication in cooperative tasks.

By structuring MARL methods within these paradigms, researchers provide a clear framework for addressing the diverse challenges of multi-agent decision-making. From autonomous driving fleets to resource allocation systems, MARL continues to push the boundaries of what distributed intelligent systems can achieve \cite{amato2024CTDEsurvey}.

% \textcolor{red}{TODO.}
\begin{figure*}[!t]
\centering
\includegraphics[width=0.9\linewidth]{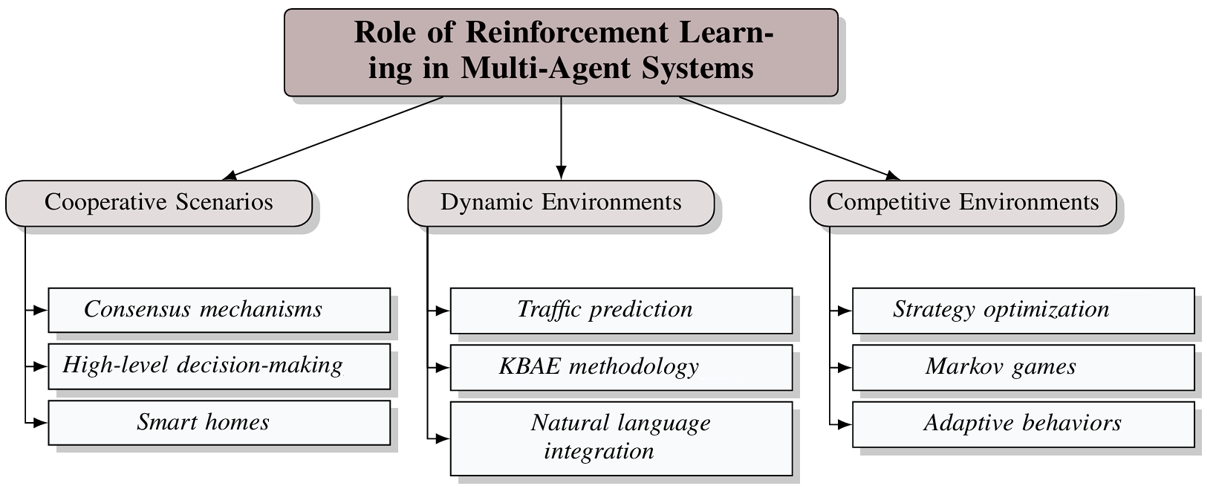}
\caption{The typical role of reinforcement learning in multi-agent systems, highlighting its applications in cooperative, dynamic, and competitive environments. It shows how MARL facilitates consensus mechanisms, optimizes strategies, and processes complex data inputs for real-time decision making.}
\label{fig_MARLadd}
\end{figure*}

\subsubsection{LLMs-based Multi-Agent Systems}
Although LLMs like GPT \cite{OpenAI2023GPT4,TrainingGPT35feedback,mao2024gptsurvey}, Llama \cite{grattafiori2024llama3herdmodels,touvron2023llama}, and Gemini \cite{geminiteam2024gemini} support very long input contexts, their ability to understand complex inputs still varies. In this context, multi-agent collaboration optimizes task execution through role assignment, enabling better performance through collaboration among multiple agents compared to a single agent. Each agent has an independent workflow, memory, and can seek help from other agents when necessary. LLMs-based Multi-Agent Systems represent a relatively new multi-agent decision-making model that leverages the powerful capabilities of language models,
% especially pre-trained models like GPT, Llama, and Gemini, 
to enhance communication and collaboration between autonomous agents. In an LLMs-based multi-agent system, agents communicate via natural language or symbolic representations, breaking down complex tasks into smaller, more manageable subtasks. 
One important feature of LLMs-based systems is the hierarchical organization of agents, typically consisting of two levels \cite{Wang2024LLMsurvey,LLMmarlsurvey}:

1) Global planning agents, responsible for high-level decisions such as task decomposition, resource allocation, and overall strategy management. 

2) Local execution agents, which are responsible for executing specific subtasks and providing feedback to the global planning agent. These agents are generally more focused on local tasks but communicate progress and challenges with the global level for adjustments. This decomposition makes distributed problem solving possible, with agents sharing information, strategies, and goals through language, thus advancing task execution together.
% In LLMs-based systems, agents are designed to continuously cooperate through natural language interactions. This communication not only aids in real-time decision-making but also enables agents to learn from each other's actions, enhancing the overall system's performance and adaptability.

For example, frameworks like AutoGen \cite{wu2023autogen,dibia2024autogen,AutoAgents2024ijcai}, Crew AI \cite{zhang2025crewai,duan2024crewaiLangGraph}, and LangGraph \cite{wang2024LangGraph,wangJ2024LangGraph} provide rich tools for building multi-agent solutions, supporting efficient cooperation and interaction between agents. Through these frameworks, developers can build virtual teams that leverage the strengths of different agents in distributed tasks. Additionally, LLMs-based multi-agent systems possess adaptive re-planning capabilities, enabling them to adjust in dynamic environments. When agents encounter changes or new information, they can quickly update strategies or re-plan tasks using language models, ensuring the system remains aligned with changing goals.

Firstly, LLMs-based multi-agent environments have emerged as pivotal platforms for advancing research in multi-agent collaboration, reasoning, and task execution. For instance, environments such as ThreeDWorld Multi-Agent Transport (TDW-MAT) \cite{ThreeDWorldv2,ThreeDWorldv1}, Communicative Watch-And-Help (C-WAH) \cite{puig2021watchandhelp}, Cuisineworld \cite{gong2024mindagent}, and AgentScope \cite{gao2024agentscope} offer diverse settings for evaluating and enhancing multi-agent systems in various contexts, from household chores to gaming interactions and beyond.
For instance, MindAgent \cite{gong2024mindagent} is a novel infrastructure for evaluating planning and coordination capabilities in gaming interactions, leveraging large foundation models (LFMs) to coordinate multi-agent systems, collaborate with human players. Communicative Watch-And-Help (C-WAH) \cite{Puig2018VirtualHomeSH,puig2021watchandhelp} is a realistic multi-agent simulation environment and an extension of the Watch-And-Help Challenge platform, VirtualHome-Social. AgentScope \cite{gao2024agentscope} is a robust and flexible multi-agent platform designed to empower developers in building advanced multi-agent systems by leveraging the potential of LLMs.

Meanwhile, LLMs-based multi-agent systems have broad applications and great prospects \cite{Wang2024LLMsurvey,LLMmarlsurvey}. They can collaborate in robotic teams to perform complex tasks, such as product assembly or joint research, ensuring seamless interaction and cooperation \cite{xu24pmlrLLMWerewolf,mordatch2017emergence,li2024rescueLLM}. In autonomous driving, LLMs help vehicles communicate, sharing real-time data and navigation strategies to achieve coordinated actions. 
% In energy management, LLMs optimize smart grids and resource allocation, improving energy efficiency and grid management capabilities. 
Moreover, LLMs can support agents (such as drones) in disaster response, transmitting critical information to help systems efficiently respond to crises.
% 
% et al. \cite{} .
% et al. \cite{} .
% et al. \cite{} .
% et al. \cite{} .
% et al. \cite{} .
% et al. \cite{} .
% et al. \cite{} .
% 
\textit{Wu et al.} \cite{wu2023autogen} proposed AutoGen, an open-source framework for developing next-generation LLM applications through multi-agent conversations, allowing customizable agent interactions and integration of LLMs, human inputs, and tools.
\textit{Xiao et al.} \cite{xiao2024chainofexperts} introduced Chain-of-Experts (CoE), a multi-agent framework that enhances reasoning in complex operations research (OR) problems using LLMs, with domain-specific roles and a conductor for guidance.
\textit{Chen et al.} \cite{chen2024agentverse} presented AgentVerse, a multi-agent framework inspired by human group dynamics, dynamically adjusting agent roles and composition to enhance complex task-solving across various domains.
\textit{Chen et al.} \cite{AutoAgents2024ijcai} developed AutoAgents, a framework that adaptively generates and coordinates multiple specialized agents for efficient task completion.
\textit{Liu et al.} \cite{liu2024dynamic} proposed Dynamic LLM-Agent Network (DyLAN), a framework that enhances LLM-agent collaboration through dynamic interactions based on task requirements.
%
% \textit{Wen et al.} \cite{MAT} introduced Multi-Agent Transformer (MAT), a novel architecture that frames cooperative multi-agent reinforcement learning (MARL) as a sequence modeling problem, achieving superior performance on various benchmarks.
%
% \textit{Wang et al.} \cite{wang2024mllmtool} developed MLLM-Tool, a multimodal tool agent system that integrates LLMs with multimodal encoders to process visual and auditory inputs, and introduced ToolMMBench for real-world multimodal multi-agent scenarios.
%
\textit{Zhang et al.} \cite{zhang2024building} introduced CoELA, a Cooperative Embodied Language Agent framework that leverages LLMs to enhance multi-agent cooperation in complex, decentralized environments.
\textit{Hong et al.} \cite{hong2024metagpt,hong2024data} proposed MetaGPT, a meta-programming framework that enhances LLMs-based multi-agent system collaboration using Standard Operating Procedures (SOPs).
\textit{XAgent Team} \cite{xagent2023} developed XAgent, an open-source, LLM-driven autonomous agent framework for solving complex tasks using a dual-loop architecture for task planning and execution.
\textit{Zheng et al.} \cite{zheng2024planagent} introduced PlanAgent, a closed-loop motion planning framework for autonomous driving using multi-modal LLMs to generate hierarchical driving commands.
\textit{Wang et al.} \cite{wang2024LangGraph,wangJ2024LangGraph} developed LangGraph, a library for building stateful, multi-actor applications with LLMs, offering fine-grained control over workflows and state management.
\textit{Zhang et al.} \cite{zhang2025crewai,duan2024crewaiLangGraph} introduced CrewAI, an open-source framework for coordinating AI agents in role-playing and autonomous operations, with a modular design for efficient collaboration.
\textit{Hou et al.} \cite{hou2024coact} proposed CoAct
% \footnote{CoAct: \url{https://github.com/dxhou/CoAct}.}
, a hierarchical multi-agent system leveraging LLMs for collaborative task execution. It features a global planning agent for task decomposition and strategy management, and a local execution agent for subtask implementation, feedback collection, and adaptive replanning, ensuring alignment with overarching goals.
% proposed CoAct, a multi-agent system based on LLMs designed for hierarchical collaboration tasks, consisting of six stages: task decomposition, subtask assignment and communication, subtask analysis and execution, feedback collection, progress evaluation, and replanning.

% \textbf{Hou et al.} \cite{hou2024coact} developed CoAct, a multi-agent system for hierarchical collaboration tasks, consisting of six stages from task decomposition to replanning.

Despite the strong capabilities of LLMs in small to medium-sized multi-agent systems, scalability remains an open issue, particularly in maintaining coherent communication between large numbers of agents in large environments. As the number of agents increases, the complexity of coordinating their behaviors through language also intensifies. Finding a balance between agent autonomy and effective collaboration is a significant challenge. Additionally, LLMs are often seen as black-box models, meaning understanding the reasoning process behind an agent's decision-making can be difficult. The lack of transparency poses challenges for trust and debugging.

In summary, LLMs-based multi-agent systems hold great potential in a variety of applications, offering an advanced way to model and solve complex decision-making problems that require high levels of coordination, adaptability, and communication between agents. By optimizing task decomposition, collaboration, and feedback mechanisms, LLMs bring unprecedented efficiency and flexibility to multi-agent systems.

% \textcolor{red}{TODO.}

\subsection{MARL-based Multi-Agent Decision-Making Taxonomies}
\label{MARLtaxonomies}
% \subsubsection{Training and Execution Ways for MARL} % 强化学习分类之二 通过是否中心和分布的分类
%% 介绍CTCE CTDE DTDE AC框架 基于策略梯度优化的框架等.
% In multi-agent systems, where multiple autonomous agents interact with a shared environment and often with each other, the complexity of decision-making increases significantly. To achieve optimal performance, agents need to learn not only how to act individually but also how to coordinate with others. 
Reinforcement Learning (RL) serves as a fundamental framework in multi-agent systems, enabling agents to learn and adapt based on environmental feedback, which is crucial in complex and dynamic settings. Advanced techniques like Multi-Agent Reinforcement Learning (MARL) model decision-making as sequence problems, optimizing agent actions through iterative learning. This MARL-based approach is vital for applications ranging from autonomous vehicles to collaborative robotics \cite{Nguyen2020survey,Zhu2024Survey,Pamul2023survey}.

As illustrated in Figure~\ref{fig_MARLadd}, the role of reinforcement learning in multi-agent systems encompasses various applications across cooperative, dynamic, and competitive environments. 
In cooperative scenarios, MARL plays a key role in achieving consensus among agents, allowing them to align their actions toward a shared goal. The applications in this category include high-level decision-making, which is crucial in scenarios such as smart homes, where agents must collaboratively optimize resources and enhance user experience.
In dynamic environments, where conditions frequently change, MARL is indispensable in managing complex and evolving tasks. Traffic prediction is a prime example of how MARL adapts to the real-time flow of information to forecast traffic patterns. Additionally, RL methods like the Knowledge-based Agent Environment (KBAE) methodology and natural language integration empower agents to process and interact with complex data inputs.
In competitive environments, RL enhances agent interactions by optimizing strategies through techniques like Markov games. Here, each agent competes or collaborates with others to improve performance, with strategies that evolve based on the observed actions of others.
This figure highlight show MARL facilitates consensus mechanisms, optimizes strategies, and processes complex data inputs for real-time decision-making, further emphasizing the versatility of MARL in enhancing agent interactions.

\begin{figure*}[!t]
\centering
\includegraphics[width=\linewidth]{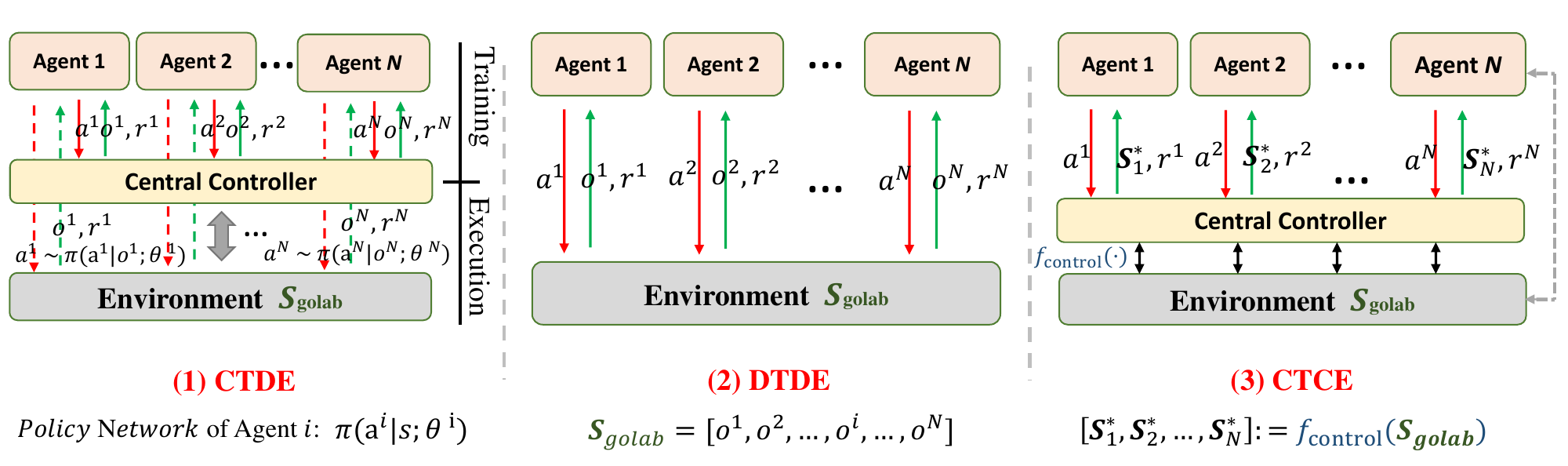}
\caption{The paradigms visualization of CTDE (left), DTDE (centre), and CTCE (right), consisting of three crucial elements: agent (i.e., algorithm or model), environment, central controller (Optional).}
\label{fig_CDTCDE}
\end{figure*}

One of the central challenges in MARL-based multi-agent systems is determining how much information should be shared among agents during different phases of learning and deployment. The MARL research is typically structured into three primary paradigms: CTCE \cite{DQNv1,DQNv2,DQNv3,DQNori,zhou2023CTEenough}, DTDE \cite{amato2024introductionDTE,dewitt2020independentPPO}, and CTDE \cite{amato2024CTDEsurvey,CTDEMAS,KRAEMER2016CTDEori}. As illustrated in Figure \ref{fig_CDTCDE}, we will next explain the principles and differences of the three methods in conjunction with this conceptual framework diagram. % \textcolor{red}{TODO.}

\subsubsection{Centralized Training with Decentralized Execution (CTDE)}
As shown in the left of Figure \ref{fig_CDTCDE}, CTDE is a hybrid MARL approach that combines the strengths of both centralized and decentralized systems \cite{amato2024CTDEsurvey}. In CTDE, each agent possesses its own policy network, which is trained under the guidance of a central controller. This approach is characterized by a two-phase process: \textit{centralized training} followed by \textit{decentralized execution}. % \textcolor{red}{TODO.}

\begin{enumerate}
% [align=parleft,leftmargin=22.7pt,labelsep=0pt,label={[}10{]}]
\item \textit{Centralized Training (Phase 1):}
During the training phase, the central controller collects data from all agents, including their observations, actions, and rewards. This centralized data aggregation allows the controller to oversee the learning process and facilitate the training of each agent's policy network. 
% The controller's role is crucial in this phase as it coordinates the training process, ensuring that the policy networks of all agents are optimized collectively.
\item \textit{Decentralized Execution (Phase 2):}
Once the training is complete, the central controller's involvement ceases, and each agent operates independently. At execution, agents make decisions based on their own observations using their trained policy networks.
% There is no need for communication with the central controller, as each agent is capable of autonomous decision-making.
\end{enumerate}

In some communication-constrained scenarios, agents often cannot share or fully share their observations of the environment. Instead, they must make decisions independently based on their own local observations and policies, which limits the applicability of fully centralized methods. To overcome this challenge, Kraemer et al. \cite{amato2024CTDEsurvey,KRAEMER2016CTDEori} proposed the CTDE learning paradigm. The CTDE agents have access to global environmental state information and the observations of other agents during the training phase, allowing them to learn a joint policy together. However, during the execution phase, each agent relies solely on its own observations and the trained policy to make independent decisions. It combines the advantages of fully decentralized and fully centralized methods, effectively mitigating issues such as learning non-stationarity and the curse of dimensionality, making it the dominant paradigm in current MARL solutions.

Under CTDE, MARL algorithms can primarily be categorized into three types based on their technical implementations: value function decomposition-based algorithms, actor-critic-based algorithms, and algorithms based on policy gradient methods, such as proximal policy optimization (PPO).

\textbf{1. Value Decomposition-based Algorithms}
Value decomposition-based algorithms mainly address the challenge of estimating the joint state-action value function (Q-value) in multi-agent systems, which is difficult due to the high dimensionality of the joint action space. Instead of directly estimating this joint value function, these algorithms decompose it into more manageable individual state-action value functions (Q-value) for each agent. During execution, each agent selects its action based on its own value function. In training, the joint value function is computed from individual value functions, and the temporal difference error of the joint value guides the learning of the individual functions. A key principle these algorithms must satisfy is the Individual-Global-Max (IGM) principle, ensuring that the actions maximizing the joint value are consistent with those maximizing individual values. Different algorithms use various methods to approximate or satisfy the IGM principle.

Value Decomposition Networks (VDN) \cite{VDN} is one of the earliest value decomposition-based algorithms in CTDE-based MARL models. VDN simplifies the estimation of the joint state-action value function by assuming that it can be represented as the sum of the individual state-action value functions of all agents. It means that the joint value function is obtained by simply adding up the individual value functions, which does not take into account the varying contributions of each agent's Q-value. However, the assumption made by VDN is a sufficient but not necessary condition for satisfying the IGM principle, which can limit its applicability. Additionally, VDN does not utilize global state information during training, further restricting its effectiveness in more complex environments.

To address this issue, Rashid et al. \cite{QMIX} proposed the QMIX algorithm within the CTDE paradigm. QMIX assumes a monotonic nonlinear relationship between the joint state-action value function and the individual state-action value functions of agents. To implement this, QMIX introduces a mixing network that computes the joint state-action value function based on the individual Q-values of all agents. This mixing network is designed with non-negative parameters to ensure that the monotonicity assumption is met. QMIX has been successfully applied in various scenarios and is considered one of the most successful value decomposition algorithms to date. By enforcing a monotonic relationship between the joint action Q-values and individual Q-values, QMIX simplifies the policy decomposition process, facilitating decentralized decision-making. However, the monotonicity assumption, while sufficient for ensuring the Individual-Global-Max (IGM) principle, is not a necessary condition. This limitation restricts the algorithm's applicability in situations where an agent's optimal action depends on the actions of other agents.

Weighted QMIX \cite{WeightedQMIX} builds upon QMIX and addresses this limitation by introducing a novel weighting mechanism during the projection of Q-values, which is widely used for cooperative MARL scenarios. In QMIX, the projection of Q-learning targets into the representable space is done with equal weighting across all joint actions, which can lead to suboptimal policy representations, even if the true optimal Q-values (Q*) are known. To overcome this, Weighted QMIX introduces two weighting schemes-Centrally-Weighted (CW) QMIX and Optimistically-Weighted (OW) QMIX-that place greater emphasis on the better joint actions during this projection process. The weighting schemes ensure that the correct maximal action is recovered for any set of joint action Q-values, effectively improving the algorithm's ability to learn optimal policies. These schemes in Weighted QMIX enhances the representational capacity of QMIX, demonstrating improved results on both predator-prey scenarios of Multi-Agent Particle Environment (MPE) \cite{MPE9619081} and the challenging StarCraft II benchmarks \cite{vinyals2017starcraftiiPYSC2,Vinyals2019sc2,samvelyan19smac}.

Since then, numerous methods building on value function decomposition have been developed. QPLEX \cite{ICLRQPLEX} introduces a novel duplex dueling network architecture for multi-agent Q-learning, designed to nonlinearly decompose the joint state-action value function while embedding the IGM principle directly into the network structure.
% This approach allows QPLEX to achieve a complete IGM function class, leading to superior performance and high sample efficiency in complex multi-agent environments, as demonstrated in StarCraft II micromanagement tasks.
Fast-QMIX \cite{FastQMIX} enhances the original QMIX by dynamically assigning virtual weighted Q-values with an additional network, improving convergence speed, stability, and overall performance in cooperative multi-agent scenarios. QTRAN \cite{QTRAN} introduces a more flexible factorization method that overcomes the structural limitations of QMIX, where the joint Q-value is constrained to be a monotonic function of the individual Q-values, thereby imposing a specific structural form on the factorization. Specifically, QTRAN introduces a necessary and sufficient condition for the IGM principle and incorporates two additional loss terms into the loss function to constrain the training of individual Q-value functions, ensuring they satisfy this IGM principle. 
% This flexibility allows QTRAN to handle a broader range of tasks and achieve superior performance in cooperative settings. 

\textbf{2. Actor-Critic-based Algorithms:}
Actor-Critic-based algorithms \cite{MADDPG,Schwartz9002707,ShenIntrinsicA3C} represent a foundational class of methods within the CTDE paradigm, offering a flexible and effective approach for tackling the challenges of multi-agent environments. These algorithms combine the strengths of policy optimization (actor) with value estimation (critic), allowing agents to learn robust and adaptive strategies in both cooperative and competitive settings. By leveraging a centralized critic during training, Actor-Critic-based methods \cite{lowe2017multi,MPE9619081,ChuMARLonTrafficSingal} address key issues such as environmental non-stationarity and credit assignment, enabling effective policy optimization in dynamic and complex multi-agent scenarios. Below, we discuss several prominent Actor-Critic-based approaches and their contributions to advancing MARL. % \textcolor{red}{TODO.}

% MADDPG \cite{lowe2017multi,MADDPG} .
MADDPG \cite{MADDPG} is a typical Actor-Critic-based CTDE approach specifically designed to address the challenges of multi-agent environments, where agents engage in both cooperative and competitive interactions.
Traditional reinforcement learning algorithms, such as Q-learning and policy gradient methods, struggle in multi-agent settings due to issues like non-stationarity-where the environment constantly changes as other agents learn-and increased variance with the growing number of agents.
MADDPG adapts the actor-critic framework by incorporating a centralized critic during training, which has access to the actions and observations of all agents. This centralized critic helps mitigate the non-stationarity problem by learning a more stable value function that considers the joint action space. During execution, however, each agent independently follows its policy (actor) based on local observations, enabling decentralized decision-making. It allows each agent to successfully learn and execute complex coordination strategies, outperforming existing methods in both cooperative and competitive multi-agent environments. % in multi-agent environment
% MADDPG also enhances training by incorporating an ensemble of policies for each agent, making the learned policies more robust and adaptable to various scenarios. 
% 这个是CTCE 还是CTDE ？？？ GPT 说是CTDE
To address the computational challenges of continuous action spaces, Li et al. \cite{li2019M3DDPG} extend the MADDPG algorithm to Multi-Agent Mutual Information Maximization Deep Deterministic Policy Gradient (M3DDPG) by incorporating a minimax approach to enhance robustness in multi-agent environments. M3DDPG introduce Multi-Agent Adversarial Learning (MAAL), which efficiently solves the minimax formulation, ensuring agents can generalize even when opponents' policies change and leading to significant improvements over existing baselines in mixed cooperative-competitive scenarios.

Counterfactual Multi-Agent Policy Gradient (COMA) \cite{COMA} is a cooperative algorithm based on the Actor-Critic framework that uses centralized learning to address the credit assignment problem in multi-agent settings. COMA employs a centralized critic to compute advantage functions for each agent, using counterfactual baselines to reduce policy dependencies among agents and improve learning efficiency. Each agent has its own policy network, but the shared centralized critic evaluates joint Q-values by considering the collective state and action information of all agents. This approach minimizes the negative impacts of policy dependencies and allows for a more comprehensive assessment of each agent's behavior, enhancing overall policy optimization. 

\textbf{3. Proximal Policy Optimization-based Algorithms:} % \textcolor{red}{TODO.}
Proximal Policy Optimization (PPO) \cite{schulman2017ppo} is a widely used CTDE reinforcement learning algorithm that has been adapted and extended to address challenges in MARL. Within the CTDE paradigm, PPO and its multi-agent variants have shown remarkable effectiveness in balancing policy optimization efficiency and stability. 
% This section delves into PPO-based algorithms in MARL, with a particular focus on Heterogeneous-Agent Trust Region Policy Optimization (HATRPO) and Heterogeneous-Agent Proximal Policy Optimization (HAPPO). 
PPO was introduced by Schulman et al. \cite{schulman2017ppo} as an efficient policy gradient algorithm designed to improve upon the trust region policy optimization (TRPO) framework \cite{SchulmanJMLRTrust}. PPO employs a clipped surrogate objective function that simplifies the trust region constraint in TRPO, allowing for stable updates without overly restrictive computational overhead. The key innovation of PPO lies in its ability to control the magnitude of policy updates through the clipping mechanism, which ensures that policies do not deviate excessively from their previous versions.

In MARL, Multi-Agent PPO (MAPPO) \cite{yu2022surprising} extends PPO to the centralized critic paradigm. MAPPO uses a centralized value function (critic) that evaluates joint states and actions during training, while agents execute independently using their decentralized policies. MAPPO has demonstrated superior performance in various cooperative and competitive multi-agent environments, such as the StarCraft II \cite{samvelyan19smac,SMACv2} and Multi-Agent Particle Environment (MPE) \cite{mordatch2017emergence,lowe2017multi,MPE9619081} benchmarks. The centralized critic allows for improved credit assignment and non-stationarity handling during training, while the decentralized execution ensures scalability.
While MAPPO leverages parameter sharing among agents, this assumption may not hold in heterogeneous-agent systems where agents differ in capabilities, objectives, or action spaces. 

To address this, Kuba et al. \cite{kuba2022trust} proposed Heterogeneous-Agent Trust Region Policy Optimization (HATRPO) and Heterogeneous-Agent Proximal Policy Optimization (HAPPO). These algorithms remove the parameter-sharing assumption, allowing for individualized policy networks for each agent. HATRPO builds upon TRPO by introducing a sequential update scheme, where only one agent updates its policy at a time while the policies of other agents remain fixed. This approach ensures monotonic improvement in joint policies, as it approximates the Nash equilibrium under certain conditions, such as full observability and deterministic environments. HAPPO extends PPO in a similar vein, replacing parameter sharing with individualized policies. Like HATRPO, HAPPO employs a sequential update mechanism, but it retains the computational efficiency and practical simplicity of PPO’s clipped objective function.

Both HATRPO and HAPPO utilize a sequential update process where one agent updates its policy while others remain fixed. This prevents conflicts during policy optimization and ensures theoretical convergence to a stable joint policy. Moreover, HATRPO and HAPPO provide monotonic improvement guarantees under specific conditions. By removing the parameter-sharing constraint, these algorithms enable agents to learn tailored policies that account for their unique roles and capabilities. Both algorithms perform competitively in benchmark tasks, demonstrating their ability to scale to high-dimensional state-action spaces while maintaining robust coordination among agents.

PPO-based algorithms, including MAPPO \cite{yu2022surprising}, HATRPO \cite{kuba2022trust}, and HAPPO \cite{kuba2022trust}, have revolutionized multi-agent reinforcement learning by combining the stability of PPO with the coordination benefits of centralized critics. These algorithms have proven effective across a wide array of cooperative and competitive MARL tasks, offering strong performance and scalability.

\textbf{3. Other Categories of Algorithms within the CTDE Paradigm:} % Alternative
In addition to the well-established categories of Value Decomposition-based, Actor-Critic-based, and Proximal Policy Optimization (PPO)-based algorithms, the MARL research has seen significant advancements through innovative optimizations and enhancements within CTDE paradigm that are not confined to these traditional classifications. These approaches aim to address the inherent challenges of multi-agent environments, such as non-stationarity and limited communication, to improve overall cooperation and policy learning efficiency.

% 还有很多工作针对CTDE的改进，并未依附于特定的某个类别，例如工作zhou2023CTDEenough
For example, Centralized Advising and Decentralized Pruning (CADP) is a novel framework  introduced by Zhou et al. \cite{zhou2023CTDEenough} to address limitations in the CTDE paradigm. CADP enhances the training process by allowing agents to explicitly communicate and exchange advice during centralized training, thus improving joint-policy exploration. To maintain decentralized execution, CADP incorporates a smooth model pruning mechanism that gradually restricts agent communication without compromising their cooperative capabilities, demonstrating its superior performance on multi-agent StarCraft II SMAC and Google Research Football benchmarks. 
CommNet \cite{CommNet} introduces a neural model where multiple agents learn to communicate continuously and collaboratively through a shared communication channel, optimizing their performance on fully cooperative tasks. The method allows agents to develop their own communication protocols during training, leading to improved coordination and task-solving capabilities.
Mao et al. \cite{MetaMARL} introduced a novel Meta-MARL framework by integrating game-theoretical meta-learning with MARL algorithms using the CTDE's framework, such as  the Actor-Critic-based COMA \cite{COMA}. This framework offers initialization-dependent convergence guarantees and significantly improves convergence rates by addressing related tasks collectively. % Meta-MARL combines the COMA algorithm based on the actor-critic framework with the MAML meta-learning algorithm.
Yun et al. \cite{quantumMetaMARL} proposed a novel approach called Quantum Meta Multi-Agent Reinforcement Learning (QM2ARL), achieving high rewards, fast convergence, and quick adaptation in dynamic environments. QM2ARL leverages the unique dual-dimensional trainability of Quantum Neural Networks (QNNs) to enhance MARL. 
% This approach initially employs angle parameter training for meta-learning, then utilizes pole parameter training for few-shot or local learning. Additionally, it incorporates an angle-to-pole regularization technique to mitigate overfitting.
Liu et al. \cite{liu2024grounded} proposed the Learning before Interaction (LBI) framework, which integrates a language-guided simulator into the multi-agent reinforcement learning pipeline to address complex decision-making problems. By leveraging a generative world model with dynamics and reward components, LBI generates trial-and-error experiences to improve policy learning, demonstrating superior performance and generalization on the StarCraft Multi-Agent Challenge benchmark \cite{samvelyan19smac,SMACv2}.

\subsubsection{Decentralized Training with Decentralized Execution (DTDE)}
As shown in the centre of Figure \ref{fig_CDTCDE}, DTDE represents a fully decentralized mechanism where each agent interacts independently with the environment and updates its own policy based on its own observations and rewards \cite{amato2024introductionDTE}.
% Agents are independent entities that do not communicate with each other \cite{zhou2023CTEenough}. 
In this framework, each agent trains and operates completely independently, relying only on its own observations and rewards to update its strategy. DTDE is particularly suited for environments with limited communication or no global coordination, offering strong scalability and robustness \cite{zhou2023CTEenough}.

The core idea behind DTDE is the independence of agents \cite{amato2024introductionDTE}. Each agent interacts with its environment and learns without requiring information from others. This makes DTDE scalable, but it also introduces challenges such as non-stationarity, where the environment appears to change as other agents adapt their strategies. This characteristic makes DTDE a valuable and challenging area of research.
The theoretical foundation of DTDE is often based on Decentralized Partially Observable Markov Decision Processes (Dec-POMDPs). As described by Amato et al. \cite{amato2024introductionDTE,amato2024CTDEsurvey}, a Dec-POMDP models a decentralized decision-making environment where agents operate independently with limited observations while aiming to maximize a shared reward. The decentralized nature of DTDE requires each agent to learn optimal strategies based on local information only.

First and foremost, one of the earliest DTDE approaches is Independent Q-Learning (IQL) by  et al. \cite{claus1998IQL}. Here, each agent applies Q-learning independently, maintaining its own Q-function and updating it based on local observations and rewards. However, IQL faces several challenges, such as the non-stationary nature of the environment caused by other agents learning simultaneously. It also struggles with credit assignment, where it is hard to determine how an individual agent contributes to the team’s success.
To address these issues, several extensions of IQL have been proposed:

\begin{itemize} 
\item \textbf{Distributed Q-Learning} \cite{LauerDistributedQLearning} optimistically assumes other agents always take optimal actions, focusing on learning from high-reward interactions. While effective in deterministic settings, it can be overly optimistic in environments with randomness.
\item \textbf{Hysteretic Q-Learning} \cite{HystereticQlearning} By introducing two learning rates—one for positive updates and another, smaller rate for negative updates—hysteretic Q-learning balances optimism with robustness in stochastic environments.
\item \textbf{Lenient Q-Learning} \cite{WeiLenient} dynamically adjusts how lenient the agent is in updating its values, depending on how frequently specific state-action pairs are encountered. It allows for more exploration in the early stages of learning while focusing on optimization later.
\end{itemize}

As MARL problems become more complex, DTDE methods have been extended to deep learning. Deep Q-Networks \cite{DQNv1,DQNv2,DQNv3,DQNori} and Deep Recurrent Q-Networks \cite{hausknecht2017DRQN} have been adapted for decentralized settings, enabling agents to handle high-dimensional state and action spaces. Independent DRQN (IDRQN) \cite{IDRQNpone}, for example, combines DRQN with independent learning, but its asynchronous experience replay can cause instability. To solve this, Concurrent Experience Replay Trajectories (CERTs) \cite{OmidshafieiCERT} synchronize experience replay among agents, reducing non-stationarity and improving learning efficiency. Other advancements include Deep Hysteretic DRQN (Dec-HDRQN) \cite{OmidshafieiCERT}, which combines hysteretic updates with deep neural networks and uses concurrent buffers to handle decentralized data. These methods have shown robust performance in partially observable environments.

% \textcolor{red}{TODO.}

In the DTDE paradigm, policy gradient methods offer an alternative to value-based approaches, particularly for scenarios involving continuous action spaces \cite{amato2024introductionDTE}. Several policy gradient DTDE methods have been proposed:

\begin{itemize} 
\item \textbf{Decentralized REINFORCE} \cite{BOWLING2002DecMARL} independently optimizes each agent’s policy using gradient ascent based on rewards observed during episodes. While simple, it is less sample-efficient.
\item \textbf{Independent Actor-Critic (IAC)} \cite{COMA} Combining value estimation (critic) and policy optimization (actor), IAC enables agents to learn faster and update more frequently, improving sample efficiency.
\item \textbf{Independent Proximal Policy Optimization (IPPO)} \cite{schulman2017ppo,dewitt2020independentPPO,yu2022surprising} Extending Proximal Policy Optimization (PPO) to decentralized settings, IPPO improves policy stability by limiting how much policies can change between updates.
\end{itemize}

Despite its advantages, DTDE still faces significant challenges \cite{amato2024introductionDTE,zhou2023CTEenough}: 1. \textit{Non-Stationarity}: As other agents learn and adapt, the environment appears dynamic and unstable to each agent, making convergence difficult; 2. \textit{Credit Assignment}: It is hard to determine how each agent’s actions contribute to the team’s overall reward in cooperative tasks; 3. \textit{Trade-Offs Between Scalability and Performance}: While DTDE scales well, its performance may be limited in tasks requiring high levels of coordination. To overcome these challenges, future research could focus on improving communication strategies during training and more robust strategies for dynamic environments. % , hybrid methods combining decentralized learning with limited centralized components,

In conclusion, the DTDE paradigm provides a powerful framework for solving distributed decision-making problems, balancing scalability, independence, and efficiency. It has been successfully applied in areas such as autonomous driving, distributed energy management, and swarm robotics. As research continues, DTDE is expected to play a larger role in real-world multi-agent systems, especially in scenarios requiring robust, independent learning.

\subsubsection{Centralized Training with Centralized Execution (CTCE)} 
% \cite{} . \textcolor{red}{TODO.}
As shown in the right of Figure \ref{fig_CDTCDE}, Centralized Training with Centralized Execution (CTCE) stands out as a fully centralized mechanism to MARL decision-making, where all agents transmit their information to a central controller \cite{DQNv1,DQNv2,DQNv3,DQNori}. This central controller has access to the observations, actions, and rewards of all agents. The agents themselves do not possess policy networks and are not responsible for making decisions. Instead, they simply execute the directives issued by the central controller \cite{Sharma2021CTE,zhou2023CTEenough}.

Multi-Agent DQN \cite{DQNori} is a classic example of the CTCE paradigm, where DQN is combined with a parameter-sharing mechanism to address tasks in multi-agent environments.
Gupta et al. \cite{DQNori} firstly extends three single-agent DRL algorithms (DQN \cite{mnih2013DQNv1,Mnih2015DQNv2}, TRPO, and A3C) to multi-agent systems, resulting in Multi-Agent-DQN, Multi-Agent-TRPO, and Multi-Agent-A3C. These approaches were designed to learn cooperative policies in complex, partially observable environments without requiring explicit communication between agents. The DQN algorithm based on multi-agent settings, also known as PS-DQN (Parameter Sharing DQN), effectively utilizes curriculum learning to handle increasing task complexity. By starting with fewer agents and gradually increasing the number, the model scales well to more complex scenarios.
Further, this foundational work has led to numerous enhancements and variants based on Multi-Agent DQN, each designed to address specific challenges in multi-agent systems, such as CoRe \cite{DQNv1}, MARL-DQN \cite{DQNv2}, and  \cite{DQNv3}.
% the variant models of Multi-Agent DQN. 
CoRe \cite{DQNv1} introduces a counterfactual reward mechanism into MARL to address the credit assignment problem in cooperative settings. By computing the difference in global rewards when an agent hypothetically changes its action while others keep theirs fixed, CoRe enhances the standard DQN framework, significantly improving learning efficiency and performance in cooperative tasks.
MARL-DQN \cite{DQNv2} optimizes energy efficiency and resource allocation in NOMA wireless systems by using MARL framework combined with Deep Q-Networks. By combining MARL with DQN, it dynamically adjusts power and time allocation to minimize energy consumption while ensuring quality of service, outperforming traditional methods in terms of efficiency and performance.
Hafiz et al. \cite{DQNv3} proposed a simplified and efficient multi-agent DQN-based multi-agent system (MAS) that addresses the challenges of complexity, resource demands, and training difficulties inherent in more advanced MARL frameworks. The work introduced a shared state and reward system while maintaining agent-specific actions, which streamlines the experience replay process. The significant improvements in tasks such as Cartpole-v1\footnote{Cartpole-v1 game: \url{https://www.gymlibrary.dev/environments/classic_control/cart_pole/}.}, LunarLander-v2\footnote{LunarLander-v2 game: \url{https://www.gymlibrary.dev/environments/box2d/lunar_lander/} and \url{https://github.com/topics/lunarlander-v2}.}, and Maze Traversal\footnote{Maze Traversal game: \url{https://github.com/vision-mini/MazeSolverLLM}.} from OpenAI Gym \cite{brockman2016openaigym} demonstrates the model's effectiveness and superiority.

% ############################################################
% \subsubsection{for MARL} % 强化学习分类之三 通过是否基于值函数还是梯度优化的分类
% \textbf{} \cite{} .
% \textbf{} \cite{} .
% \textbf{} \cite{} .
% ############################################################

\subsubsection{Addition Taxonomies and Efforts of Communication-based MARL Algorithms}
As outlined above, three primary paradigms—CTCE, DTDE, and CTDE—have emerged in the MARL domain to tackle the challenges associated with training and execution in multi-agent systems. Each of these paradigms has its strengths and limitations, yet all face inherent difficulties in handling communication among agents, which is critical for effective collaboration and decision-making. 

Specifically, the CTCE paradigm, while providing a fully integrated framework for learning and execution, struggles with scalability as the system size grows. The DTDE paradigm, on the other hand, allows for independent agent training and execution, but often lacks the necessary coordination required for global task optimization. The CTDE paradigm has emerged as a widely adopted approach due to its ability to leverage centralized information during training to learn effective policies, while enabling decentralized execution to operate efficiently in distributed environments. However, even in CTDE, the communication between agents during execution is a bottleneck, prompting researchers to focus on improving communication strategies to enhance system performance.

\begin{figure*}[!t]
\centering
\includegraphics[width=0.95\linewidth]{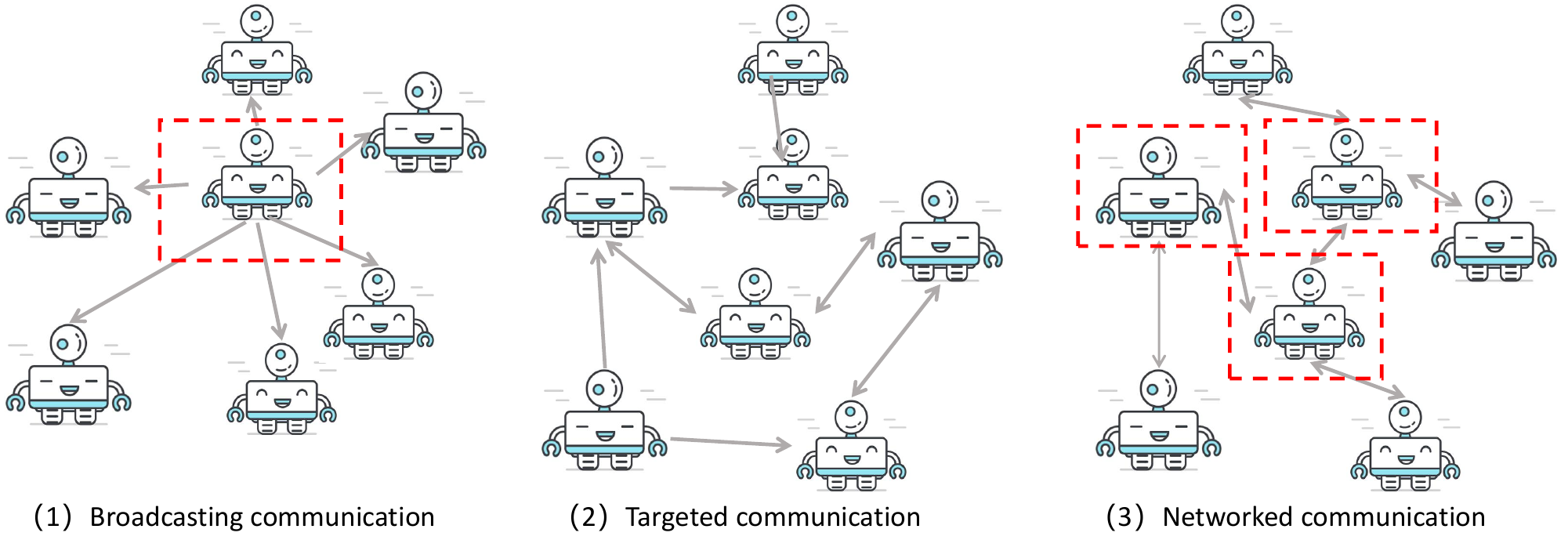}
\caption{A schematic representation of three distinct communication methods among agents, with arrows indicating the direction of message transmission. (a) \textit{Broadcasting communication}: The activated agent transmits messages to all other agents within the communication network. (b) \textit{Targeted communication}: Agents selectively communicate with specific target agents based on a supervisory mechanism that regulates the timing, content, and recipients of the messages. (c) \textit{Networked communication}: Agents engage in localized interactions with their neighboring agents within the network.}
\label{figcommways}
\end{figure*}

Communication-based MARL algorithms have made significant progress in overcoming these challenges. From the perspective of communication protocols and languages, communication-based MARL methods can be categorized into three types: broadcasting communication, targeted communication, and networked communication, as shown in Figure \ref{figcommways}. From the technical angle, we provide an overview of these communication-based MARL advancements, categorizing the algorithms into three main groups based on their focus: (1) value function-based Communication-based MARL, (2) policy search-based Communication-based MARL, and (3) Communication-based MARL algorithms designed to improve communication efficiency. These approaches represent the forefront of research in enabling agents to effectively share information, coordinate actions, and optimize performance in complex environments. Here, we provide a detailed introduction to these approaches.

\textbf{Value Function-Based Communication-based MARL:}
For Value Function-Based Communication-based MARL Algorithms, several notable works include Differentiable Inter-Agent Learning (DIAL) \cite{foerster2016learningDIAL} and Deep Distributed Recurrent Q-Networks (DDRQN) \cite{FoersterDDRQN,DDRQN8612583}. Among them, DIAL facilitates effective collaboration and optimization of joint action policies by enabling the exchange of gradients of Q-functions between agents. On the other hand, DDRQN leverages recurrent neural networks to address partially observable environments, allowing agents to share critical Q-values or hidden states and make adaptive decisions in dynamic settings.

\textbf{Policy Search-Based Communication-based MARL:}
For Policy Search-Based Communication-based MARL Algorithms, significant progress has been made with approaches such as Communication Networks (CommNet) \cite{sukhbaatar2016learning}, Bidirectional Coordinated Network (BiCNet) \cite{peng2017BidirectionalBiCNet,WangBiCDDPG}, Multi-Agent Distributed MADDPG (MD-MADDPG) \cite{FanMDMADDP,MADDPG}, Intrinsic A3C \cite{mnih2016asynchronous,ShenIntrinsicA3C}, and Multi-Agent Communication and Coordination (MACC) \cite{ijcai2022p85,Vanneste2025}. Among them, CommNet \cite{sukhbaatar2016learning} proposes a centralized but differentiable communication framework where agents share encoded signals to form a global context, improving policy decisions. BiCNet \cite{peng2017BidirectionalBiCNet,WangBiCDDPG} enhances coordination among agents by employing bidirectional recurrent layers, making it suitable for complex tasks. MD-MADDPG \cite{FanMDMADDP,MADDPG} combines centralized training and decentralized execution, enabling agents to exchange critical state-action information during training for robust policy learning. Intrinsic A3C \cite{mnih2016asynchronous,ShenIntrinsicA3C} introduces intrinsic motivation to encourage effective exploration in sparse-reward scenarios, with agents sharing intrinsic rewards through communication to boost performance. Finally, Multi-Agent Communication and Coordination (MACC) \cite{ijcai2022p85,Vanneste2025} focuses on adaptive communication mechanisms, providing stable and secure coordination to enhance training and execution in dynamic multi-agent environments.

\textbf{Communication-based MARL Enhancing Communication Efficiency:}
For algorithms aimed at enhancing communication efficiency, several outstanding approaches include Attentional Communication (ATOC) \cite{jiang2018learning}, Targeted Multi-Agent Communication (TarMAC) \cite{SunaaaiT2MAC}, Inter-Agent Centralized Communication (IC3Net) \cite{singh2019ic3net}.
%, and Implicit and Independent Communication (I2C) \cite{}
Attentional Communication (ATOC) \cite{jiang2018learning} employs an attention mechanism to dynamically determine when communication is necessary, achieving a balance between efficiency and coordination. Targeted Multi-Agent Communication (TarMAC) \cite{SunaaaiT2MAC} introduces targeted attention mechanisms to direct messages to relevant teammates, minimizing redundant communication, and improving overall performance. Inter-Agent Centralized Communication (IC3Net) \cite{singh2019ic3net} incorporates a gating mechanism that allows agents to learn when and how to communicate, optimizing both the frequency and quality of interactions. % Implicit and Independent Communication (I2C) \cite{} further refines this concept by enabling agents to independently infer their communication needs, sharing only the most critical information to enhance scalability and robustness in complex systems.

These research advances in Communication-based MARL methods demonstrate significant strides in enabling agents to share information and achieve coordinated decision-making in MAS. These advancements will pave the way for deploying MARL in real-world scenarios where efficient and effective communication is essential. % indispensable

\subsection{LLMs-based Multi-Agent System Taxonomies}
\label{LLMtaxonomies}
% In the LLM multi-agent survey of \textit{Wang et al.} \cite{Wang2024LLMsurvey,LLMmarlsurvey}, they provide a comprehensive review of the rapidly evolving field of LLMs-based autonomous agents, which are seen as a promising approach towards achieving artificial general intelligence (AGI). The survey is organized around the construction, application, and evaluation of LLM agents: 1. addressing two main problems: (1) designing agent architectures that effectively leverage LLMs, and (2) enhancing the capabilities of agents to perform various tasks; 2. exploring the diverse applications of LLMs-based autonomous agents across social sciences, natural sciences, and engineering, demonstrating the broad utility and potential of these agents in different domains; 3. Delving into both subjective and objective evaluation methods to assess the performance of LLMs-based autonomous agents. 
% In addition to organizing existing work into comprehensive taxonomies, the survey identifies ongoing challenges in the field and discusses potential future directions, inspiring further groundbreaking studies in the field of LLMs-based autonomous agents.

% \textcolor{red}{TODO. lack of reference.}

The field of LLMs-based multi-agent systems has seen significant advancements, with researchers exploring various aspects of these systems to enhance their capabilities and applications \cite{LLMmarlsurvey,Wang2024LLMsurvey}. A comprehensive taxonomy can help categorize and understand the different dimensions of LLMs-based multi-agent systems, including architectural design, application domains, evaluation methods, and future research directions.

\subsubsection{Architectural Design}
The design of architectures for LLMs-based multi-agent systems is a critical component in harnessing the full potential of LLMs to enhance the capabilities of autonomous agents. Architectural design encompasses the framework and mechanisms that enable agents to interact, adapt, and make decisions in complex and dynamic environments. This section explores two primary levels of autonomy within these systems: Adaptive Autonomy and Self-Organizing Autonomy. 
% Understanding these levels is essential for developing multi-agent systems that can effectively leverage the strengths of LLMs while addressing the challenges of real-world applications.

\begin{itemize}  % surveys
\item 
\textbf{Adaptive Autonomy:} \cite{wu2023autogen,dibia2024autogen,zheng2024planagent,xagent2023}
Adaptive autonomy refers to systems where agents can adjust their behavior within a predefined framework. These agents are designed to operate within the constraints set by the system architects but can adapt their actions based on the specific requirements of the task at hand. For example, in a task-specific adaptation scenario, an agent might adjust its search strategy in an information retrieval task based on the relevance of the results. In a context-aware adaptation scenario, an agent might change its communication style based on the social context of the interaction. This level of autonomy is crucial for agents that need to operate in dynamic environments where the task requirements can change over time.
\item 
\textbf{Self-Organizing Autonomy:} \cite{xiao2024chainofexperts,wu2023autogen,chen2024agentverse,wang2024mllmtool,MAT,vinyals2017starcraftiiPYSC2}
Self-organizing autonomy represents a higher level of autonomy where agents can dynamically adapt their behavior without predefined structures. This allows for more flexible and context-aware interactions among agents. For instance, in dynamic task allocation, agents can assign tasks to each other based on the current state of the environment and their individual skills. Emergent behavior is another key feature at this level, where agents can form coalitions or develop new strategies to solve complex problems. This level of autonomy is essential for multi-agent systems that need to operate in highly dynamic and unpredictable environments.
\end{itemize}

\subsubsection{Suitable Scenarios} % Applications
In the \textbf{social sciences} \cite{ThreeDWorldv1,puig2021watchandhelp,wangJ2024LangGraph}, LLMs-based agents have been used to simulate various social phenomena, providing insights into human behavior and social dynamics. 
\begin{itemize}
\item 
\textit{1) Economic Agents:} \cite{LangChainCustom,xiao2024chainofexperts}
LLMs can be used to model economic agents, similar to how economists use the concept of homo economicus. Experiments have shown that LLMs can produce results qualitatively similar to those of traditional economic models, making them a promising tool for exploring new social science insights. For example, in market simulation, LLMs can predict market trends and the impact of economic policies. In behavioral economics, LLMs can model individual and group decision-making processes, providing a more nuanced understanding of economic behavior. 
\item 
\textit{2) Social Network Simulation:} \cite{ThreeDWorldv2,ThreeDWorldv1,puig2021watchandhelp,sun2023corex}
The Social-network Simulation System (S3) uses LLMs-based agents to simulate social networks, accurately replicating individual attitudes, emotions, and behaviors. This system can model the propagation of information, attitudes, and emotions at the population level, providing valuable insights into social dynamics. For example, it can simulate how information spreads through social networks and identify influential nodes, or model the evolution of social norms and behaviors over time. 
\item 
\textit{3) User Behavior Analysis:} \cite{hou2024coact,wang2024LangGraph,wangJ2024LangGraph}
LLMs are employed for user simulation in recommender systems, demonstrating superiority over baseline simulation systems. They can generate reliable user behaviors, improving the accuracy of recommendations. For example, in personalized recommendations, LLMs can generate user profiles and behaviors to optimize recommendation algorithms. In user engagement, LLMs can simulate user interactions to optimize user retention and engagement.
\end{itemize}

In the \textbf{natural sciences} \cite{gao2024agentscope,gong2024mindagent,ParkSocialSimulacra}, LLMs-based agents have been used to simulate complex systems and processes, providing insights into natural phenomena and scientific theories.
\begin{itemize}
\item 
\textit{1) Macroeconomic Simulation:}
LLMs-based agents are used for macroeconomic simulation, making realistic decisions and reproducing classic macroeconomic phenomena. These agents can simulate the impact of economic policies on the macroeconomy, providing a more accurate and dynamic model of economic behavior. For example, they can simulate the interactions between different economic sectors and their impact on the overall economy, helping policymakers make more informed decisions.
\item 
\textit{2) Generative Agent-Based Modeling:}
This approach couples mechanistic models with generative artificial intelligence to unveil social system dynamics, such as norm diffusion and social dynamics. By combining the strengths of both approaches, researchers can model complex social systems and predict their behavior over time. For example, they can model the spread of diseases in a population, the impact of environmental changes on ecosystems, or the evolution of social norms in a community.
\end{itemize}

In \textbf{engineering} \cite{wu2023autogen,dibia2024autogen,zhang2025crewai,duan2024crewaiLangGraph}, LLMs-based agents have been used to develop and optimize complex systems, improving efficiency and performance.
\begin{itemize}
\item 
\textit{1) Software Development:}
LLMs-based agents are used for software development, facilitating sophisticated interactions and decision-making in a wide range of contexts. These agents can assist in code generation, bug detection, and system optimization, improving the productivity and quality of software development. For example, they can generate code snippets based on natural language descriptions, detect bugs in code, and suggest optimizations to improve performance.
\item 
\textit{2) Multi-Robot Systems:}
LLMs-based multi-agent systems are used to simulate complex real-world environments effectively, enabling interactions among diverse agents to solve various tasks. These systems can coordinate the actions of multiple robots, optimizing their behavior to achieve common goals. For example, they can be used in search and rescue operations, where multiple robots need to coordinate their actions to locate and rescue victims.
\end{itemize}

\section{Simulation Environments of Multi-Agent Decision-Making}
\label{simenvs}
% 讲述相关的模拟仿真环境：场景与平台 progress
First and foremost, the designs and implementations of multi-agent cooperative simulation environments are crucial in the historical research of multi-agent decision-making, which are widely utilized in practical applications and production.
These simulation environments form the foundation for conducting efficient and effective studies in multi-agent cooperative decision-making.
Specifically, a dynamic multi-agent cooperative decision-making environment refers to predetermined scenarios and platforms where multiple agents collaborate to solve problems, complete tasks, and achieve goals.
Such environments provide not only a platform for testing and validating various intelligent decision-making algorithms but also help us better understand the behaviors and interactions of agents in dynamic settings.
By simulating these interactions, researchers can gain insights into how agents coordinate and adapt to changing conditions, thereby improving the robustness and efficiency of multi-agent systems in real-world applications.
Consequently, the importance of these simulation environments cannot be overstated. They serve as a testing ground for theoretical models, allowing researchers to observe the practical implications of their intelligent algorithms.
Additionally, these platforms help in identifying potential issues and refining strategies before deployment in actual scenarios, ensuring that the agents are well-prepared to handle the complexities of real-world environments. In Table  \ref{table_multi_agent_environments}, a wide range of simulated environments is listed. Next, we will delve into these environments one by one, emphasizing their significance and features for future development. % prospects

% % ##########################################
% .
% .
% .
% % 这里添加大量的现有的多智能体仿真平台
% \%  \textcolor{red}{Here we need to add much simulation platforms.}

\begin{table*}[ht]
\renewcommand{\arraystretch}{2.0}
\caption{Diverse MARL-based and LLMs-based Simulated Environments for Multi-Agent Systems.}
\centering
\setlength{\tabcolsep}{1.0mm}{
\scalebox{0.85}{
\begin{tabular}{cp{11.5cm}} % l
\hline
\textbf{Categories} & \multicolumn{1}{c}{\textbf{Multi-Agent System Environments}} \\ \hline % Simulation
\textbf{MARL-based} & Multi-Agent Particle Environment (MPE) \cite{mordatch2017emergence,lowe2017multi,MPE9619081}, PettingZoo \cite{terry2021pettingzoo}, SMAC \cite{samvelyan19smac}, SMAC-v2 \cite{SMACv2}, GFootball \cite{Kurach2020gfootball}, Gym-Microrts \cite{huang2021gymmurts}, MAgent \cite{zheng2018magent}, Dexterous Hands \cite{YufnbotDexterous,OpenAIdexterous}, OpenAI Gym \cite{brockman2016openaigym}, Gym-MiniGrid \cite{boisvert2023minigrid}\footnote{Gym-MiniGrid: \url{https://minigrid.farama.org/}.}, Melting Pot \cite{leibo2021scalable}\footnote{Melting Pot: \url{https://github.com/google-deepmind/meltingpot}.}, Capture The Flag\footnote{Capture The Flag: \url{https://github.com/gavit21/Capture-The-Flag-CTF-}.} \cite{mehlman2024cat}, VillagerAgent \cite{dong2024villageragent}, Minecraft \cite{li2024autom,zhu2023ghost,MinecraftB}, Unity ML-Agents \cite{juliani2020unity}, SUMO\footnote{SUMO: \url{https://github.com/OkYongChoi/sumo-marl}.} \cite{HU202599}, Hanabi Learning \cite{anonymous2024a,bredell2024augment}, Predator-Prey \cite{kölle2024aquarium,Chatterjee2024}  \\ \hline
% SUMO (Simulation of Urban Mobility) , \cite{}, \cite{}

\textbf{LLMs-based} & TDW-MAT \cite{ThreeDWorldv2,ThreeDWorldv1}, C-WAH \cite{puig2021watchandhelp}, Cuisineworld \cite{gong2024mindagent}, AgentScope \cite{gao2024agentscope}, RoCoBench \cite{mandi2023roco}, Generative Agents \cite{ParkSocialSimulacra,park2023generative}, SocialAI school \cite{kovac2023the,kovac2024socialai}, Welfare Diplomacy \cite{mukobi2024welfare} \\ \hline
% , \cite{}, \cite{}, \cite{}, \cite{}, \cite{}, \cite{}, \cite{}, \cite{}
\end{tabular}}}
\label{table_multi_agent_environments}
\end{table*}

\subsection{MARL-based Simulation Environments}
\label{marlsimenvs}
This section provides an overview of several widely-used simulation environments designed for MARL. These platforms, such as Multi-Agent Particle Environment \cite{mordatch2017emergence,lowe2017multi,MPE9619081}, and PettingZoo \cite{terry2021pettingzoo}, offer diverse scenarios and functionalities for exploring cooperative and competitive agent interactions in both simple and complex tasks. % Google Research Football, 

\begin{figure}[h]
\centering
\includegraphics[width=3.5in]{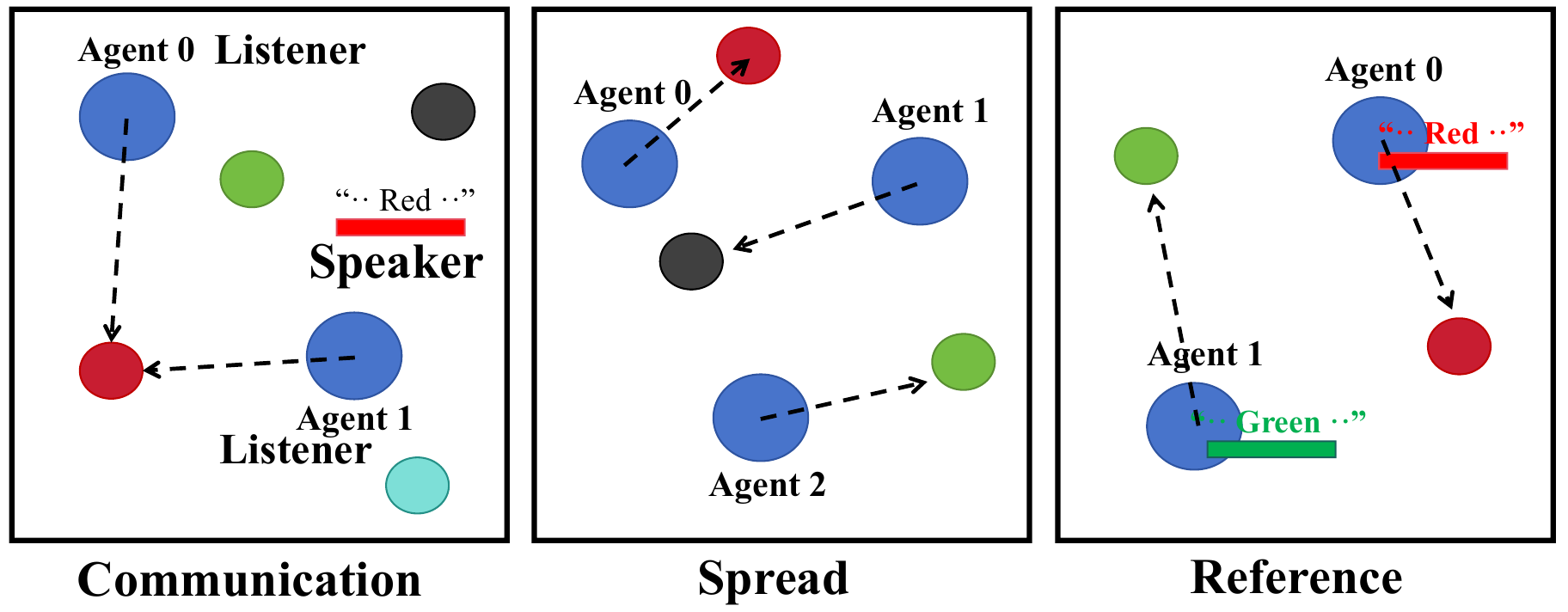}  % MPE
\caption{Typical Scenarios in Multi-Agent Particle Environment (MPE).}
\label{fig_MPE}
\end{figure}

\subsubsection{Several Widely-used Environments on MARL}
% ### 例举出做多智能体强化学习的人论文用的baseline和模拟环境 （主流的）

\textbf{Multi-Agent Particle Environment (MPE)} \cite{mordatch2017emergence,lowe2017multi,MPE9619081} is a versatile and widely-used MARL platform designed for research in both cooperative and competitive settings. Developed by OpenAI, it is primarily known for being the testing environment of the MADDPG algorithm \cite{lowe2017multi}. % , which was introduced at \texttt{NeurIPS} 2017. 
MPE is a time-discrete, space-continuous 2D platform designed for evaluating MARL algorithms. 

Figure \ref{fig_MPE}, initially derived from Malloy et al. \cite{MPE9619081}, illustrates various scenarios within the Multi-Agent Particle Environment (MPE), including tasks such as adversarial interactions, cooperative crypto, object pushing, and team-based world navigation. Compatible with the widely-used Gym interface, it supports a variety of tasks ranging from fully cooperative to mixed cooperative-competitive scenarios, such as \texttt{simple\_adversary, simple\_crypto, simple\_spread, simple\_speaker\_listener, and simple\_world\_comm}\footnote{Multi-Agent Particle Environment: \url{https://github.com/openai/multiagent-particle-envs/tree/master/multiagent/scenarios}.}. Each scenario highlights distinct cooperative and competitive dynamics among agents. MPE allows agents to interact and strategize within a visually simplistic UI where particles represent different entities. MPE is a open-source platform that widely adopted in the multi-agent system research, enabling extensive customization and contributing to its role as a standard tool for studying complex multi-agent dynamics.

Overall, MPE is a pivotal resource in the MARL community, offering a well-rounded platform for experimentation and algorithm comparison. Its design and functionality have made it an indispensable tool for researchers seeking to push the boundaries of what is possible in multi-agent systems.

\begin{figure}[h]
\centering
\includegraphics[width=3.5in]{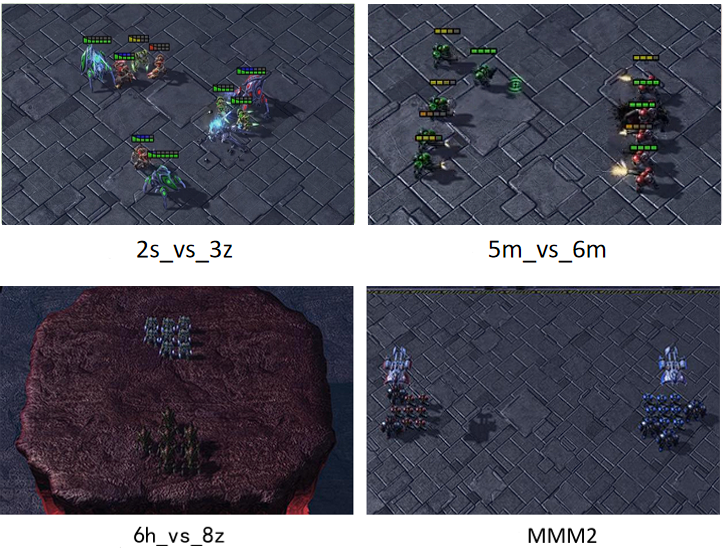} % SMAC
\caption{Several Typical Scenarios in StarCraft Multi-Agent Challenge (SMAC).}
\label{fig_SMAC}
\end{figure}

\textbf{StarCraft Multi-Agent Challenge (SMAC)}\footnote{StarCraft Multi-Agent Challenge (SMAC): \url{https://github.com/oxwhirl/smac}.} \cite{samvelyan19smac} is a widely-used benchmark for MARL that focuses on decentralized micromanagement tasks in the popular real-time strategy game StarCraft II\footnote{StarCraft II: \url{https://starcraft2.blizzard.com/}.}. In SMAC, multiple agents control individual units and must learn to cooperate and coordinate actions based on local, partial observations. The agents face complex challenges, including coordinating combat techniques like focus fire, kiting, and positioning, while the opponent is controlled by the built-in StarCraft II AI. SMAC emphasizes problems such as partial observability, decentralized decision-making, and multi-agent credit assignment. The environment is structured to simulate real-world scenarios where agents must learn to collaborate without full knowledge of the global state. Agents' observations are restricted to a limited field of view, forcing them to rely on local information for decision-making. As shown in Figure \ref{fig_SMAC}, these multi-agent cooperative decision-making environments are respectively \texttt{2s vs 3z}, \texttt{5m vs 6m}, % \texttt{3s vs 5z}, \texttt{Corridor}
\texttt{6h vs 8z}, \texttt{MMM2}, where the inside numbers represent the number of units and the letters represent the unit types in general. In recent years, SMAC has become a standard benchmark for evaluating MARL algorithms, offering a rigorous and challenging environment for advancing the field.

\begin{figure}[h]
\centering
\includegraphics[width=3.5in]{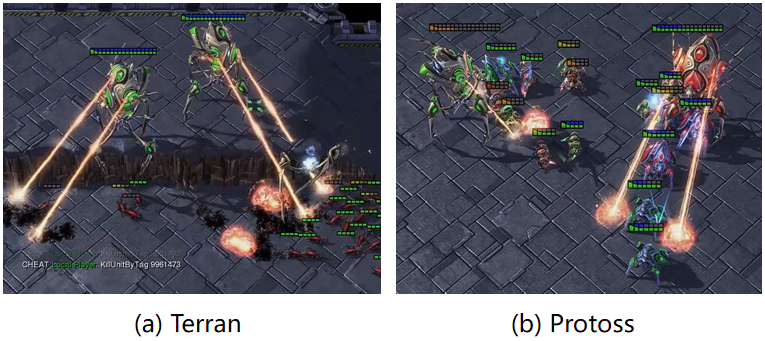}  % SMACv2
\caption{Several scenarios from SMACv2 showing agents battling the built-in AI.}
\label{fig_SMACv2}
\end{figure}

\textbf{StarCraft Multi-Agent Challenge 2 (SMACv2)}\footnote{StarCraft Multi-Agent Challenge 2 (SMACv2): \url{https://github.com/oxwhirl/smacv2}.} \cite{samvelyan19smac,vinyals2017starcraftiiPYSC2,Vinyals2019sc2} However, SMAC \cite{samvelyan19smac} has limitations, including insufficient stochasticity and partial observability, which allows agents to perform well with simple open-loop policies. To address these shortcomings, SMACv2 introduces \textit{procedural content generation (PCG)}, randomizing team compositions and agent positions, ensuring agents face novel, diverse scenarios. Several multi-agent decision-making scenarios are depicted in Figure \ref{fig_SMACv2}, which are from Benjamin et al. \cite{SMACv2}. This requires more sophisticated, closed-loop policies that condition on both ally and enemy information. Additionally, SMACv2 includes the \textit{Extended Partial Observability Challenge (EPO)}, where enemy observations are masked stochastically, forcing agents to adapt to incomplete information and communicate more effectively.
SMACv2 thus represents a major evolution of the original benchmark, addressing key gaps such as the lack of stochasticity and meaningful partial observability.
These changes make SMACv2 a more challenging environment, requiring agents to generalize across varied settings and improve coordination, communication, and decentralized decision-making. Overall, SMACv2 provides a more rigorous testbed for advancing the field of cooperative MARL.

% \textcolor{red}{TODO.}

\textbf{Google Research Football Environment (GFootball)} \cite{Kurach2020gfootball} is a state-of-the-art multi-agent simulation environment developed by the Google Research Brain Team. It is specifically designed for reinforcement learning research and is built on top of the open-source football game, GamePlay Football. GFootball is compatible with the OpenAI Gym API, making it a versatile tool not only for training intelligent agents but also for allowing players to interact with the built-in AI or trained agents using a keyboard or game controller. GFootball features an advanced, physics-based 3D football simulator where agents can be trained to play football, offering a challenging yet highly customizable platform for testing novel reinforcement learning algorithms and ideas. GFootball is tailored for multi-agent experiments and multiplayer scenarios, enabling the exploration of more complex interactions and strategies. GFootball supports various scenarios, including full-game simulations with varying difficulty levels, as well as simpler tasks in the Football Academy that focus on specific skills like passing or scoring. 

\begin{figure}[h]
\centering
\includegraphics[width=3.5in]{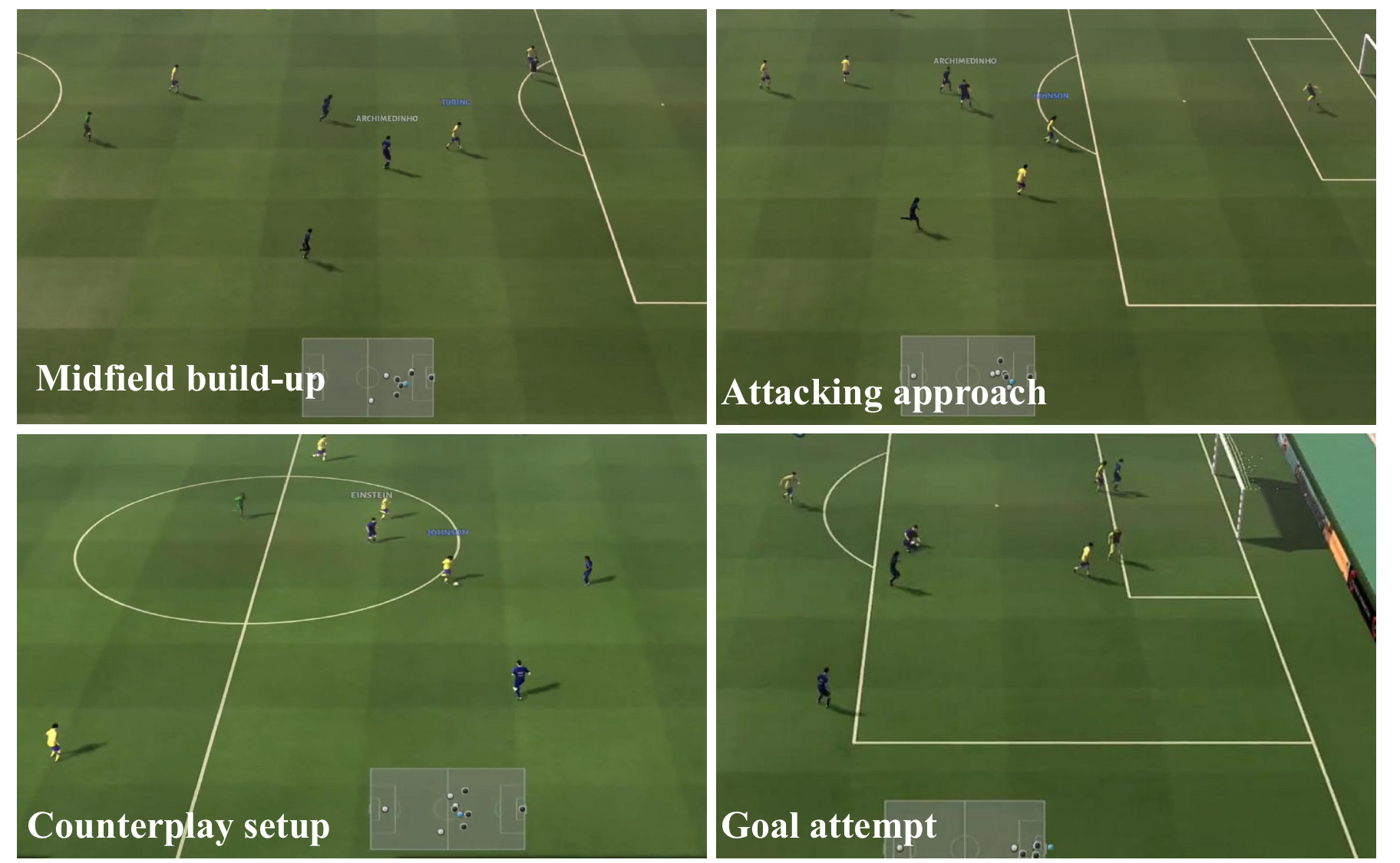} % gfootball.png
\caption{Typical examples of Training Scenarios in Football Academy.}
\label{fig_gfootball}
\end{figure}

Moreover, training agents for the "\textit{Football Benchmark}" can be quite challenging. To help researchers efficiently test and iterate on new ideas, researchers provide a toolset called "Football Academy", as illustrated in Figure \ref{fig_gfootball}, which includes a series of scenarios with varying levels of difficulty. These scenarios range from simple setups, such as a single player scoring against an open goal (e.g., approaching an open goal, scoring in an open goal, or scoring while running), to more complex team-based setups, where a controlled team must break through specific defensive formations (e.g., scoring while running against a goalkeeper, passing and shooting against a goalkeeper, and 3v1 against a goalkeeper). % , or performing a combination of running, passing, and shooting against a goalkeeper
Additionally, the toolset covers common situations in football matches, such as corner kicks, simple counterattacks, and complex counterattacks.
Lastly, as an famous open-source GitHub project\footnote{Google Research Football: \url{https://github.com/google-research/football}.}, it offers a unique opportunity for researchers and pushes the boundaries of AI research in a reproducible and scalable manner.

\textbf{Unity Machine Learning-Agents Toolkit}\footnote{Unity ML-Agents Toolkit: \url{https://github.com/Unity-Technologies/ml-agents}.} \cite{juliani2020unity} is an open-source platform designed to enable games and simulations to serve as environments for training intelligent agents. Built on Unity's powerful game engine, it supports a wide range of AI and machine learning methods, including reinforcement learning, imitation learning, and neuroevolution, through an intuitive Python API. The platform includes state-of-the-art algorithm implementations (based on PyTorch), allowing researchers and developers to train agents for 2D, 3D, and VR/AR applications.

\begin{figure}[h]
\centering
\includegraphics[width=3.5in]{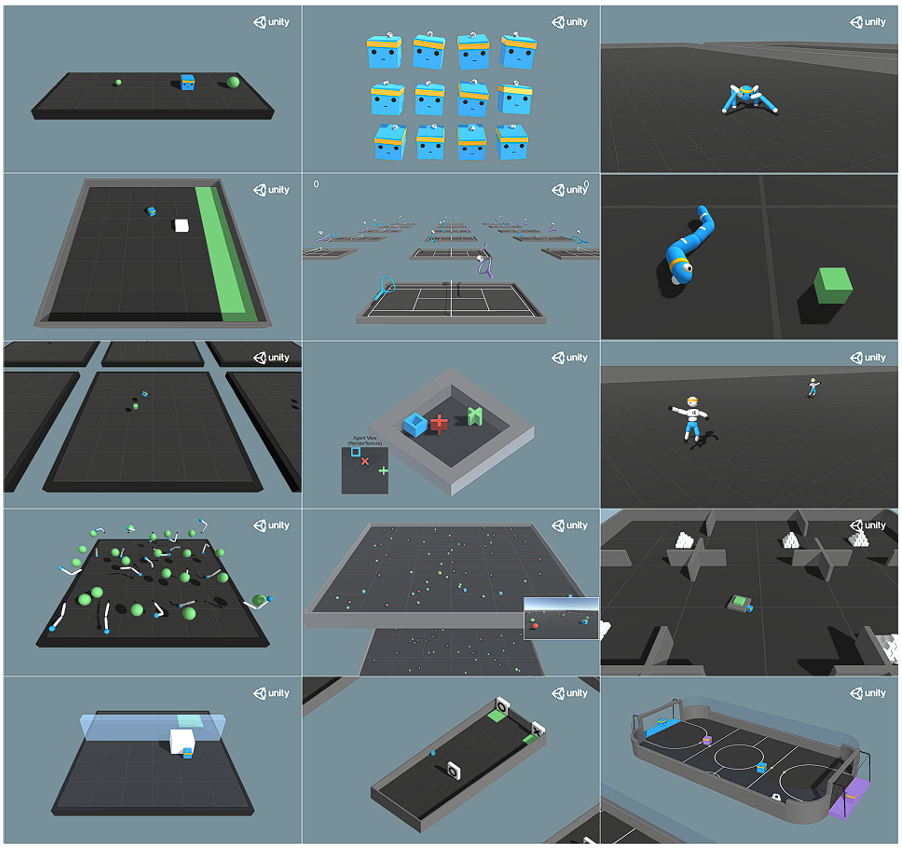}
\caption{Typical Training Scenarios in Unity Machine Learning-Agents Toolkit (released version: v0.11). From Left-to-right, up-to-down: (a) Basic, (b) 3DBall, (c) Crawler, (d) Push Block, (e) Tennis, (f) Worm, (g) Bouncer, (h) Grid World, (i) Walker, (j) Reacher, (k) Food Collector, (l) Pyramids, (m) Wall Jump, (n) Hallway, (o) Soccer Twos \cite{juliani2020unity}.}
\label{fig_Unity}
\end{figure}

ML-Agents is particularly useful for training NPC behaviors in diverse scenarios, automated testing of game builds, and evaluating game design decisions. It features a highly flexible simulation environment with realistic visuals, physics-driven interactions, and rich task complexity. By integrating tools for creating custom environments and supporting multi-agent and adversarial settings, the toolkit bridges the gap between AI research and practical applications in game development. 

As seen from Figure \ref{fig_Unity}, it depicts several typical multi-agent environments from the previous work of Juliani et al. \cite{juliani2020unity}. The platform also provides key components such as a Python API, Unity SDK, and pre-built environments, enabling users to customize and evaluate their algorithms in Unity's interactive and visually rich settings. With its versatility and accessibility, Unity ML-Agents Toolkit has become an indispensable resource for both AI researchers and game developers, driving innovation in artificial intelligence and simulation-based learning.

\begin{figure}[h]
\centering
\includegraphics[width=2.75in]{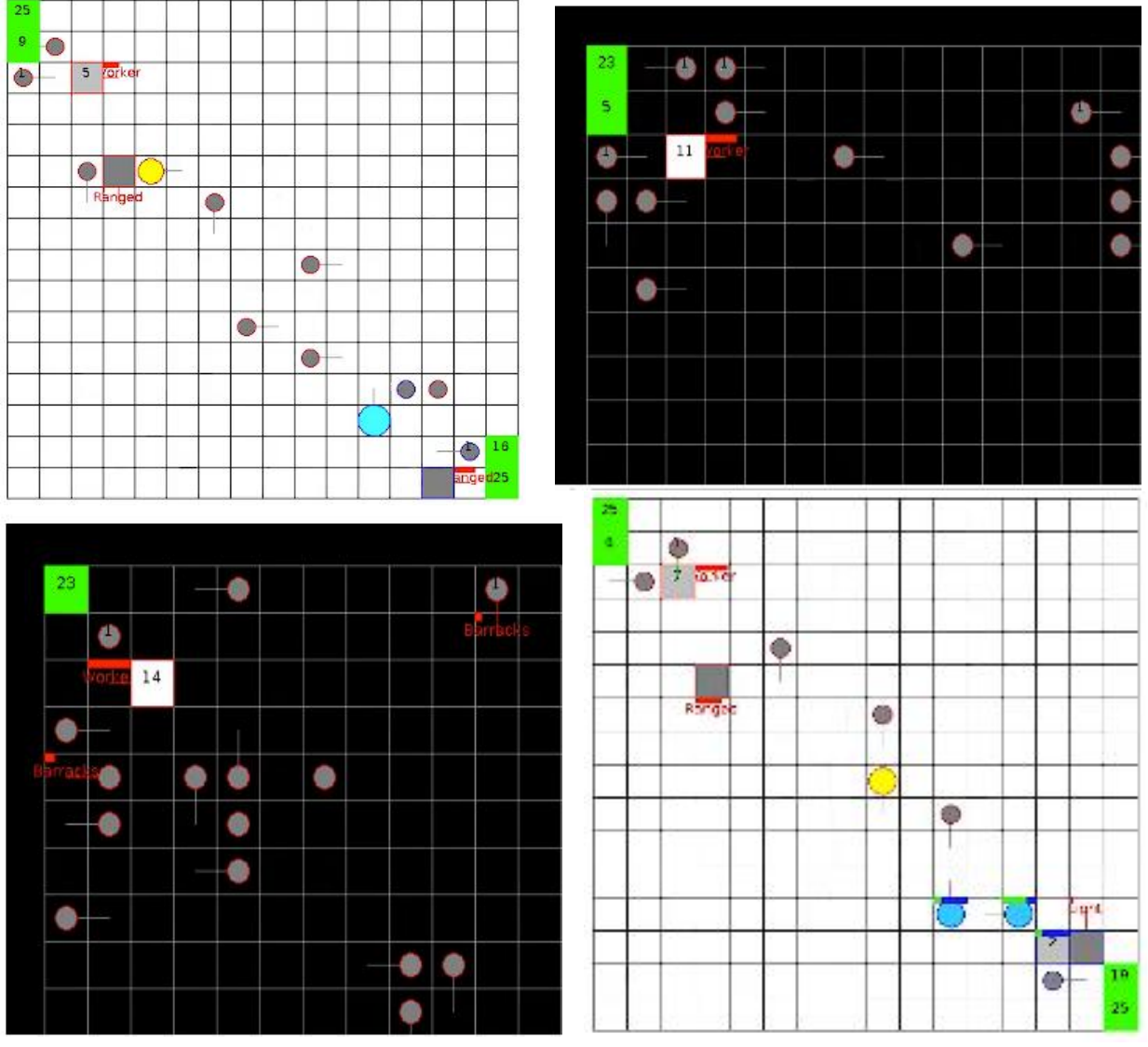} % GymRTS.png
\caption{Screenshot of our best-trained agent (top-left) playing against CoacAI (bottom-right), the 2020 $\mu$RTS AI competition champion \cite{huang2021gymmurts}.}
\label{fig_GymRTS}
\end{figure}

\textbf{Gym-Microrts}\footnote{Gym-Microrts: \url{https://github.com/kered9/gym-microrts}.} \cite{huang2021gymmurts} (pronounced "\textit{Gym-micro-RTS}") is a fast and affordable reinforcement learning (RL) platform designed to facilitate research in full-game Real-Time Strategy (RTS) games. Unlike traditional RTS research that demands extensive computational resources, Gym-µRTS allows training advanced agents using limited hardware, such as a single GPU and CPU setup, within reasonable timeframes. Figure \ref{fig_GymRTS} showcases a match between our best-trained agent (top-left) and CoacAI (bottom-right), the 2020 µRTS AI competition champion. The agent employs an efficient strategy, starting with resource harvesting and a worker rush to damage the enemy base, transitioning into mid-game combat unit production to secure victory.

The platform offers a simplified RTS environment that captures the core challenges of RTS games, including combinatorial action spaces, real-time decision-making, and partial observability. Gym-µRTS employs a low-level action space, enabling fine-grained control over individual units without AI assistance, which poses unique challenges and opportunities for RL algorithms. It supports Proximal Policy Optimization (PPO) and incorporates techniques like invalid action masking, action composition, and diverse training opponents to enhance training efficiency and agent performance.

Gym-µRTS has demonstrated its effectiveness by producing state-of-the-art DRL agents capable of defeating top competition bots, such as CoacAI. The platform is open-source and provides all necessary tools for researchers to explore and advance RL techniques in RTS games, making it a valuable resource for both AI researchers and gaming enthusiasts.

\textbf{MAgent}\footnote{MAgent: \url{https://github.com/geek-ai/MAgent}.} \cite{zheng2018magent} is an open-source platform specifically designed to support large-scale MARL research, with a focus on exploring Artificial Collective Intelligence (ACI). Unlike traditional MARL platforms, MAgent excels in handling scenarios involving hundreds to millions of agents, making it ideal for studying complex interactions and emergent behaviors in large populations. 

\begin{figure}[h]
\centering
\includegraphics[width=3.5in]{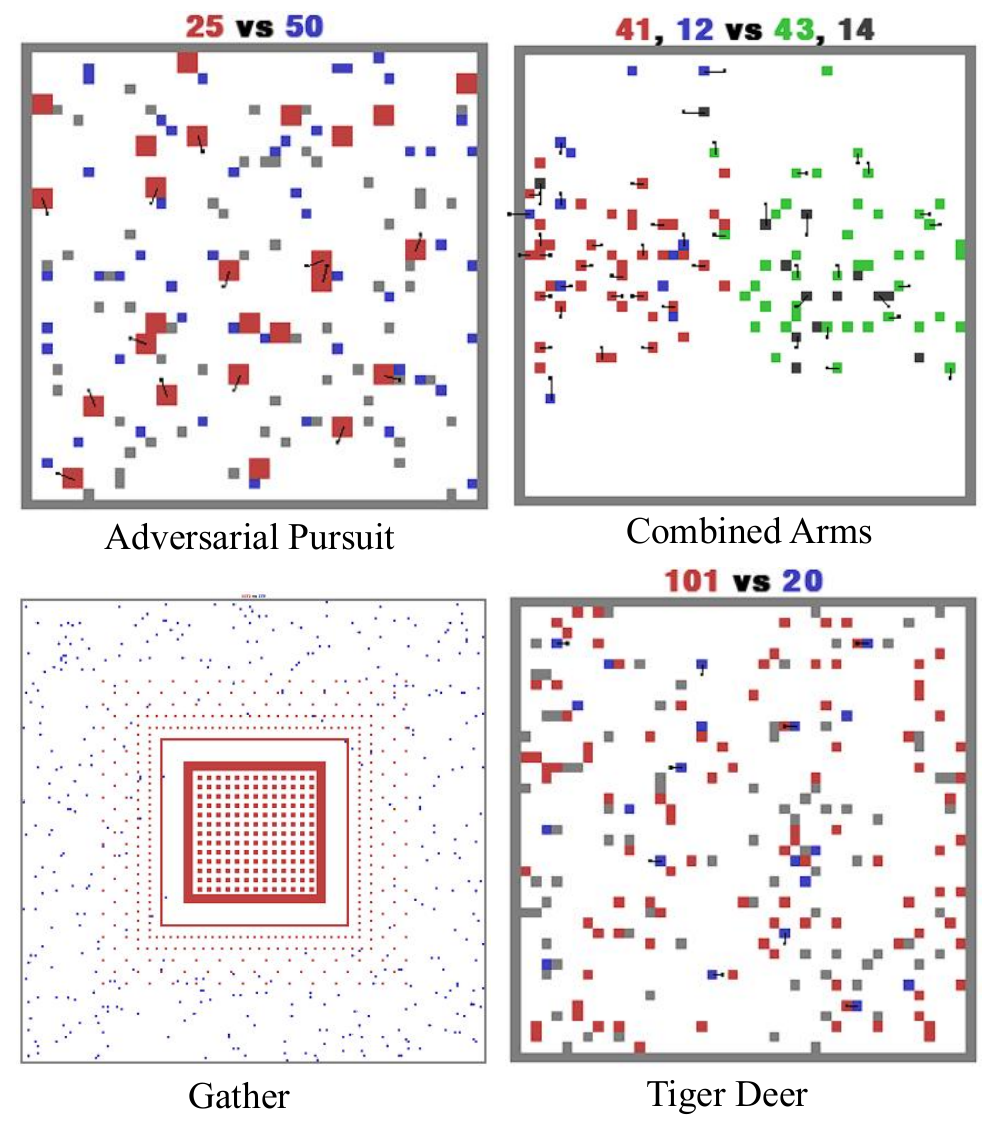} % MAgent.png
\caption{Illustrations of one of the typical running example in MAgent, called "\textit{Pursuit}" \cite{zheng2018magent}.}
\label{fig_MAgent}
\end{figure}

For instance, as shown in Figure \ref{fig_MAgent}, the "\textit{Pursuit}" scenario is a classic example designed to showcase the emergent cooperative behaviors of agents in a predator-prey environment. In this setup, predators work together to capture preys while the preys attempt to evade capture. Each predator receives rewards for successfully attacking a prey, while preys are penalized if caught. Over time, predators learn to form collaborative strategies, such as surrounding and trapping preys, highlighting the emergence of local cooperation.

The platform is based on a grid-world model where agents can perform actions such as moving, turning, or attacking, while perceiving both local and global information. Through a flexible Python interface, researchers can easily customize the state space, action space, and reward mechanisms, enabling the rapid creation of complex cooperative or competitive environments. MAgent comes with several built-in scenarios, such as pursuit, resource gathering, and team-based battles, which highlight emergent social behaviors like cooperative strategies, competitive dynamics, and resource monopolization.

MAgent is highly scalable, leveraging GPU-based parallelism to simulate large-scale interactions efficiently. It also provides intuitive visualization tools for real-time observation of agent behaviors, facilitating analysis and debugging. Its flexibility and scalability make MAgent a powerful tool for MARL research, enabling the study of large-scale agent interactions, emergent behaviors, and the dynamics of artificial societies.

\subsection{LLMs Reasoning-based Simulation Environments}
\label{llmsimenvs}
% LLM 多智能体的仿真模拟环境
% \textcolor{red}{TODO.}
LLMs-based multi-agent systems have become an essential tool for enhancing the collaboration, reasoning, and decision-making capabilities of autonomous agents \cite{LLMmarlsurvey}. By integrating LLMs with simulation platforms, researchers can create complex test environments to explore the interactions of multi-agent systems in various tasks and scenarios. These simulation environments not only provide rich dynamic testing scenarios but also promote the widespread application of LLMs in task planning, coordination, and execution. The following will introduce several widely used simulation platforms for LLM multi-agent systems.
% , which demonstrate how LLMs drive the co-evolution of intelligent systems and provide valuable experimental platforms for future multi-agent research.
% We list several widely-used environments based on LLMs Reasoning in the following:

\begin{figure}[h]
\centering
\includegraphics[width=3.5in]{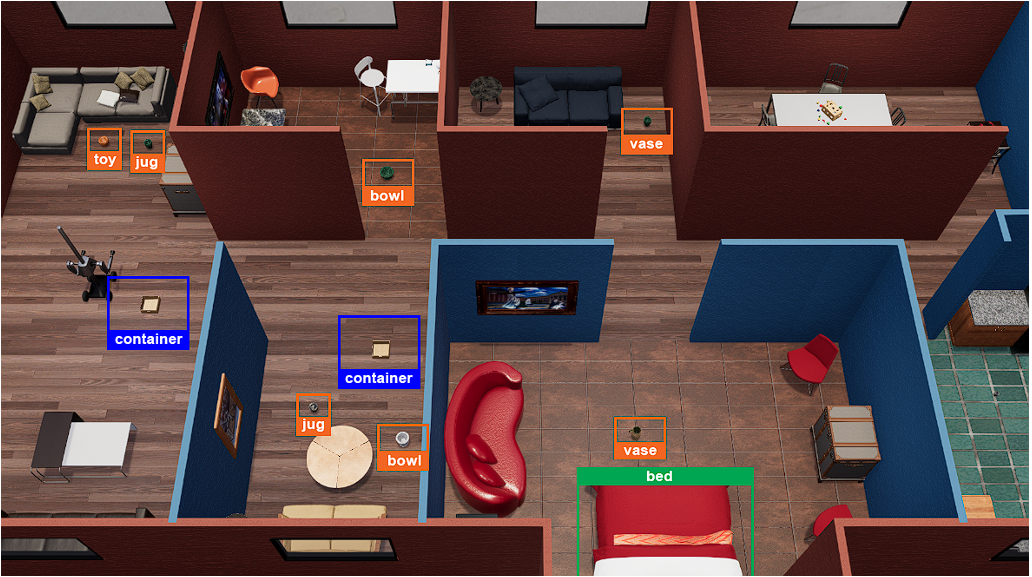}
\caption{An overview of one of the example task in ThreeDWorld Transport Challenge \cite{ThreeDWorldv2,ThreeDWorldv1}.}
\label{fig_ThreeDWorldTrans}
\end{figure}

\textbf{ThreeDWorld Multi-Agent Transport (TDW-MAT)}\footnote{ThreeDWorld Multi-Agent Transport: \url{https://github.com/threedworld-mit/tdw}.} \cite{ThreeDWorldv2,ThreeDWorldv1} is a simulation environment for multi-agent embodied task, which is extended from the ThreeDWorld Transport Challenge \cite{ThreeDWorldv2} and is designed for visually-guided task-and-motion planning in physically realistic settings. It operates within the ThreeDWorld (TDW) platform, which offers high-fidelity sensory data, real-time physics-driven interactions, and near-photorealistic rendering. In TDW-MAT, embodied agents are tasked with transporting objects scattered throughout a simulated home environment using containers, emphasizing the need for coordination, physics awareness, and efficient planning. For instance, in the common scenario shown in Figure \ref{fig_ThreeDWorldTrans}, the agent must transport objects scattered across multiple rooms and place them on the bed (marked with a green bounding box) in the bedroom.
% The environment supports complex multi-agent interactions and provides a robust platform for developing and evaluating advanced AI algorithms in tasks that require both physical realism and strategic cooperation.

\begin{figure}[h]
\centering
\includegraphics[width=3.5in]{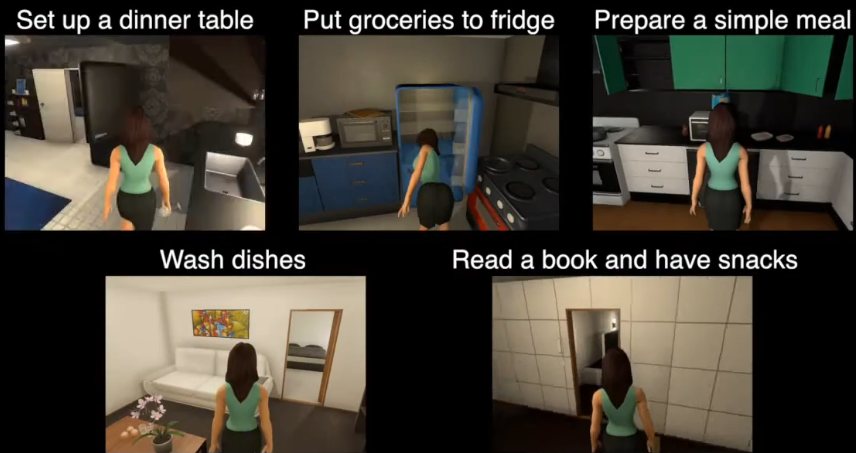}  % CWAH.png
\caption{An typical object-moving task leveraging LLMs-based embodied agents within the Communicative Watch-And-Help \cite{puig2021watchandhelp}.}
\label{fig_CWAH}
\end{figure}

\textbf{Communicative Watch-And-Help (C-WAH)}\footnote{Communicative Watch-And-Help: \url{https://github.com/xavierpuigf/watch_and_help}.} \cite{puig2021watchandhelp} is a realistic multi-agent simulation environment and an extension of the Watch-And-Help Challenge platform, VirtualHome-Social \cite{Puig2018VirtualHomeSH}. C-WAH places a greater emphasis on cooperation and enhances communication between agents compared to VirtualHome-Social. Built on the VirtualHome-Social, C-WAH simulates common household activities where agents must collaborate to complete tasks such as preparing meals, washing dishes, and setting up a dinner table. As shown in Figure \ref{fig_CWAH}, C-WAH supports both symbolic and visual observation modes, allowing agents to perceive their surroundings either through detailed object information or egocentric RGB and depth images. 
% The action space includes typical navigation and interaction commands, along with a new "send message" action, enabling agents to coordinate their efforts more effectively. 
% The goal is to accomplish all subgoals within 250 time steps, making it a robust testbed for exploring cooperative behaviors in a realistic, multi-agent setting

\begin{figure}[h]
\centering
\includegraphics[width=3.5in]{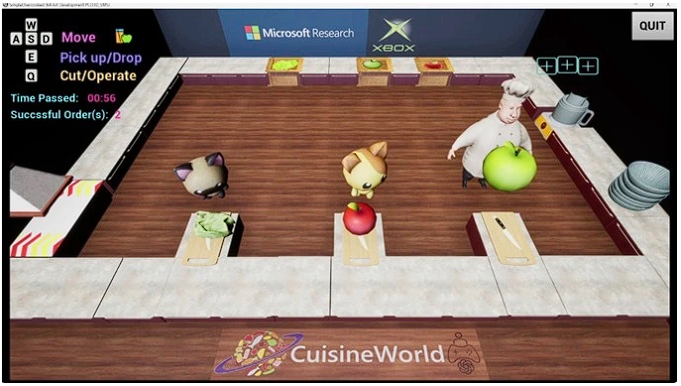}
\caption{An typical multi-agent cooperative scenario in the CuisineWorld platform \cite{gong2024mindagent}.}
\label{fig_CuisineWorld}
\end{figure}

\textbf{Cuisineworld}\footnote{Cuisineworld: \url{https://mindagent.github.io/}.} \cite{gong2024mindagent} is a virtual kitchen environment designed to evaluate and enhance multi-agent collaboration and coordination (i.e., the working efficiency) in a gaming context. As shown in Figure \ref{fig_CuisineWorld}, in this scenario, multiple agents work together to prepare and complete dish orders within a limited time frame. The tasks range from simple preparations, like chopping ingredients, to complex cooking processes that involve multiple tools and steps. CuisineWorld is equipped with a textual interface, and it supports various levels of difficulty, making it a flexible and robust testbed for assessing the planning and scheduling capabilities of Large Foundation Models (LFMs). The environment also introduces a "Collaboration Score" (CoS) metric to measure the efficiency of agent coordination as task demands increase, providing a comprehensive benchmark for multi-agent system performance in dynamic and cooperative settings.

% \subsubsection{Several Widely-used Environments on LLMs Reasoning}
% % TO MANY CONTENTS NEED WRITE.
% .
% \textcolor{red}{TODO.}
 % multi-agent platform designed to empower developers to build
\textbf{AgentScope}\footnote{AgentScope: \url{https://github.com/modelscope/agentscope}.} \cite{gao2024agentscope} is a innovative, robust and flexible multi-agent platform designed to empower developers in building advanced multi-agent systems by leveraging the potential of LLMs. At its core, the platform employs a process-based message exchange mechanism, simplifying the complexities of agent communication and collaboration. This approach ensures smooth and efficient agent interaction, enabling developers to focus on designing workflows rather than low-level details. The platform stands out for its comprehensive fault-tolerance infrastructure, which includes retry mechanisms, rule-based corrections, and customizable error-handling configurations. AgentScope also excels in multi-modal support, seamlessly integrating text, images, audio, and video into its workflows. By decoupling data storage and transfer, it optimizes memory usage and enhances scalability, making it ideal for applications requiring rich multi-modal interactions. Additionally, its actor-based distributed framework enables efficient parallel execution and supports hybrid deployments, bridging the gap between local and distributed environments with ease.

\begin{figure}[h]
\centering
\includegraphics[width=3.5in]{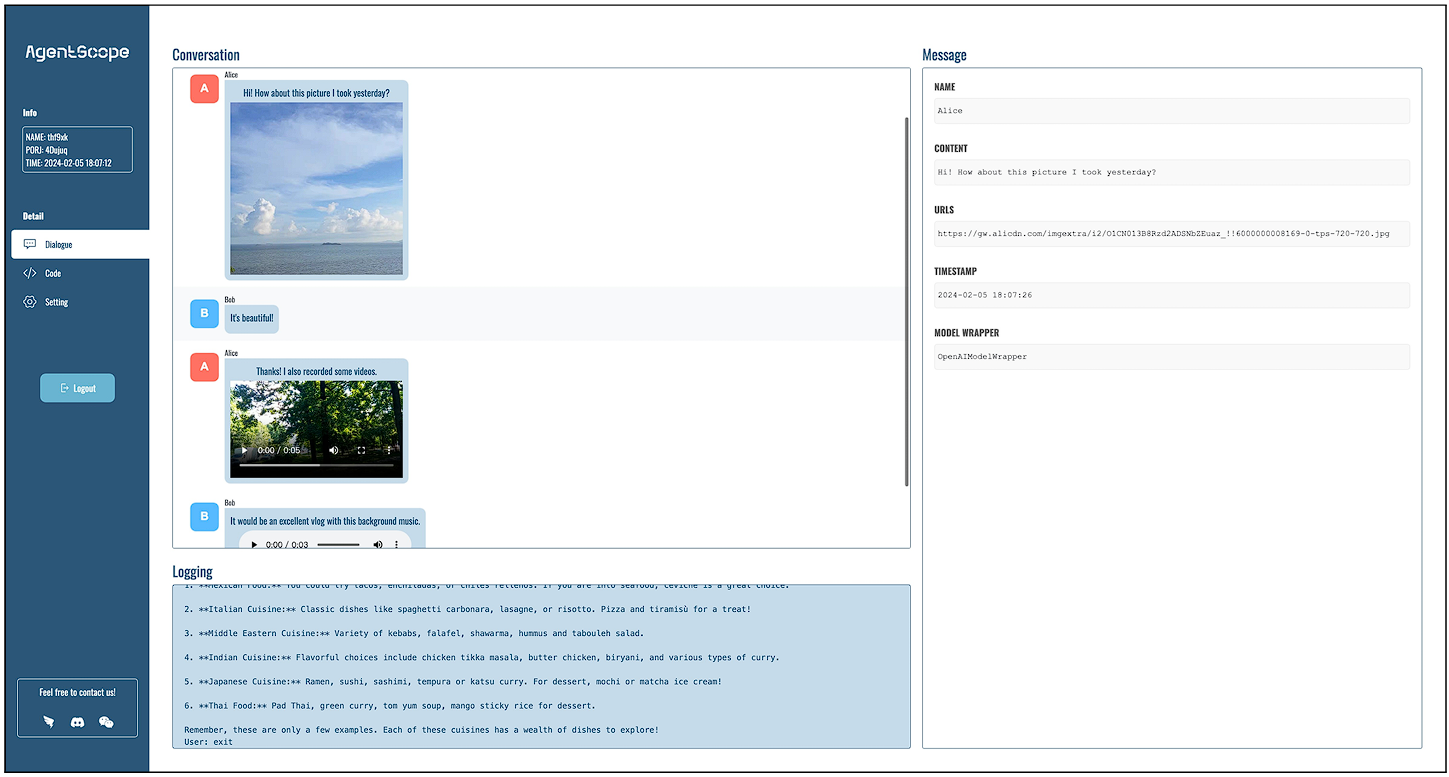}
\caption{The official multi-modal interaction Web UI page between agents in the AgentScope platform \cite{gao2024agentscope}.}
\label{fig_AgentScope}
\end{figure}

Moreover, to improve user interaction with multiple agents, AgentScope assigns distinct colors and icons to each agent, as shown in Figure \ref{fig_AgentScope}, providing clear visual differentiation in both the terminal and web interface. Designed with user accessibility in mind, AgentScope provides intuitive programming tools, including pipelines and message centers, which streamline the development process. Its interactive user interfaces, both terminal- and web-based, allow developers to monitor performance, track costs, and engage with agents effectively. These features position AgentScope as a state-of-the-art platform for advancing multi-agent systems, combining ease of use with cutting-edge technology.

\begin{figure}[h]
\centering
\includegraphics[width=3.5in]{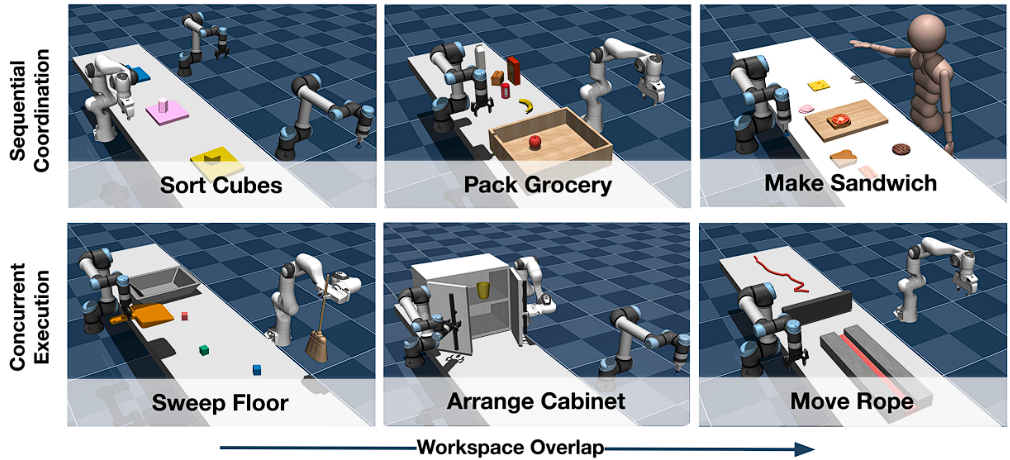}
\caption{An overview of RoCoBench, a collection of six multi-robot collaboration tasks set in a tabletop manipulation environment. The scenarios encompass a diverse range of collaborative challenges, each demanding distinct communication and coordination strategies between robots, incorporating familiar, intuitive objects designed to align with the semantic understanding capabilities of LLMs \cite{mandi2023roco}.}
\label{fig_RoCoBench}
\end{figure}

\textbf{RoCoBench}\footnote{RoCoBench: \url{https://project-roco.github.io/}.} RoCoBench is a benchmark platform, proposed by \textit{Mandi et al.} \cite{mandi2023roco}, designed to evaluate and enhance the collaborative capabilities of multi-robot systems powered by LLMs. Built as an extension to the RoCo project\footnote{RoCo Project: \url{https://project-roco.github.io/}.}, RoCoBench provides a realistic simulation environment where robotic agents interact and collaborate to complete complex tasks, as shown in Figure \ref{fig_RoCoBench}, such as sorting packages, assembling components, or preparing a workspace.
RoCoBench places a strong emphasis on communication-driven collaboration, integrating both symbolic and visual interaction modes to enable robots to perceive and respond to their environment effectively. Each robot is equipped with LLMs-powered reasoning, facilitating real-time dialogue and coordination. 
% As shown in Figure \ref{fig_RoCoBench}, t
Correspondingly, the benchmark introduces a "Collaboration Efficiency Metric" (CEM) to evaluate the effectiveness of multi-robot teamwork, taking into account factors like task completion time, resource allocation, and the quality of inter-robot communication. RoCoBench serves as a comprehensive platform for evaluating the potential of LLMs in driving dialectic multi-robot collaboration, offering a scalable and flexible environment for developers and researchers alike

\textbf{Generative Agents}\footnote{Generative Agents: \url{https://youmingyeh.github.io/cs-book/papers/generative-agents}.} \textit{Park et al.} \cite{ParkSocialSimulacra,park2023generative} introduces Generative Agents, a groundbreaking framework for simulating human behavior in interactive virtual worlds. These agents exhibit realistic individual and collective behaviors by incorporating dynamic memory, self-reflection, and action planning capabilities. The system leverages LLMs to store, retrieve, and synthesize memories into higher-level reasoning, enabling agents to adapt their actions based on personal experiences and evolving environmental changes.
As illustrated in Figure \ref{fig_RoCoBench}, they present an interactive sandbox environment called \textit{Smallville}, akin to "The Sims," where 25 distinct virtual agents live, interact, and carry out daily activities. Each agent has a detailed initial profile, including personal traits, relationships, and goals, stored as "seed memories." Agents engage in natural language-based dialogues and demonstrate social behaviors such as hosting events, making new acquaintances, and responding to user interventions. 
Generative Agents enable interactive applications in fields such as simulating realistic social dynamics for games and training simulations; designing dynamic, non-scripted virtual worlds for interactive systems; and exploring theories and behaviors in a controlled yet realistic virtual setting.
The evaluations revealed the critical role of \textit{memory retrieval, self-reflection, and action planning} in achieving coherent agent behaviors. Common issues, such as exaggerated responses and overly formal communication, were identified as areas for improvement.
Generative Agents push the boundaries of human behavior simulation, offering a robust framework for creating autonomous, memory-driven virtual agents.

\begin{figure}[h]
\centering
\includegraphics[width=3.5in]{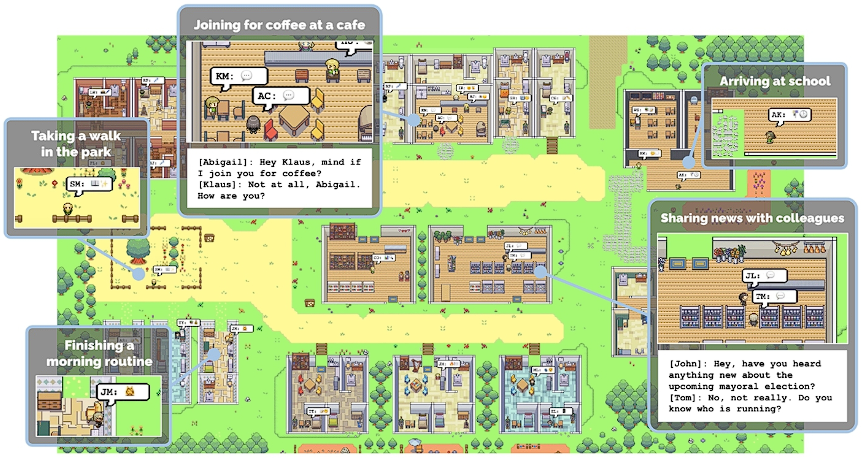}
\caption{Generative agents serve as realistic simulations of human behavior, designed for interactive applications. In a sandbox environment inspired by The Sims, twenty-five agents engage in activities such as planning their routines, sharing updates, building relationships, and collaborating on group events, while allowing users to observe and interact with them. \cite{ParkSocialSimulacra,park2023generative}.}
\label{fig_GenerativeAgents}
\end{figure}

\begin{figure}[h]
\centering
\includegraphics[width=3.5in]{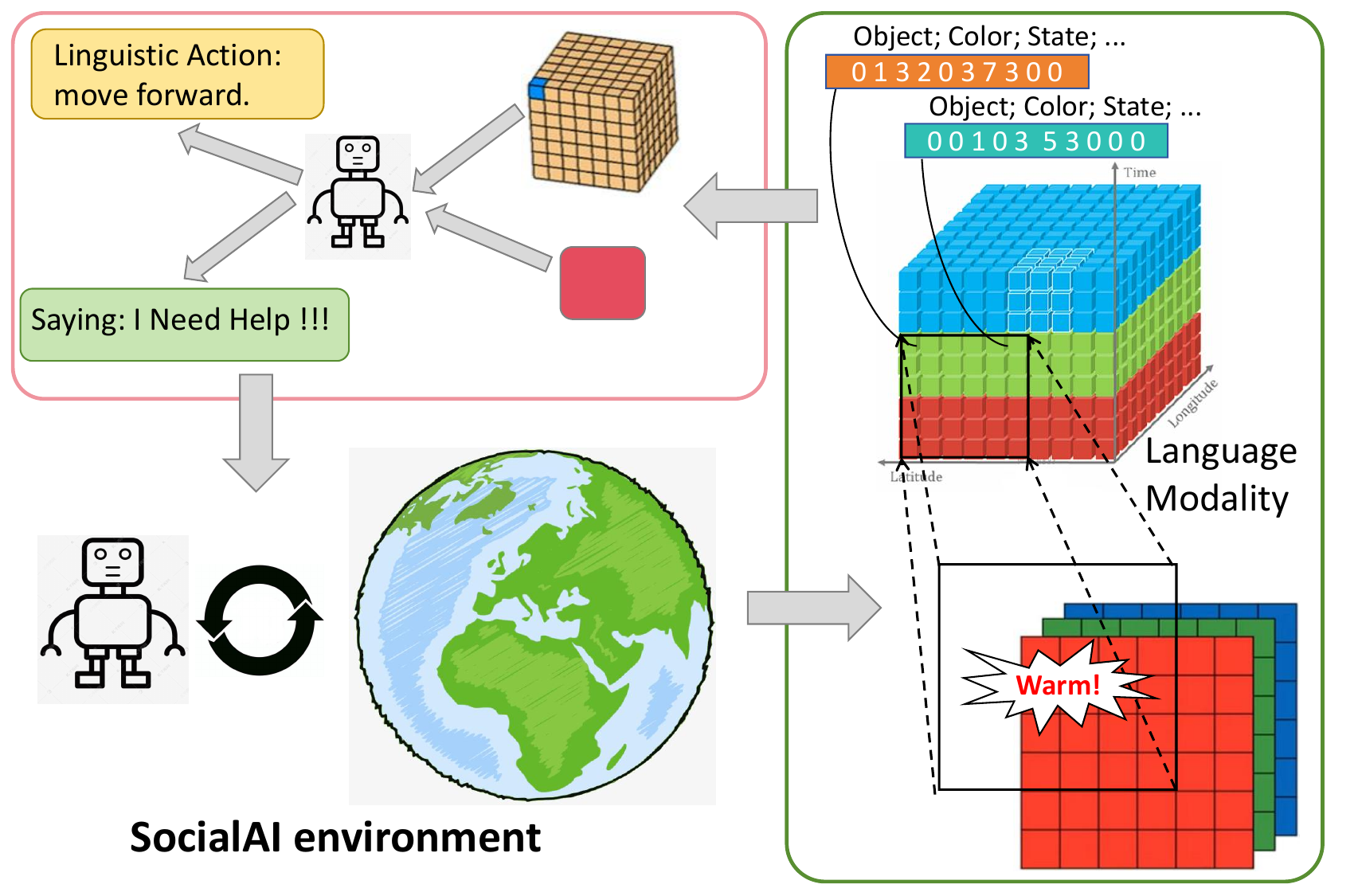} % png
\caption{A clear workflow of an agent acting in the SocialAI school environment \cite{kovac2023the,kovac2024socialai}.}
\label{fig_SocialAIschool}
\end{figure}

\textbf{SocialAI school}\footnote{SocialAI school project: \url{https://sites.google.com/view/socialai-school}.} \textit{Kovač et al.} \cite{kovac2023the,kovac2024socialai} introduces The SocialAI School, a novel framework designed to explore and develop socio-cognitive abilities in artificial agents.
% Drawing from developmental psychology theories, particularly the work of Vygotsky, Bruner, and Tomasello, t
The study emphasizes the importance of socio-cognitive skills as foundational to human intelligence and cultural evolution. As shown in Figure \ref{fig_SocialAIschool}, the SocialAI School provides a customizable suite of procedurally generated environments that enable systematic research into the socio-cognitive abilities required for artificial agents to interact with and contribute to complex cultures. 
% The SocialAI School leverages Tomasello's socio-cognitive concepts—social cognition, communication, and cultural learning—and Bruner's formats and scaffolding to design social environments. 
Built on MiniGrid, it provides procedural environments for RL and LLMs-based agents to study social skills like joint attention, imitation, and scaffolding. Open-source and versatile, it enables diverse research, including generalizing social inferences, role reversal studies, and scaffolded learning.
The SocialAI School represents a significant step toward enriching AI systems with socio-cognitive abilities inspired by human development.

\textbf{Welfare Diplomacy}\footnote{Welfare Diplomacy: \url{https://github.com/mukobi/welfare-diplomacy}.} \cite{mukobi2024welfare} is an innovative variant of the zero-sum game Diplomacy, designed to evaluate the cooperative capabilities of multi-agent systems. Unlike the original game, which focuses on a single winner, Welfare Diplomacy introduces a general-sum framework where players balance military conquest with investments in domestic welfare. Players accumulate Welfare Points (WPs) throughout the game by prioritizing welfare over military expansion, and their total utility at the end of the game is determined by these points, removing the notion of a single "winner". Welfare Diplomacy enables clearer assessments of cooperation and provides stronger incentives for training cooperative AI agents. Players take on the roles of European powers, negotiating, forming alliances, and strategizing to compete for key supply centers. Orders are submitted and resolved simultaneously, with the goal of controlling a specified number of SCs to achieve victory, emphasizing a balance between cooperation and betrayal. Based on these rules, Welfare Diplomacy implements themselves via an open-source platform, and develops zero-shot baseline agents using advanced language models like GPT-4 \cite{TrainingGPT35feedback,OpenAI2023GPT4}. Experiments reveal that while these agents achieve high social welfare through mutual demilitarization, they remain vulnerable to exploitation, highlighting room for future improvement.

\begin{figure}[h]
\centering
\includegraphics[width=2.75in]{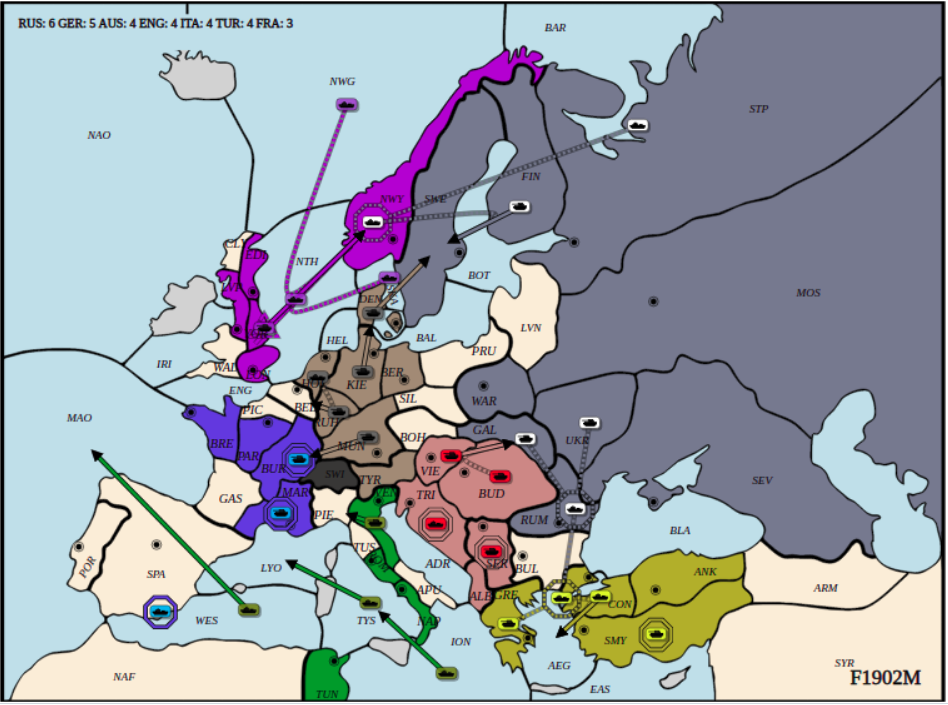} %WelfareDiplomacy
\caption{The Balkans in the Diplomacy map in Welfare Diplomacy
% In WD, there are likely bargaining problems such as between Austria (red), Russia (grey), Turkey (yellow), and Italy (green) over the allocation of neutral SCs SER, RUM, BUL, and GRE 
\cite{mukobi2024welfare}.}
\label{fig_WelfareDiplomacy}
\end{figure}

% \textbf{}\footnote{: \url{}.} \cite{} .

% \textbf{}\footnote{: \url{}.} \cite{} .

% \textbf{}\footnote{: \url{}.} \cite{} .

In summary, these cutting-edge LLMs-powered simulation environments—ranging from task-specific platforms like TDW-MAT \cite{ThreeDWorldv1,ThreeDWorldv2} and CuisineWorld \cite{gong2024mindagent} to socially-driven frameworks such as Generative Agents \cite{park2023generative} and the SocialAI School \cite{kovac2024socialai}—highlight the transformative potential of integrating advanced AI reasoning and multi-agent systems. By fostering research on collaboration, social cognition, and cooperative decision-making, these tools not only advance our understanding of AI's capabilities but also pave the way for practical applications in dynamic, real-world scenarios.

\section{Practice Applications of Multi-Agent Decision-Making}
\label{praappsMADM}
Multi-agent cooperative decision-making has a wide range of practical applications across various domains. In this section, we delve into the practical applications of multi-agent decision-making, focusing on how advanced methods, particularly multi-agent MARL, are employed to address complex challenges in dynamic and evolving environments. We explore the contributions of advanced multi-agent systems across domains such as agriculture, disaster rescue, military simulations, traffic management, autonomous driving, and multi-robot collaboration. A broad array of applications applications is presented in Table \ref{table_multi_agent_applications}. In the following, we will provide a detailed introduction to these applications, highlighting their impact and potential for future advancements. % are summarized % comprehensive set of

\begin{table*}[ht]
\renewcommand{\arraystretch}{1.5}
\caption{Categorized Applications of MARL and LLMs \\ in Diverse Domains.} % Applications of Multi-Agent Decision-Making Techniques
\centering
\scalebox{0.8}{
\begin{tabular}{lll}
\hline
\textbf{Category} & \textbf{Application Area} & \textbf{Works / References} \\ \hline
\multirow{10}{*}{\begin{tabular}[c]{@{}l@{}}MARL- \\ based MAS\end{tabular}} & \begin{tabular}[c]{@{}l@{}} Smart Precious Agriculture \\ \& Disaster Rescue \end{tabular} & \begin{tabular}[c]{@{}l@{}}Seewald et al. \cite{seewald2024multiagentagri}, Qazzaz et al. \cite{qazzaz2024rescueUAV}, Samad et al. \cite{samad2018multi}, Boubin  et al. \cite{boubin2021programming}, \\ Li et al. \cite{li2023multi}, Mahajan et al. \cite{mahajan2024comparative}\end{tabular} \\ 
% & Disaster Rescue & Qazzaz et al. \cite{qazzaz2024rescueUAV}, Samad et al. \cite{samad2018multi} \\ 
& Military Confrontations & \begin{tabular}[c]{@{}l@{}}Qi et al. \cite{qimilitaryMARL}, Benke et al. \cite{BenkeAdversarialMAS}, Sutagundar et al. \cite{SutagundarMilitary}, Vangaru et al. \cite{vangaru2024multi}, \\ Wang et al. \cite{wang2024marl}, MaCA \cite{gaoMaCA}, SMAC \cite{samvelyan19smac}, SMAC-v2 \cite{SMACv2}\end{tabular} \\ 

% & \begin{tabular}[c]{@{}l@{}}Efficient Limited- \\ Bandwidth Communication\end{tabular} & Wang et al. \cite{pmlrwang20iIMAC} \\ 

% & \begin{tabular}[c]{@{}l@{}}UAV Swarm Communications \\ Against Jamming\end{tabular} & Lv et al. \cite{LVMARLonUAVswarm} \\ 

& \begin{tabular}[c]{@{}l@{}}UAV Pursuit-Evasion \\ \& Swarm Communications \\ \& Navigation\end{tabular} & \begin{tabular}[c]{@{}l@{}}Kouzeghar \cite{kouzeghar2023UAVpursuit}, Alexopoulos et al. \cite{AlexanderUAVPE}, Luo et al. \cite{LUO2024MultiUAVPE}, Lv et al. \cite{LVMARLonUAVswarm}, \\ Xue et al. \cite{UAVNavigation1}, Rezwan et al. \cite{UAVNavigation2}, Baroomi et al. \cite{UAVNavigation3}\end{tabular} \\ 

% & Multiple UAVs Navigation & \cite{UAVNavigation1,UAVNavigation2,UAVNavigation3} \\ 

& Traffic Signal/Flow Control & \begin{tabular}[c]{@{}l@{}}Wang \cite{WangMARLautoDriving}, Chu et al. \cite{ChuMARLonTrafficSingal}, Aboueleneen et al. \cite{aboueleneen2024traffic}, Yu et al. \cite{yu2024traffic}, \\ Sun et al. \cite{sun2024traffic}, Azfar et al. \cite{azfar2024traffic}, Bokade et al. \cite{bokade2024offlight}, Kwesiga et al. \cite{kwesiga2024traffic}, \\ Zhang et al. \cite{Zhang2024traffic}\end{tabular} \\ 

& Autonomous Driving & \begin{tabular}[c]{@{}l@{}}Xue et al. \cite{xue2023twostageautodrive}, Liu et al. \cite{liu2024diversedriving}, Wen et al. \cite{LuMARLautoDriving}, Jayawardana et al. \cite{JayawardanaMBLautodrive}, \\ Liu et al. \cite{liu2024cooperative}, Formanek et al. \cite{formanek2024putting}, Zhang et al. \cite{zhang2024multiagent}, Kotoku et al. \cite{kotoku2024decentralized}, \\ Hua et al. \cite{hua2024communication}\end{tabular} \\ % , et al. \cite{}, et al. \cite{}, et al. \cite{}

& Multiple Robots Collaborative & \begin{tabular}[c]{@{}l@{}}Georgios et al. \cite{collabrobots9431107}, Silva et al. \cite{collabrobots10039365}, Huang et al. \cite{collabrobots9107997}, Cena et al. (SMART) \\ \cite{MASCENA20134737}, Kevin (SCRIMMAGE) \cite{SCRIMMAGE}, Liu et al. \cite{liu2024grounded}\end{tabular} \\ \hline

\multirow{5}{*}{\begin{tabular}[c]{@{}l@{}}LLMs- \\ based MAS\end{tabular}} & Multi-Agent Collaboration & \begin{tabular}[c]{@{}l@{}}Wu et al. (AutoGen) \cite{wu2023autogen}, Xiao et al. (CoE) \cite{xiao2024chainofexperts}, Chen et al. (AgentV- \\ erse) \cite{chen2024agentverse}, Liu et al. (DyLAN) \cite{liu2024dynamic}, Zhang et al. (CoELA) \cite{zhang2024building}\end{tabular} \\ 
& Gaming Interaction & \begin{tabular}[c]{@{}l@{}}Xu et al. (LLM-Werewolf) \cite{xu24pmlrLLMWerewolf}, Gong et al. (MindAgent) \cite{gong2024mindagent}, Xie et al. \\ \cite{xie2024differentai}, Lin et al. \cite{lin2024outcomes}, Jia et al. (GameFi) \cite{jia2024decentralized}, Yin et al. (MIRAGE)  \\ \cite{yin2025mirage}, Zhang et al. (DVM) \cite{zhang2025dvm}, Bonorino et al. \cite{gonzalezbonorino2025llm}\end{tabular} \\ 

& Autonomous Driving & \begin{tabular}[c]{@{}l@{}}Zheng et al. (PlanAgent) \cite{zheng2024planagent}, Luo et al. (SenseRAG) \cite{luo2025senserag}, Mahmud et al. \cite{mahmud2025integrating}, \\ Peng et al. (LearningFlow) \cite{peng2025learning}, Karagounis et al. \cite{karagounis2024leveraging}, Luo et al. \cite{luo2025whatshappening}, \\ Gao et al. \cite{gao2025application}, Hegde et al. \cite{hegde2025distilling}\end{tabular} \\ 

& Multi-Modal Application & \begin{tabular}[c]{@{}l@{}}Wang et al. (LangGraph) \cite{wang2024LangGraph,wangJ2024LangGraph}, Zhang et al. (CrewAI) \cite{zhang2025crewai,duan2024crewaiLangGraph}, \\ Zheng et al. (PlanAgent) \cite{zheng2024planagent}, Wang et al. (MLLM-Tool) \cite{wang2024mllmtool}\end{tabular} \\ 

% & \begin{tabular}[c]{@{}l@{}}Role-Playing Multi- \\ Agent Systems\end{tabular} & CrewAI \cite{zhang2025crewai,duan2024crewaiLangGraph} \\ 
\hline
\end{tabular}}
\label{table_multi_agent_applications}
\end{table*}

% ,  et al. \cite{}

\subsection{MARL-based Intelligent Applications}
Recently, a variety of MARL methods have been developed, fostering efficient collaboration, strategic learning, and adaptive problem-solving in multi-agent systems \cite{wai2018MARL,Du2021Survey,NING2024Survey,Zhu2024Survey}. Below, we highlight notable contributions that demonstrate the application of MARL in enhancing multi-agent collaboration and performance

% exploration
In \textit{smart precious agriculture and continuous pest disease detection}, Seewald et al. \cite{seewald2024multiagentagri} addressed the challenge of continuous exploration for multi-agent systems with battery constraints by integrating ergodic search methods with energy-aware coverage. In disaster rescue, Qazzaz et al. \cite{qazzaz2024rescueUAV} proposed a novel technique using a reinforcement learning multi Q-learning algorithm to optimize UAV connectivity operations in challenging terrain. Samad et al. \cite{samad2018multi} presents a cloud-based multi-agent framework for efficiently managing aerial robots in disaster response scenarios, aiming to optimize rescue efforts by autonomously processing real-time sensory data to locate and assist injured individuals. % field emergency 
% They introduces a Strategic Planning Agent for efficient path planning and collision awareness, and a Real-time Adaptive Agent to maintain optimal connection with cellular base stations.

% ##################### 这里还得加点rescue task 的方法

In \textit{military confrontations}, Qi et al. \cite{qimilitaryMARL} designed a distributed MARL framework based on the actor-work-learner architecture, addressing the issues of slow sample collection and low training efficiency in MARL within the MaCA \cite{gaoMaCA} and SMAC 3D realtime gaming  \cite{samvelyan19smac,SMACv2} military simulation environments. Benke et al. \cite{BenkeAdversarialMAS} proposed a computational model for agent decision-making that incorporates strategic deception, enhancing the representation of deceptive behaviors in multi-agent simulations for military operations research. Sutagundar et al. \cite{SutagundarMilitary} proposed a Context Aware Agent based Military Sensor Network (CAMSN) to enhance multi-sensor image fusion, using node and sink-driven contexts, forming an improved infrastructure for multi-sensor image fusion.

In \textit{efficient limited-bandwidth communication} field, Wang et al. \cite{pmlrwang20iIMAC} proposed a method called IMAC (Informative Multi-Agent Communication) to address the problem of limited-bandwidth communication in MARL. 
% This approach learns efficient communication protocols and a weight-based scheduler to achieve effective communication under bandwidth constraints. 

In the research of \textit{UAV swarm communications against jamming}, Lv et al. \cite{LVMARLonUAVswarm} proposed a MARL-based scheme to optimize relay selection and power allocation. This strategy leverages network topology, channel states, and shared experiences to improve policy exploration and stability, ultimately enhancing anti-jamming performance. 

In \textit{UAV pursuit-evasion} \cite{UAVNavigation1,UAVNavigation2,UAVNavigation3}, Kouzeghar \cite{kouzeghar2023UAVpursuit} proposed a decentralized heterogeneous UAV swarm approach for multi-target pursuit using MARL technique and introduced a variant of the MADDPG \cite{MADDPG} to address pursuit-evasion scenarios in non-stationary environments with random obstacles. Alexopoulos et al. \cite{AlexanderUAVPE} addressed the challenge of pursuit-evasion games involving two pursuing and one evading unmanned aerial vehicle (UAV) by introducing a hierarchical decomposition of the game.  % At a higher collaboration level, the pursuers select their optimal behavioral strategies (e.g., pursue and herd), forming a three-player non-cooperative dynamic game. This game is then solved at a subordinate level, enabling the pursuers to change behaviors intelligently based on game-theoretical solutions.
Luo et al. \cite{LUO2024MultiUAVPE} proposed a cooperative maneuver decision-making method for multi-UAV pursuit-evasion scenarios using an improved MARL approach, which incorporates an enhanced CommNet network with a communication mechanism to address multi-agent coordination.

% #####################  加点UAVs navigatio 方法
% In \textit{multiple UAVs navigation},  .

In large-scale traffic signal/flow control, Wang \cite{WangMARLautoDriving} proposed a curiosity-inspired algorithm to optimize safe and smooth traffic flow in autonomous vehicle on-ramp merging; Chu et al. \cite{ChuMARLonTrafficSingal} proposed a fully scalable and decentralized multi-agent deep reinforcement learning algorithm based on the advantage actor-critic (A2C) method. 
% This approach addresses the challenges of observability and learning difficulty, demonstrating optimality and robustness in managing complex urban traffic networks. 

In \textit{autonomous driving} area, a large number of superior multi-agent decision-making algorithms and models are continuously being explored and devised. Xue et al. \cite{xue2023twostageautodrive} developed a two-stage system framework for improving Multi-Agent Autonomous Driving Systems (MADS) by enabling agents to recognize and understand the Social Value Orientations (SVOs) of other agents. Liu et al. \cite{liu2024diversedriving} proposed the Personality Modeling Network (PeMN), which includes a cooperation value function and personality parameters to model diverse interactions in highly interactive scenarios, addressing the issue of diverse driving styles in autonomous driving. Wen et al. \cite{LuMARLautoDriving} proposed a safe reinforcement learning algorithm called Parallel Constrained Policy Optimization (PCPO) based on the actor-critic architecture to address the issues of unexplainable behaviors and lack of safety guarantees in autonomous driving. Jayawardana et al. \cite{JayawardanaMBLautodrive} proposed enabling socially compatible driving by leveraging human driving data to learn a social preference model, integrating it with reinforcement learning-based AV policy synthesis using Social Value Orientation theory.

In \textit{multiple robots collaborative} fields, Georgios et al. \cite{collabrobots9431107} introduces a novel cognitive architecture for large-scale multi-agent Learning from Demonstration (LfD), leveraging Federated Learning (FL) to enable scalable, collaborative, and AI-driven robotic systems in complex environments. Silva et al. \cite{collabrobots10039365} address the challenges and limitations in evaluating intelligent collaborative robots for Industry 4.0. The review emphasizes the urgent need for improved evaluation methods and standards to account for the complexities posed by human variability, AI integration, and advanced control systems in industrial environments. Huang et al. \cite{collabrobots9107997} presents a multi-agent reinforcement learning approach using the MADDPG algorithm, enhanced with an experience sample optimizer, to train swarm robots for autonomous, collaborative exploration on Mars. This approach outperforms traditional DRL algorithms in efficiency as the number of robots and targets increases. 
The SMART multi-agent robotic system \cite{MASCENA20134737} is a comprehensive and advanced platform designed for executing coordinated robotic tasks. It integrates both hardware components, such as robots and IP-Cameras, and software agents responsible for image processing, path planning, communication, and decision-making. By utilizing Work-Flow Petri Nets for modeling and control, the system effectively ensures coordination and successful task execution even in unstructured environments.

Furthermore, the well-known project, Simulating Collaborative Robots in a Massive Multi-agent Game Environment (SCRIMMAGE)\footnote{SCRIMMAGE Web: \url{http://www.scrimmagesim.org/}.} \cite{SCRIMMAGE}, tackles the high costs of field testing robotic systems by offering a flexible and efficient simulation environment specifically designed for mobile robotics research. Unlike many existing simulators that are primarily designed for ground-based systems with high-fidelity multi-body physics models, SCRIMMAGE focuses on simulating large numbers of aerial vehicles, where precise collision detection and complex physics are often unnecessary. SCRIMMAGE is designed to be highly adaptable, with a plugin-based architecture that supports various levels of sensor fidelity, motion models, and network configurations. This flexibility allows the simulation of hundreds of aircraft with low-fidelity models or a smaller number with high-fidelity models on standard consumer hardware. Overall, SCRIMMAGE\footnote{SCRIMMAGE project: \url{https://github.com/gtri/scrimmage}.} provides a robust and scalable solution for testing and refining robotic algorithms in a controlled virtual environment, significantly reducing the risks and costs associated with physical testing.

% \textbf{Learning before Interaction (LBI):} 
Liu et al. \cite{liu2024grounded} proposed the Learning before Interaction (LBI) framework, a novel approach designed to enhance multi-agent decision-making through generative world models. Traditional generative models struggle with trial-and-error reasoning, often failing to produce reliable solutions for complex multi-agent tasks. To address this limitation, LBI integrates a language-guided simulator into the MARL pipeline, enabling agents to learn grounded reasoning through simulated experiences. 
LBI consists of a world model composed of a dynamics model and a reward model. The dynamics model incorporates a vector quantized variational autoencoder (VQ-VAE) \cite{VQVAE} for discrete image representation and a causal transformer to autoregressively generate interaction transitions. Meanwhile, the reward model employs a bidirectional transformer trained on expert demonstrations to provide task-specific rewards based on natural language descriptions. LBI further distinguishes itself by generating explainable interaction sequences and reward functions, providing interpretable solutions for multi-agent decision-making problems. By addressing challenges such as the compositional complexity of MARL environments and the scarcity of paired text-image datasets, LBI represents a significant advancement in the field.

% % ######################################################
% % 除了实际应用，还要很多原理性的支持工作.
% In addition to practical applications, there are many fundamental theoretical works that support multi-agent cooperative decision-making.
% % ######################################################
% For instance, d
Ye et al. \cite{YemultiUAVsys} proposed an adaptive genetic algorithm (AGA) that dynamically adjusts crossover and mutation populations, leveraging the Dubins car model and state-transition strategies to optimize the allocation of heterogeneous UAVs in suppression of enemy air defense missions.
Radac \textit{et al.} combine two model-free controller tuning techniques linear virtual reference feedback tuning (VRFT) and nonlinear state-feedback Q-learning as a novel mixed VRFT-Q learning method \cite{reinforcementDQNIJSS}. VRFT is initially employed to determine a stabilizing feedback controller using input-output experimental data within a model reference tracking framework. Subsequently, reinforcement Q-learning is applied in the same framework, utilizing input-state experimental data gathered under perturbed VRFT to ensure effective exploration. Extensive simulations on position control of a two-degrees-of-motion open-loop stable multi input-multi output (MIMO) aerodynamic system (AS) demonstrates the mixed VRFT-Q's significant performance improvement over the Q-learning controllers and the VRFT controllers.

% mutual information (MI) 主要针对智能体之间的信息交流，和参数共享的时机
To address the lack of a general metric for quantifying policy differences in MARL problems, Hu et al. \cite{hu2024pdmultiagent} proposed the Multi-Agent Policy Distance (MAPD), a tool designed to measure policy differences among agents. Additionally, they developed a Multi-Agent Dynamic Parameter Sharing (MADPS) algorithm based on MAPD, demonstrating its effectiveness in enhancing policy diversity and overall performance through extensive experiments. 
To addresses the challenge of cooperative MARL in scenarios with dynamic team compositions, Wang et al. \cite{Wang2023Mutual} propose using mutual information as an augmented reward to encourage robustness in agent policies across different team configurations. They develop a multi-agent policy iteration algorithm with a fixed marginal distribution and demonstrate its strong zero-shot generalization to dynamic team compositions in complex cooperative tasks.
Progressive Mutual Information Collaboration (PMIC)\footnote{PMIC code: \url{https://github.com/yeshenpy/PMIC}.} is a novel framework that leverages mutual information (MI) to guide collaboration among agents, thereby enhancing performance 
in mult-agent cooperative tasks \cite{li2022pmic}. The key innovation of is its dual MI objectives: maximizing MI associated with superior collaborative behaviors and minimizing MI linked to inferior ones, ensuring more effective learning and avoiding sub-optimal collaborations.
Wai et al. \cite{wai2018MARL} proposes a novel double averaging primal-dual optimization algorithm for MARL, specifically targeting decentralized applications like sensor networks and swarm robotics. The MARL algorithm enables agents to collaboratively evaluate policies by incorporating neighboring gradient and local reward information, achieving fast finite-time convergence to the optimal solution in decentralized convex-concave saddle-point problems.
To address the challenge of sparse rewards in MARL, Kang et al. \cite{kang2024dpm} introduce the Dual Preferences-based Multi-Agent Reinforcement Learning (DPM) framework. DPM extends preference-based reinforcement learning (PbRL) by incorporating dual preference types-comparing both trajectories and individual agent contributions-thereby optimizing individual reward functions more effectively. DPM also leverages LLMs to gather preferences, mitigating issues associated with human-based preference collection. Experimental results in the StarCraft Multi-Agent Challenge (SMAC) \cite{SMACv2} demonstrate that DPM significantly outperforms existing baselines, particularly in sparse reward settings.

% Large-Scale Multi-Agent Systems
Traditional methods like soft attention struggle with scalability and efficiency in LMAS due to the overwhelming number of agent interactions and large observation spaces. To address these challenges of large-scale multi-agent systems (LMAS) involving hundreds of agents, University of Chinese Academy of Sciences \cite{Fu2022ConcNet} introduces the Concentration Network (ConcNet), a novel reinforcement learning framework. ConcNet mimics human cognitive processes of concentration by prioritizing and aggregating observations based on motivational indices, such as expected survival time and state value. It allows the system to focus on the most relevant entities, enhancing decision-making efficiency in complex environments. In ConcNet, a novel concentration policy gradient architecture was further proposed, demonstrating its superior performance and scalability in large-scale multi-agent scenarios, such as decentralized collective assault simulations. This research represents a significant advancement in the field, providing a scalable solution for effective decision-making in large-scale multi-agent environments.

% \subsubsection{Outstanding Achievements}
% .

% Next, we will provide a detailed introduction to some outstanding achievements in the MARL application for multi-agent collaborative task execution.

% \textbf{:} \cite{} . 

% \textbf{:} \cite{} . 

% \textbf{:} \cite{} .

% \textbf{:} \cite{} .

% \textcolor{red}{TODO.}

In conclusion, MARL-based intelligent applications have shown exceptional adaptability across diverse domains such as autonomous driving, UAV systems, disaster response, and collaborative robotics \cite{pmlrwang20iIMAC,kouzeghar2023UAVpursuit,UAVNavigation1,liu2024diversedriving,collabrobots9431107,MASCENA20134737}. Key innovations, including communication-enhanced learning \cite{foerster2016learningDIAL,CommNet,FanMDMADDP,SunaaaiT2MAC}, adaptive policy optimization, and mutual information \cite{li2022pmic} frameworks, have significantly advanced the field. While challenges like sparse rewards and scalability remain, these advancements highlight MARL's potential to address dynamic and complex multi-agent environments effectively, paving the way for further impactful developments.

\subsection{LLMs reasoning-based Intelligent Applications}
To address diverse and complex challenges, a variety of frameworks leveraging LLMs have been developed, enabling advanced reasoning, collaboration, and task execution in multi-agent systems \cite{LLMmarlsurvey,Wang2024LLMsurvey,li2024rescueLLM}. Below, we highlight notable contributions that demonstrate the application of LLMs in enhancing multi-agent decision-making and coordination.

% LLMs-based Mullti-agent
Wu et al. \cite{wu2023autogen} introduced AutoGen, an open-source framework designed to enable the development of next-generation LLM applications through multi-agent conversations. AutoGen allows for customizable agent interactions and the integration of LLMs, human inputs, and tools to collaboratively solve complex tasks. 
Xiao et al. \cite{xiao2024chainofexperts} proposed Chain-of-Experts (CoE), a novel multi-agent framework designed to enhance reasoning in complex operations research (OR) problems using LLMs.
% CoE assigns domain-specific roles to each agent and incorporates a conductor to guide the agents through a forward thought construction and backward reflection mechanism. The framework aims to reduce the reliance on domain experts and offers a robust solution for various industry sectors.
Chen et al. \cite{chen2024agentverse} presented AgentVerse, a multi-agent framework designed to facilitate collaboration among autonomous agents, inspired by human group dynamics. AgentVerse dynamically adjusts the composition and roles of agents throughout the problem-solving process, enhancing their ability to tackle complex tasks across various domains, including text understanding, reasoning, coding, and embodied AI. The framework consists of four stages: Expert Recruitment, Collaborative Decision-Making, Action Execution, and Evaluation. 
% Extensive experiments demonstrate AGENTVERSE's effectiveness, revealing emergent social behaviors such as volunteering, conformity, and occasional destructive actions, highlighting both the benefits and potential risks of multi-agent collaboration.
Chen et al. \cite{AutoAgents2024ijcai} introduced AutoAgents, a framework capable of adaptively generating and coordinating multiple specialized agents based on different tasks, thereby constructing efficient multi-agent teams to accomplish complex tasks.
Liu et al. \cite{liu2024dynamic} proposed the Dynamic LLM-Agent Network (DyLAN), a framework designed to enhance LLM-agent collaboration by enabling agents to interact dynamically based on task requirements, rather than within a static architecture. 
% DyLAN includes an automatic agent team optimization method that selects the most effective agents based on their contributions, leading to significant improvements in performance and efficiency for tasks like reasoning and code generation.
Xu et al. \cite{xu24pmlrLLMWerewolf} proposed a novel multi-agent framework that combines LLMs with reinforcement learning to enhance strategic decision-making and communication in the Werewolf game\footnote{Werewolf game: \url{https://sites.google.com/view/strategic-language-agents/}.}, effectively overcoming intrinsic biases and achieving human-level performance.
Wen et al. \cite{MAT} introduce the Multi-Agent Transformer (MAT), a novel architecture that frames cooperative MARL as a sequence modeling problem. Experiments on StarCraftII \cite{vinyals2017starcraftiiPYSC2,Vinyals2019sc2,samvelyan19smac}, Multi-Agent MuJoCo (MAMuJoCo) \cite{MAMuJoCo}, Dexterous Hands Manipulation \cite{YufnbotDexterous,OpenAIdexterous}, and Google Research Football \cite{Kurach2020gfootball} benchmarks demonstrate that it achieves superior performance and data efficiency by leveraging modern sequence models in an on-policy learning framework.
Wang et al. \cite{wang2024mllmtool} introduced MLLM-Tool\footnote{MLLM-Tool: \url{https://github.com/MLLM-Tool/MLLM-Tool}.}, a multimodal tool agent system that integrates open-source LLMs with multimodal encoders, enabling it to process visual and auditory inputs for selecting appropriate tools based on ambiguous multimodal instructions. Moreover, they introduced ToolMMBench, a novel benchmark with multi-modal inputs and multi-option solutions, demonstrating its effectiveness in addressing real-world multimodal multi-agent scenarios.
Zhang et al. \cite{zhang2024building} introduce CoELA, a Cooperative Embodied Language Agent framework that leverages LLMs to enhance multi-agent cooperation in complex, decentralized environments. CoELA integrates LLMs with cognitive-inspired modules for perception, memory and execution, allowing agents to plan, communicate, and collaborate effectively on long-horizon tasks, outperforming traditional planning-based methods such as Multi-Agent Transformer(MAT) \cite{MAT}, and showing promising results in human-agent interaction simulation environments, Communicative Watch-And-Help (C-WAH) \cite{puig2021watchandhelp} and ThreeDWorld Multi-Agent Transport (TDW-MAT) \cite{ThreeDWorldv2,ThreeDWorldv1}.
Gong et al. \cite{gong2024mindagent} from Team of Li.FeiFei. introduce MindAgent, a novel infrastructure for evaluating planning and coordination capabilities in gaming interactions, leveraging large foundation models (LFMs) to coordinate multi-agent system (MAS), collaborate with human players, and enable in-context learning. Their team also present "Cuisineworld"\footnote{Cuisineworld: \url{https://mindagent.github.io/}.}, a new gaming scenario and benchmark for assessing multi-agent collaboration efficiency.
Despite LLMs' success in various collaborative tasks, they struggle with spatial and decentralized decision-making required for flocking. Li et al. \cite{li2024rescueLLM} explored the challenges faced by LLMs in solving multi-agent flocking tasks, where agents strive to stay close, avoid collisions, and maintain a formation.
Sun et al. \cite{sun2023corex} proposed Corex, a novel framework that enhances complex reasoning by leveraging multi-model collaboration. Inspired by human cognitive processes, Corex employs three collaborative paradigms-Discuss, Review, and Retrieve-where different LLMs act as autonomous agents to collectively solve complex tasks. Corex empowers LLM agents to "think outside the box" by facilitating collaborative group discussions, effectively mitigating the cognitive biases inherent in individual LLMs. This approach not only enhances performance but also improves cost-effectiveness and annotation efficiency, offering a significant advantage in complex tasks.

Next, we will provide a detailed introduction to some outstanding achievements in the application of LLMs for multi-agent collaborative task execution.

% \subsubsection{Outstanding Achievements}
\textbf{MetaGPT:} Existing LLMs-based multi-agent systems often struggle with complex tasks due to logical inconsistencies and cumulative hallucinations, leading to biased results.
Hong et al. \cite{hong2024metagpt,hong2024data} from DeepWisdom\footnote{DeepWisdom: \url{https://www.deepwisdom.ai/}.} proposed MetaGPT\footnote{MetaGPT: \url{https://github.com/geekan/MetaGPT}.}, an innovative meta-programming framework designed to enhance the collaboration capabilities of LLMs-based multi-agent systems.
MetaGPT integrates Standard Operating Procedures (SOPs) commonly used in human workflows, thereby constructing a more efficient and coherent multi-agent collaboration system.
% As illustrated in Figure \ref{fig_MetaGPT}, 
MetaGPT employs an assembly-line approach, breaking down complex tasks into multiple subtasks and assigning them to agents with specific domain expertise. These agents collaborate during task execution through clearly defined roles and structured communication interfaces, reducing the risk of information distortion and misunderstanding.
% Moreover, MetaGPT introduces an executable feedback mechanism, a self-correction process during runtime.  After generating initial code, agents perform unit tests and iteratively debug and optimize based on the results, continuously improving code quality.
In summary, MetaGPT offers a flexible and powerful platform for developing LLMs-based multi-agent systems. Its unique meta-programming framework and rigorous workflow design enable it to excel in handling complex tasks, greatly advancing the field of multi-agent collaboration research.

% \begin{figure}[!t]
% \centering
% \includegraphics[width=3.25in]{MetaGPT.png}
% \caption{A diagram illustrating the software development process in MetaGPT, highlighting its strong reliance on SOPs \cite{hong2024metagpt}.}
% \label{fig_MetaGPT}
% \end{figure}

\textbf{CoAct:} Hou et al. \cite{hou2024coact} proposed CoAct\footnote{CoAct: \url{https://github.com/dxhou/CoAct}.}, a multi-agent system based on LLMs designed for hierarchical collaboration tasks. % As shown in Figure \ref{fig_CoAct}, t
The framework consists of six stages: task decomposition, subtask assignment and communication, subtask analysis and execution, feedback collection, progress evaluation, and replanning when necessary. The global planning agent plays a critical role in managing complex tasks. 
% It begins by decomposing tasks into multiple phases (e.g., "Phase 1, Phase 2... Phase N"), defining clear outcomes for each phase, and assigning subtasks to the local execution agent. This agent oversees the entire plan, reviews progress based on feedback from the local execution agent, and adjusts strategies when required to maintain consistency with the overall plan. 
The local execution agent is responsible for executing specific subtasks. 
% It analyzes each task, performs sequential operations, and verifies the results against the global strategy. By systematically collecting feedback (e.g., execution results and error feedback), the local execution agent evaluates task progress. 
This hierarchical framework demonstrates strong adaptability and performance, particularly in complex real-world tasks requiring dynamic replanning and collaborative execution.

% \begin{figure}[h]
% \centering
% \includegraphics[width=3.5in]{CoAct.png}
% \caption{An overview of CoAct: A multi-agent system based on LLMs, designed for hierarchical collaboration among agents. It features task decomposition, subtask assignment and communication, subtask analysis and execution, feedback collection, progress evaluation, and dynamic replanning when necessary \cite{hou2024coact}.}
% \label{fig_CoAct}
% \end{figure}

% \textcolor{red}{TODO.}

\textbf{AutoGen:} Microsoft \cite{wu2023autogen,dibia2024autogen} introduced AutoGen\footnote{AutoGen: \url{https://github.com/microsoft/autogen}.}, a flexible framework for creating and managing multiple autonomous agents to collaboratively complete complex tasks, particularly in programming, planning, and creative writing domains. AutoGen allows users to define distinct agent roles, including specialists, general assistants, and decision-makers, ensuring clear task division and effective coordination. Agents interact in a structured conversational environment, exchanging messages to resolve tasks iteratively. 
% Specifically, AutoGen simplifies multi-agent collaboration into three steps. First, \textit{Agent Creation:} Developers can create and manage various agents, such as specialists, general assistants, and strategists, assigning specific roles, tasks, and permissions to ensure clear division of labor. Second, \textit{Dialogue Environment:} A virtual workspace enables agents to communicate and collaborate through text, audio, or video. All conversations and decisions are automatically recorded for review. Third, \textit{Dialogue Management:} Tools guide agent discussions to stay goal-focused, monitor progress, address issues in real-time, and enforce rules to maintain quality and efficiency. Search and filtering functions allow quick access to relevant information. The framework employs a ConversableAgent class, enabling agents to communicate, execute actions, and adapt their behavior based on task requirements. Two primary subclasses are \textit{AssistantAgent}, a general-purpose AI assistant capable of generating Python code, and \textit{UserProxyAgent}, which acts as a proxy for human input while executing code or providing feedback. 
AutoGen introduces feedback loops where agents analyze outputs, refine strategies, and optimize task completion autonomously. Notably, it supports integration with various LLMs, offering developers the flexibility to replace APIs without altering code significantly. In summary, AutoGen facilitates scalable, efficient, and robust multi-agent collaboration, demonstrating potential for applications ranging from enhanced ChatGPT systems to real-world industrial workflows.

% \begin{figure}[!t]
% \centering
% \includegraphics[width=3.25in]{AutoGen.png}
% \caption{Illustration of using AutoGen to program a multi-agent conversation. The top section highlights AutoGen's built-in, customizable agents with unified conversation interfaces. The middle section demonstrates developing a two-agent system with a custom reply function. The bottom section shows the automated chat between the agents during execution \cite{wu2023autogen,dibia2024autogen}.}
% \label{fig_AutoGen}
% \end{figure}

\begin{figure}[!t]
\centering
\includegraphics[width=3.5in]{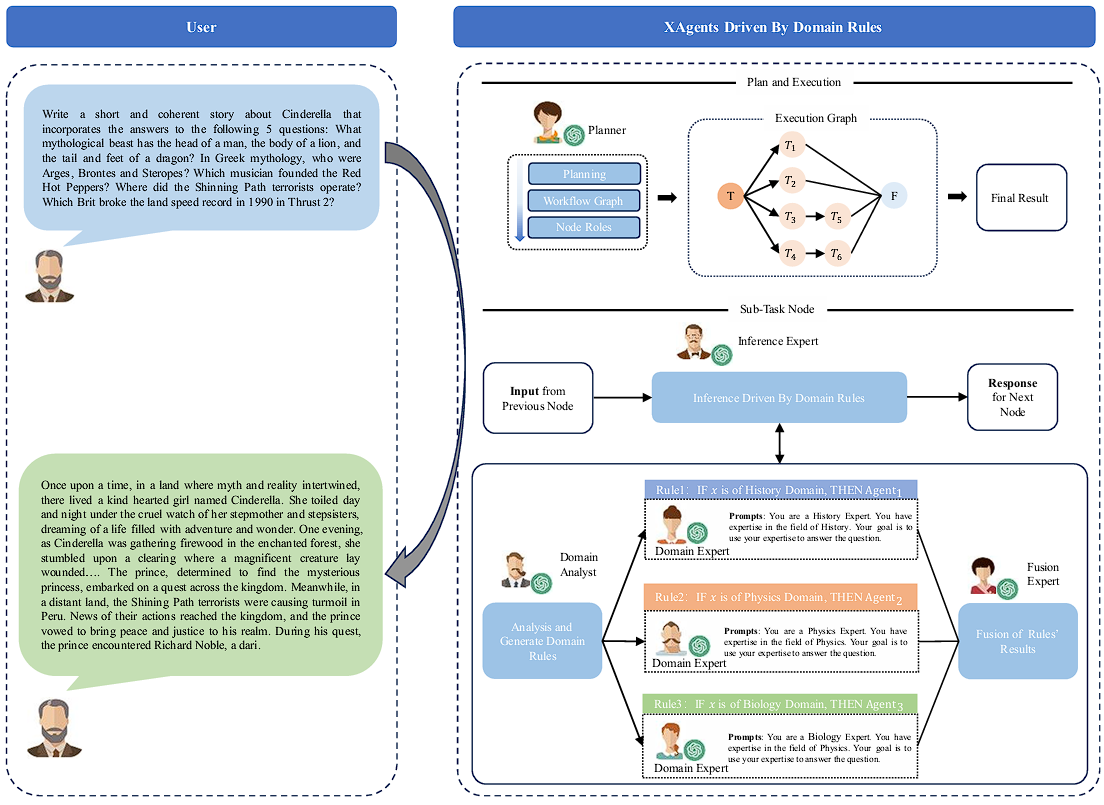}
\caption{An structure overview of the XAgents framework, highlighting the Task Node as the starting point, the sequence of Sub-Task Nodes forming the Task Execution Graph (TEG), and the Fusion Node integrating outputs for the final result \cite{xagent2023}.}
\label{fig_XAgent}
\end{figure}

\textbf{XAgent:} XAgent Team\footnote{XAgent Team: \url{https://blog.x-agent.net/}.} \cite{xagent2023} developed XAgent\footnote{XAgent: \url{https://github.com/OpenBMB/XAgent}.} is an open-source, LLMs-driven autonomous agent framework designed for solving complex tasks automatically and efficiently. As shown in Figure \ref{fig_XAgent}, it employs a dual-loop architecture: the outer loop for high-level task planning and coordination, and the inner loop for executing subtasks. The PlanAgent in the outer loop generates an initial plan by breaking a complex task into manageable subtasks, organizing them into a task queue. It iteratively monitors progress, optimizes plans based on feedback from the inner loop, and continues until all subtasks are completed. The inner loop utilizes ToolAgents, which employ various tools like file editors, Python notebooks, web browsers, and shell interfaces within a secure docker environment to execute subtasks. XAgent emphasizes autonomy, safety, and extensibility, allowing users to add new agents or tools to enhance functionality. Its GUI facilitates user interaction while supporting human collaboration, enabling real-time guidance and assistance for challenging tasks.

\textbf{PlanAgent:} PlanAgent\footnote{PlanAgent: \url{http://www.chinasem.cn/planagent}.}, developed by Chinese Academy of Sciences and Li Auto \cite{zheng2024planagent}, introduces a closed-loop motion planning framework for autonomous driving by leveraging multi-modal large language models (MLLMs). The system utilizes MLLM's multi-modal reasoning and commonsense understanding capabilities to generate hierarchical driving commands based on scene information.
% , achieving superior performance on the nuPlan benchmark in both standard and long-tail scenarios
% As depicted in FIgure \ref{fig_PlanAgent}, PlanAgent consists of three modules. First, the \textit{Scene Information Extraction Module} transforms environmental data into bird’s-eye-view (BEV) representations and graph-based text prompts. This approach reduces token usage by two-thirds compared to traditional methods while providing comprehensive contextual understanding. Second, the \textit{Reasoning Module} incorporates hierarchical chain-of-thought reasoning to generate safe and efficient planner codes, embedding MLLM's contextual learning abilities into motion planning tasks. Finally, the \textit{Reflection} Module employs a simulation-based mechanism to validate planning solutions, filtering out unsafe outputs and iteratively refining the planner.
PlanAgent addresses key limitations of traditional rule-based and learning-based methods, including overfitting in long-tail scenarios and inefficiencies in scene representation. Its novel integration of MLLM-driven reasoning into autonomous driving planning establishes a new benchmark for safety and robustness in real-world applications.

% \begin{figure}[!t]
% \centering
% \includegraphics[width=3.5in]{PlanAgent.png}
% \caption{The PlanAgent framework pipelines with three modules: \textit{Environment Transformation}, \textit{Reasoning Engine}, and \textit{Reflection Module}. The \textit{Environment Transformation} module extracts key environmental information, creating a BEV map and lane-graph representation, which is converted into textual scenario prompts. The \textit{Reasoning Engine} module uses the MLLM to generate planner codes through hierarchical chain-of-thought reasoning, combining scenario prompts with predefined system prompts. The \textit{Reflection} module simulates and evaluates the generated plan, determining whether to execute or revise it based on performance \cite{zheng2024planagent}.}
% \label{fig_PlanAgent}
% \end{figure}

\textbf{LangGraph:} LangChain Inc\footnote{LangChain Inc: \url{https://langchain.ac.cn/langgraph}.} \cite{wang2024LangGraph,wangJ2024LangGraph} introduced LangGraph\footnote{LangGraph: \url{https://www.langchain.com/langgraph}}, a library designed for building stateful, multi-actor applications with LLMs, enabling the creation of complex agent and multi-agent workflows. Inspired by frameworks like Pregel and Apache Beam, LangGraph provides fine-grained control over workflows and state management while offering advanced features like persistence and human-in-the-loop capabilities.
LangGraph stands out for its support of iterative workflows with cycles and branching, which differentiates it from DAG-based frameworks. Each graph execution generates a state, dynamically updated by node outputs, enabling reliable and adaptive agent behavior. Its built-in persistence allows workflows to pause and resume at any point, facilitating error recovery and advanced human-agent interactions, including "time travel" to modify past actions.
LangGraph integrates seamlessly with LangChain \cite{pandya2023automatingLangChain,LangChainCustom} but functions independently, offering flexibility for various applications, from dialogue agents and recommendation systems to natural language processing and game development. With streaming support, it processes outputs in real-time, making it suitable for tasks requiring immediate feedback. Its low-level architecture and customizable workflows make LangGraph a powerful tool for creating robust, scalable, and interactive LLMs-based systems.

\begin{figure}[h]
\centering
\includegraphics[width=3.5in]{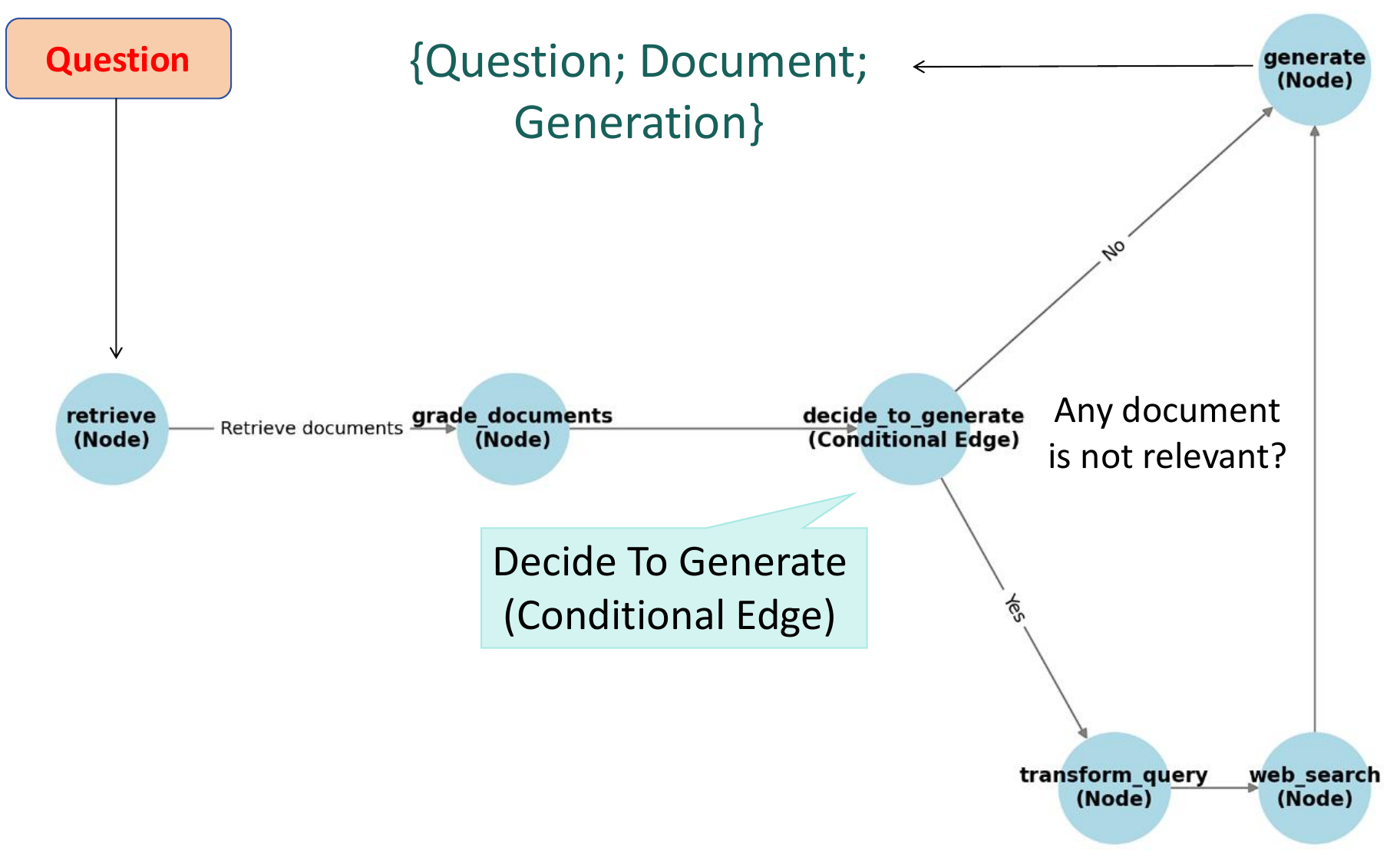} % LangGraph.png
\caption{A LangGraph workflow representation demonstrating conditional branching and iterative loops for document retrieval, grading, query transformation, and web search before generating a final output \cite{wang2024LangGraph,wangJ2024LangGraph}.}
\label{fig_LangGraph}
\end{figure}

\begin{figure}[!t]
\centering
\includegraphics[width=3.5in]{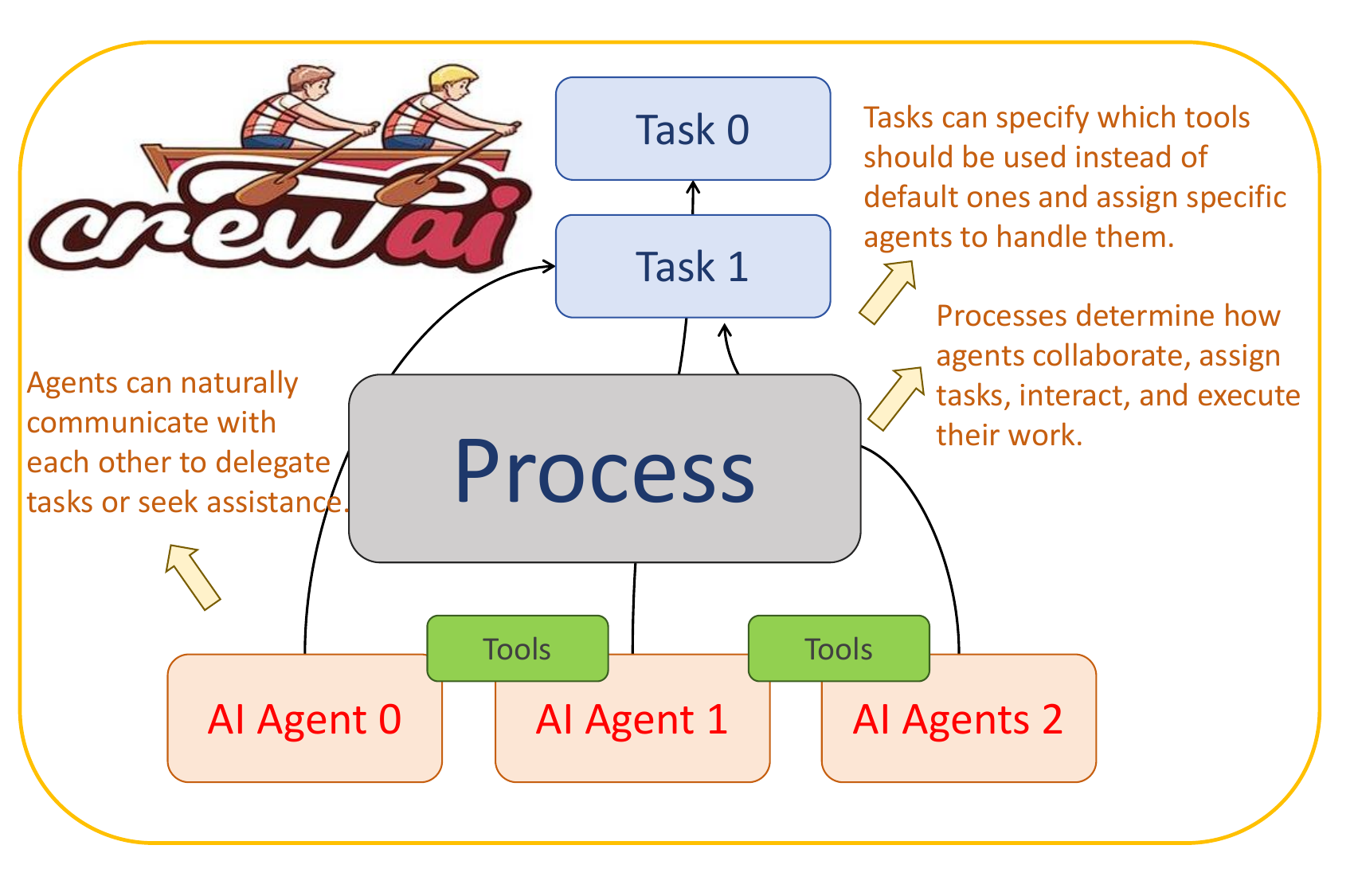} % CrewAI.png
\caption{An overview of the processing workflow for the role-playing multi-agent framework, CrewAI \cite{zhang2025crewai,duan2024crewaiLangGraph}.}
\label{fig_CrewAI}
\end{figure}

\textbf{CrewAI:} CrewAI\footnote{CrewAI: \url{https://github.com/crewAIInc/crewAI}.} \cite{zhang2025crewai,duan2024crewaiLangGraph} is an open-source framework designed to coordinate AI agents that specialize in role-playing and autonomous operations, enabling efficient collaboration to achieve complex goals. The framework’s modular design allows users to create AI teams that operate like real-world teams, with agents assigned specific roles and tasks to ensure clear division of labor and shared objectives. As seen from \ref{fig_CrewAI}, this framework operates in three primary stages: Agent Creation, where developers define roles with specific goals and tools; Task Management, enabling flexible task assignment and multi-view knowledge enhancement; and Execution and Collaboration, where agents interact in workflows to resolve tasks, with outputs parsed into reusable formats. CrewAI integrates seamlessly with the LangChain ecosystem, leveraging its tools and LLM capabilities, such as OpenAI and Google Gemini. The framework supports real-time decision-making and task adaptation, with future versions planned to include more advanced collaboration processes, such as consensus-driven workflows and autonomous decision-making. Its innovative features, such as role-based design, dynamic rule generation, and modular task workflows, position CrewAI as a robust and scalable framework for multi-agent collaboration across creative and industrial domains. Overall, CrewAI \footnote{CrewAI Multi-Agent System platform: \url{https://www.deeplearning.ai/short-courses/multi-ai-agent-systems-with-crewai/}.} offers a cutting-edge approach to multi-agent systems by integrating role-specific autonomy, flexible workflows, and advanced AI toolsets, making it a versatile framework for collaborative AI applications.

In summary, these frameworks and applications \cite{hong2024metagpt,hou2024coact,wu2023autogen,xagent2023} highlight the rapid advancements in leveraging LLMs for multi-agent collaboration, reasoning, and task execution. Each system introduces unique innovations—ranging from dynamic agent coordination to enhanced reasoning and human-in-the-loop workflows—demonstrating their potential to tackle complex, real-world challenges across various domains \cite{zheng2024planagent,wang2024LangGraph,zhang2025crewai,duan2024crewaiLangGraph}. These developments pave the way for more flexible, scalable, and efficient AI-driven solutions. % adaptable

% % ####################################################
% \section{Potential Challenges}
% .
% % ####################################################
\section{Challenges in MARL-based and LLMs-based approaches}  % Problems and 
% Potentials
\label{issuesMADM}
% 讲述多智能体决策系统面临的挑战
% 讲述相关的决策方法，以及这些方法面临的困难挑战
The extension of single-agent decision-making into multi-agent cooperative contexts introduces several challenges, including developing effective training schemes for multiple agents learning and adapting simultaneously, managing increased computational complexity due to the more sophisticated and stochastic environments compared to single-agent settings, and addressing the foundational role of strategic interaction among agents. Additionally, ensuring the scalability of algorithms to handle larger observation and action spaces, facilitating coordination and cooperation among agents to achieve consistent goals, and dealing with non-stationary environments where agents' behaviors and strategies continuously evolve are also inevitable and critical challenges.

\begin{figure*}[!t]
\centering
\includegraphics[width=0.825\linewidth]{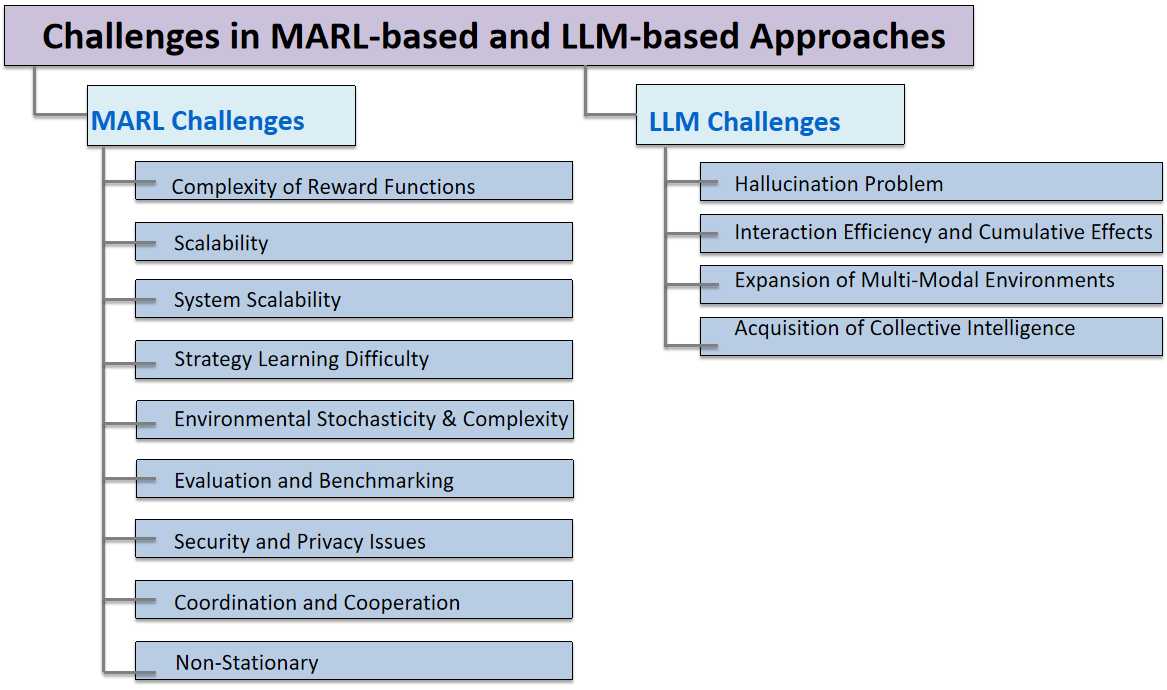} % challsumm challsummV1
\caption{A tree diagram of the challenges in MARL and LLMs-based multi-agent decision-making approaches.}
\label{fig_challsumm}
\end{figure*}

Applying multi-agent decision-making techniques to real-world problems, which often involve complex and dynamic interactions, further underscores the need for sophisticated and advanced approaches to effectively adapt these ever-increasing complexities. Multi-agent cooperative decision making significantly surpasses single-agent decision-making in terms of environmental stochasticity, complexity, the difficulty of strategy optimization, and so on.
As shown in Figure \ref{fig_challsumm}, we present a tree diagram summarizing the existing challenges in MARL-based and LLMs-based multi-agent decision-making approaches.

% The core challenge in multi-agent learning is to develop strategies that allow agents to adapt to the non-stationary behaviors of other collaborative agents. As each agent learns and adapts, the overall environment changes, making the learning target a moving one. This non-stationarity is a fundamental issue that differentiates multi-agent learning from single-agent learning. Addressing this problem requires sophisticated algorithms that can anticipate and adapt to the strategies of other agents.
% To date, various methodologies have been proposed to tackle the challenges in multi-agent cooperative decision-making.

\subsection{Challenges in MARL-based multi-agent systems}  % decision-making
% MARL-based multi-agent systems have demonstrated significant potential in addressing complex decision-making tasks across diverse domains \cite{Pamul2023survey,Gronauer2022Survey,James2023Survey,NING2024Survey,Zhu2024Survey}. However, t
The advancement of MARL remains in its formative stages, with its potential for enabling effective multi-agent coordination and achieving scalability in dynamic environments yet to be fully unlocked \cite{Pamul2023survey,Gronauer2022Survey,James2023Survey,NING2024Survey,Zhu2024Survey}. Challenges such as environmental stochasticity, strategy learning difficulty, non-stationarity, scalability, and reward complexity have emerged as major bottlenecks. This section provides an in-depth analysis of these challenges, exploring the current state, technical limitations, and potential solutions in MARL-based multi-agent systems to enable more robust, efficient, and scalable decision-making frameworks.

% non-stationarity, scalability, and partial observability.

% Typically, in the research field of multi-agent reinforcement learning (MARL): 
\begin{enumerate}
\item 
\textbf{Environmental Stochasticity and Complexity:} \cite{du2024surveycontext,NING2024Survey} In MARL-based decision-making systems, environmental dynamics are influenced not only by external factors but also by the behaviors and decisions of individual agents. This complex interaction results in high levels of stochasticity and complexity in the environment, making prediction and modeling significantly more difficult. For example, in autonomous driving scenarios \cite{liu2024diversedriving,WangMARLautoDriving,LuMARLautoDriving,xue2023twostageautodrive}, the behavior of each vehicle affects the decisions of surrounding vehicles, thereby increasing the overall complexity of the system; 
\item 
\textbf{Strategy Learning Difficulty:} \cite{HU2024128068,quantumMetaMARL,LLMmarlsurvey} Strategy learning in MARL-based decision-making systems involves multidimensional challenges. Firstly, agents must consider the behaviors of other agents, and this interdependence increases the difficulty of strategy learning. Each agent not only has to optimize its own strategy but also predict and adapt to the strategy changes of others. Additionally, the vast joint action space of multiple agents makes it challenging for any single agent to learn effective joint strategies. The vast joint action space means that each agent needs to explore and learn within a larger decision space, which significantly increases the demands on computational resources and time; 
% During the strategy learning process, agents must consider the other agents' behaviors, not only optimizing their own strategies but also adapting to the strategy changes of others. This interdependence increases the difficulty of strategy learning. Additionally, the vast joint action space of multiple agents makes it challenging for any single agent to learn effective joint strategies. The vast joint action space means that each agent needs to explore and learn within a larger decision space, making it challenging for multi-agent decision-making methods based on reinforcement learning to converge
\item 
\textbf{Complexity of Reward Functions:} \cite{du2024surveycontext} In MARL-based decision-making systems, reward functions become more complex \cite{VDN,DQNv1}. The rewards received from the environment in multi-agent cooperative techniques are influenced not only by an individual agent's actions and the environment but also by the actions of other agents, which makes the stable policy learning and modeling process more difficult. In other words, an agent's reward depends not only on its own actions but also on the actions of other agents, making it challenging for the reinforcement learning-based multi-agent decision-making policies to converge. This intricate reward mechanism complicates the design and optimization of reward functions. Agents need to evaluate their behaviors' impact on the overall system through complex interactions to learn effective strategies;
% In MARL-based methods, the reward functions become more complex since an agent's reward depends not only on its own actions but also on the actions of other agents. This intricate reward mechanism significantly complicates the design and optimization of reward functions. The rewards received from the environment in multi-agent cooperative techniques are influenced not only by an individual agent's actions and the environment but also by the actions of other agents. It makes the stable policy learning and modeling process more difficult. For example, in an autonomous driving scenario, the behavior of each car affects the decisions of surrounding vehicles
\item 
\textbf{Coordination and Cooperation:} \cite{tran2025multiagentcollab,qimilitaryMARL,talebirad2023multiagent,du2024surveycontext} Furthermore, in MARL-based decision-making systems, agents need to coordinate and cooperate to achieve common goals. This requires establishing effective communication and coordination mechanisms among agents to ensure that their actions are globally consistent and complementary \cite{Zhang2024Evolutionary}. For example, in disaster rescue scenarios \cite{li2024rescueLLM,qazzaz2024rescueUAV}, multiple drones need to coordinate their actions to cover the maximum area and utilize resources most efficiently;
\item 
\textbf{Non-Stationary:} \cite{hernandezleal2019survey,NING2024Survey} The environment of MARL-based decision-making systems is non-stationary because each agent's behavior dynamically changes the state of the environment. 
This non-stationarity increases the difficulty of strategy learning, as agents must continually adapt to changes in the environment and the behaviors of other agents. 
% This requires more advanced learning algorithms to handle the dynamic changes in the environment and strategies.
\item 
\textbf{Scalability:} \cite{MetaMARL,HU2024128068,NING2024Survey} Addressing scalability in MARL demands innovative approaches to tackle the exponential growth in complexity as the number of agents increases. Techniques that leverage \textit{knowledge reuse} \cite{peng2017BidirectionalBiCNet,jiang2018learning,foerster2016learningDIAL}, such as parameter sharing and transfer learning \cite{silva2018autonomously,Silvasurveyjair}, have proven indispensable. When agents share structural similarities, \textit{information sharing} can streamline the training process, enabling systems to scale more effectively. \textit{Transfer learning}, in particular, allows agents to adapt knowledge from previous tasks to new, related ones, significantly accelerating learning in dynamic environments.
% Another impactful method is the \textit{mean-field approach}, which reduces computational costs by approximating the interactions between agents, demonstrating its potential for scalable MARL. 
Moreover, \textit{curriculum learning} \cite{CurriculumLearning,WangCurriculum} plays a pivotal role in tackling the increased complexity of training multiple agents. It enables gradual learning by exposing agents to progressively more challenging tasks, thereby improving policy generalization and accelerating convergence.
Robustness remains critical for scalability, as learned policies must withstand environmental disturbances. Techniques like \textit{policy ensembles} and \textit{adversarial training} \cite{pinto2017robust,jwqscirep} enhance resilience by fostering diversity and adaptability in policies. 
% Similarly, \textit{self-play}, especially in scenarios involving imperfect information, encourages the development of robust strategies by continuously challenging agents with evolving tactics. In dynamic and large-scale multi-agent settings, challenges like non-stationarity arise due to agents’ interactions. 
The DTDE paradigm addresses these issues but introduces new complexities \cite{amato2024CTDEsurvey,zhou2023CTDEenough}, such as environmental instability. One promising solution is employing \textit{Independent Deep Q-networks (IDQNs)} \cite{DQNori,FoersterJMLR,claus1998IQL}, which adapt traditional single-agent approaches to multi-agent contexts.
\end{enumerate}

Overall, the interplay between robustness and scalability in MARL is a key area for future exploration. While existing techniques provide strong foundations, integrating methods like \textit{meta-learning} could offer a way for agents to rapidly adapt to new tasks and environments. Additionally, leveraging recent advances in \textit{graph neural networks} might enhance the scalability of MARL by modeling agent interactions more efficiently. These directions hold promise for tackling the dynamic and large-scale nature of multi-agent environments.

% \textcolor{red}{TODO.} 
% .

\subsection{Challenges in LLMs reasoning-based multi-agent systems}  % decision-making
% As a frontier artificial intelligence research direction, Large Language Models-based Multi-Agent Systems have demonstrated strong potential. However, t
The development of LLMs-based multi-agent systems is still in its early stages, and its advantages in multi-agent collaboration in cross-domain applications have not been fully realized. In this process, technical bottlenecks, design limitations, and imperfect evaluation methods have revealed numerous challenges. This section provides a comprehensive analysis of these challenges, exploring the current status, bottlenecks, and potential breakthrough directions of LLMs-based multi-agent systems in key areas such as multi-modal interaction, system scalability, hallucination control, evaluation, and privacy protection.

\begin{enumerate}
\item \textbf{Expansion of Multi-Modal Environments:} \cite{wang2024mllmtool,zheng2024planagent,ThreeDWorldv1}
Current LLMs-based multi-agent systems primarily focus on text processing and generation, particularly excelling in language-based interactions. However, applications in multi-modal environments remain insufficient. Multi-modal environments require agents to handle various inputs from images, audio, video, and physical sensors, while also generating multi-modal outputs, such as descriptions of visual scenes or simulations of physical actions. This cross-modal interaction not only demands stronger model processing capabilities but also requires efficient information fusion between agents. For example, in practical applications, one agent may need to extract visual features from an image and collaborate with other agents through language to accomplish complex tasks, posing new technical challenges. Future research should focus on building unified multi-modal frameworks that enable agents to efficiently understand and collaboratively process various types of data.

\item \textbf{Hallucination Problem:} \cite{guo2024llmmultiagents,Wang2024LLMsurvey,hong2024metagpt,Huang2024Hallucination}
The hallucination in LLMs, which involves generating false or inaccurate information, becomes more complex in multi-agent environments. This issue may be triggered within a single agent and further propagated through multi-agent interactions, ultimately negatively impacting the decision-making of the entire system. Because the information flow in multi-agent systems is interconnected, any misjudgment at one node can trigger a chain reaction. This characteristic makes the hallucination problem not only confined to the behavior of individual agents but also poses challenges to the stability of the entire system. Therefore, addressing this issue requires a dual approach: on one hand, improving model training methods to reduce the probability of hallucinations in individual agents; on the other hand, designing information verification mechanisms and propagation management strategies to minimize the spread of hallucinated information within the agent network.

\item \textbf{Acquisition of Collective Intelligence:} \cite{Wu1999,ijcai2022p85,tran2025multiagentcollab}
Current LLMs-based multi-agent systems rely more on real-time feedback for learning rather than offline data, unlike traditional multi-agent systems \cite{SpringerAdaptive,Tuyls2005}. This real-time learning approach imposes higher requirements on the interactive environment \cite{guo2024llmmultiagents,chen2025surveyllm}. Since designing and maintaining a reliable real-time interactive environment is not easy, it limits the scalability of the system. Additionally, existing research mostly focuses on optimizing individual agents, neglecting the potential overall efficiency improvements that could arise from agent collaboration. For example, knowledge sharing and behavioral coordination among agents may create advantages of collective intelligence in certain complex tasks. Future research needs to explore how to fully leverage the potential of collective intelligence by optimizing multi-agent interaction strategies and collaboration mechanisms.

\item \textbf{System Scalability:} \cite{guo2024llmmultiagents,dibia2023multiagent,tran2025multiagentcollab}
As the number of agents in LLMs-based multi-agent systems increases, the demand for computational resources grows exponentially, posing challenges in resource-constrained environments. A single LLM agent already requires substantial computational power, and when the system scales to hundreds or thousands of agents, existing hardware and software architectures may not be able to support it. Furthermore, scaling the system introduces new complexities, such as how to efficiently allocate tasks, coordinate, and communicate among numerous agents. Studies have shown that the more agents there are, the more difficult it becomes to coordinate their operations, especially in reducing redundancy and conflicts. Therefore, future work needs to optimize resource utilization through the development of lightweight models and efficient communication protocols, while also exploring the scaling laws for agent expansion.

\item \textbf{Evaluation and Benchmarking:} \cite{guo2024llmmultiagents,tran2025multiagentcollab}
Current evaluation methods and benchmark tests for LLMs-based multi-agent systems are still incomplete. Most research focuses solely on the performance of individual agents in specific tasks, neglecting the overall system performance in complex scenarios. Evaluating group behavior is more challenging because the dynamics and complexity of multi-agent systems are difficult to measure with a single metric. Additionally, the lack of a unified testing framework and benchmark data is a major obstacle when comparing the capabilities of different LLMs-based multi-agent systems across domains. Future research needs to develop comprehensive evaluation standards and universal benchmark tests, especially in key fields such as scientific experiments, economic analysis, and urban planning, to provide a basis for system performance comparison and improvement.

% \item \textbf{Limitations of Generative Agents:} \cite{guo2024llmmultiagents}
% Generative agents are one of the core components of LLMs-based multi-agent systems, but their capabilities are also constrained by the underlying models. In scenarios involving long text processing, generative agents often tend to forget contextual information, thereby reducing the precision of task execution. They also face shortcomings in simulating diverse features and real-time reasoning efficiency. For example, due to the autoregressive structure of models, generative agents typically require multiple queries to complete a single action, which severely affects the system's response speed. Therefore, improving the inference architecture and memory management mechanisms is necessary to enhance the capabilities of generative agents, thus meeting the demands of more complex tasks.

\item \textbf{Interaction Efficiency and Cumulative Effects:} \cite{guo2024llmmultiagents,AlanHarmsAgentic,chen2024agentverse,talebirad2023multiagent}
The complexity of multi-agent systems leads to prominent issues of low interaction efficiency and cumulative effects. Low interaction efficiency is mainly reflected in the need for generative agents to frequently query models, making the system inefficient in large-scale applications. On the other hand, because the system state highly depends on the results of the previous round, when an error occurs in one round, it may accumulate and propagate to subsequent operations, ultimately degrading the system's overall performance. Future efforts should focus on designing more efficient communication protocols and intermediate result correction mechanisms to reduce interaction costs and the impact of cumulative errors.

\item \textbf{Security and Privacy Issues:} \cite{Li2022Survey,NekhaiFuzzy,Amirkhani2022,BEYDOUN2009832,Xu2021Congestion}
Context sharing within multi-agent systems poses risks of introducing noise and privacy leaks. For example, sensitive information shared between agents (such as identities or locations) may be misused by untrusted nodes, thereby threatening the system's security. Addressing this issue requires a two-pronged approach: first, establishing clear organizational structures to restrict information access permissions; second, introducing more advanced trust management mechanisms, such as distributed trust systems based on consensus algorithms, to enhance the system's security and reliability.
\end{enumerate}

In summary, LLMs-based multi-agent systems face challenges across multiple domains, including multi-modal adaptation, scalability, evaluation methods, collective intelligence development, and privacy protection. These challenges not only reveal the current technological limitations but also provide ample space for future research. With advancements in technology and the deepening of interdisciplinary studies, LLMs-based multi-agent systems are expected to achieve significant breakthroughs both theoretically and in applications.

% \textcolor{red}{TODO.}

\section{Future Research Prospects and Emerging Trends}
\label{futresMADM}
% \subsection{}
% \cite{} . 
\textit{Multi-Agent Decision-Making Systems} are entering a new era where LLMs are combined with MARL \cite{LLMmarlsurvey}. This combination can improve learning efficiency in complex dynamic environments. It also enables better multi-modal information processing, multi-task collaboration, and long-term planning \cite{Pamul2023survey,NING2024Survey,Wang2024LLMsurvey,Zhu2024Survey,chen2025surveyllm}. In this section, we discuss future prospects and challenges of multi-agent decision-making system (MAS) research from theoretical, technical, application, and ethical perspectives.

\subsection{Theoretical Development: From Traditional RL to LLMs-Enhanced MARL Framework}
% \cite{} . 
LLMs-enhanced MARL redefines collaboration in multi-agent systems by introducing natural language understanding and reasoning \cite{collabrobots10039365,tran2025multiagentcollab}. Traditional MARL requires agents to learn control strategies in dynamic environments with limited data \cite{James2023Survey,Li2022Survey,hernandezleal2019survey,Du2021Survey}. However, this approach often faces challenges like low sample efficiency, difficult reward design, and poor generalization. LLMs, with their strong reasoning and knowledge representation capabilities, offer solutions \cite{stooke2021decoupling,Wang2024LLMsurvey}. For example, they can process multi-modal information such as natural language and vision \cite{zheng2024planagent,ijcai2019p880,wang2024mllmtool,jwqscirep}, helping agents understand tasks and environments more effectively. This improves learning speed and generalization. Furthermore, LLMs can act as reasoning tools, providing additional context and knowledge to optimize long-term planning.

\begin{figure*}[ht]
\centering
\includegraphics[width=\linewidth]{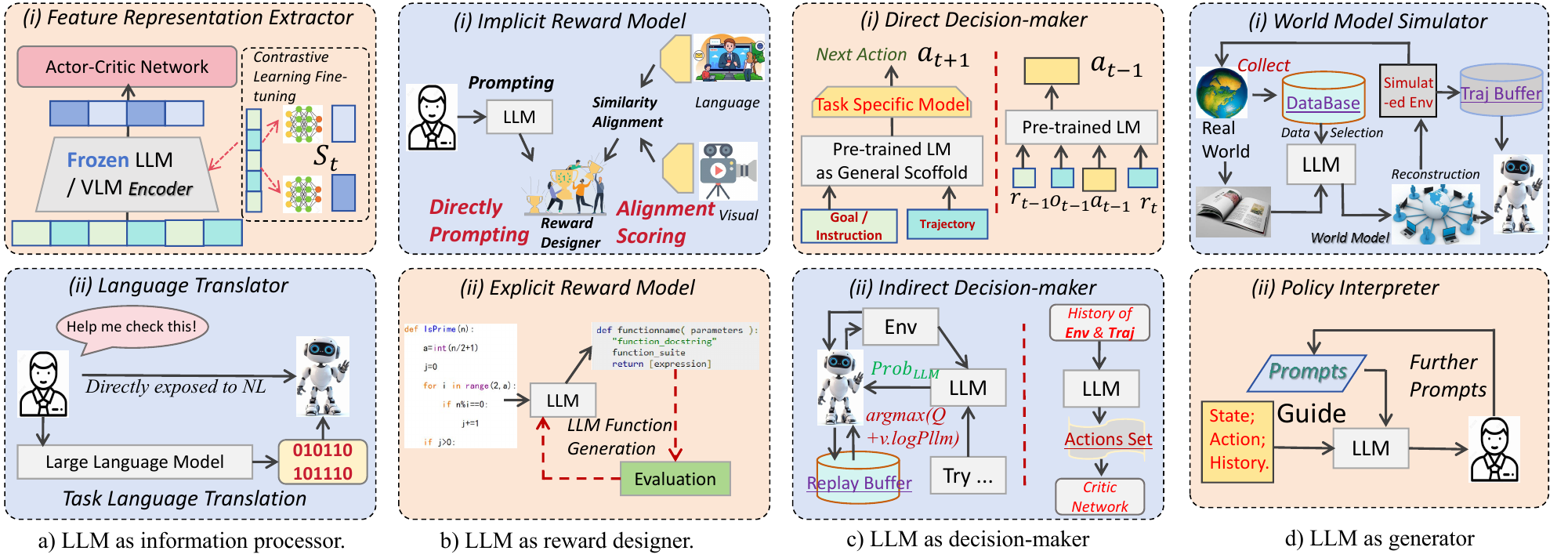}
\caption{Schematic diagram of the LLMs-enhanced MARL framework based on \textit{Cao et al.} \cite{LLMmarlsurvey}, showcasing its core roles: \textit{information processor} (a), \textit{reward designer} (b), \textit{decision-maker} (c), and \textit{generator} (d).}  % extensive and
\label{fig_overallLLMMARL} % from
\end{figure*}

The LLMs-enhanced MARL framework is a groundbreaking integration of LLMs and MARL techniques, which includes roles such as \textit{information processor, reward designer, decision-maker,} and \textit{generator} \cite{LLMmarlsurvey}. 
Figure \ref{fig_overallLLMMARL} presents a flowchart illustrating the structure of the LLMs-enhanced MARL framework, highlighting its four key roles.
These roles work together to streamline task complexity and improve learning. For instance, LLMs can translate unstructured task descriptions into formal task semantics, reducing learning difficulty. They can also design advanced reward functions to accelerate learning in sparse-reward environments.
% ##################################
%
% : \textit{information processor} (a), \textit{reward designer} (b), \textit{decision-maker} (c), and \textit{generator} (d). 
These roles collectively address the challenges of task complexity, data efficiency, and generalization in MARL \cite{Nguyen2020survey,Du2021Survey,du2024surveycontext}, while streamlining processes like reward design and policy generation. 
As shown in Table \ref{table_LLMMARL}, we summarize recent advancements in LLMs-enhanced MARL methods across these four roles into a comprehensive table for clarity and comparison.

\begin{table}[ht]
\renewcommand{\arraystretch}{1.5}
\caption{Summary of recent studies categorized by the four key roles of LLMs in MARL: Information Processor, Reward Designer, Decision-Maker, and Generator, highlighting their respective contributions and applications.}
\label{table_LLMMARL} 
\centering
\setlength{\tabcolsep}{0.5mm}{
\scalebox{0.75}{
\begin{tabular}{ll}
\hline
\textbf{\begin{tabular}[c]{@{}c@{}}Method Types \\ LLM as ...\end{tabular}} & \multicolumn{1}{c}{\textbf{Researchers. / Works. / Refs.}} \\ \hline
\textbf{\begin{tabular}[c]{@{}l@{}}Information \\ Processor\end{tabular}} &
\begin{tabular}[c]{@{}l@{}} Poudel et al. (ReCoRe) \cite{ReCoRe}, Choi et al. (ConPE) \cite{ChoiEfficient}, \\ Paischer et al. (HELM) \cite{pmlrv162paischer22a} and (Semantic HELM) \cite{paischer2023semantic}, \\ Radford et al. (CLIP) \cite{pmlrv139radford21a}, Oord et al. (CPC) \cite{oord2019representation}, \\  Michael et al. (CURL) \cite{CURL}, Schwarzer et al. (SPR) \cite{schwarzer2021dataefficient} \end{tabular} \\ \hline

\textbf{\begin{tabular}[c]{@{}l@{}}Reward \\ Designer\end{tabular}} & \begin{tabular}[c]{@{}l@{}} Kwon et al. (LLMrewardRL) \cite{kwon2023reward}, Song et al. (Self- \\ Refined LLM) \cite{song2023selfrefined}, Wu et al. (Read \& Reward) \cite{NEURIPS2023}, \\ Carta et al. (GLAM) \cite{GLAM}, Chu et al. (Lafite-RL) \cite{chu2024accelerating}, \\  Kim et al. (ARP) \cite{KimARP}, Yu et al. \cite{yu2023languagerewardsroboticskill}, Adeniji \\ et al. (LAMP) \cite{adeniji2024language}, Madaan et al. (Self-Refine) \cite{madaan2023selfrefine}, \\ Ma et al. (Eureka) \cite{ma2024eureka}, Xie et al. (Text2Reward) \cite{xie2024textreward} \end{tabular} \\ \hline

\textbf{\begin{tabular}[c]{@{}l@{}}Decision \\ -Maker\end{tabular}} & \begin{tabular}[c]{@{}l@{}} Janner et al. (TT-Offline RL) \cite{JannerOffline}, Shi et al. (LaMo) \cite{shi2024unleashing}, \\ Li et al. (LLM scaffold) \cite{Liinteractive}, Mezghani et al. \\ (text BabyAI) \cite{mezghani2023think}, Grigsby et al. (AMAGO) \cite{grigsby2024amago}, \\ Zitkovich et al. (RT-2) \cite{pmlrv229zitkovich23a}, Yao et al. (CALM) \cite{yao2020keepCALM}, \\ Hu et al. (instructRL) \cite{Huinstructed}, Zhou et al. (LLM4Teach) \cite{ZhouLLM4Teach} \end{tabular} \\ \hline

\textbf{Generator} & 
\begin{tabular}[c]{@{}l@{}} Chen et al. (TransDreamer) \cite{chen2021transdreamer}, Das et al. (S2E) \cite{NEURIPS2023Das}, \\ Lin et al. (Dynalang) \cite{lin2024learningmodelworldlanguage} , Robine et al. (TWM) \cite{WorldiclrHappy}, \\ Poudel et al. (LanGWM) \cite{poudel2023langwm}, Lin et al. (HomeGrid) \cite{lin2024learning}
% , \\ et al. () \cite{}, et al. () \cite{}, \\ et al. () \cite{}, et al. () \cite{} 
\end{tabular} \\ \hline
\end{tabular}}}
\end{table}

% Contrastive Predictive Coding CPC
% Self-Predictive Representations (SPR)
% Language Models for Motion Control (LaMo)
% TrajectoryTransformer
% transformer-based MBRL TransDreamer
% State2Explanation (S2E)
% Lafite-RL (Language agent feedback interactive Reinforcement Learning)
% Adaptive Return-conditioned Policy (ARP)
%  LAnguage reward Modulated Pretraining (LAMP)
% generating text-augmented expert trajectories on BabyAI
% WORLDALIGNING HomeGrid

\subsection{Technical Integration: From Multi-Modal to Multi-Task Optimization}
% \cite{} . 
Combining LLMs and MARL significantly improves the ability to handle multi-modal information, multi-task learning, and long-term task planning \cite{ijcai2019p880,wang2024mllmtool,stooke2021decoupling,Wang2024LLMsurvey}. Traditional MARL often requires separate modules to process visual, textual, or other forms of data. In contrast, LLMs can unify this processing, enabling comprehensive environment understanding. For example, in a robot task involving voice commands and visual inputs, LLMs can process both types of data simultaneously and generate actions directly.
Additionally, LLMs provide a distinct advantage in multi-task learning due to their pre-trained knowledge \cite{Wang2024LLMsurvey,li2024rescueLLM}. Through knowledge transfer, they help agents share experiences across different tasks, improving adaptability \cite{silva2018autonomously,Maruyama9494454}. For long-term planning, LLMs can break down complex tasks into subtasks, addressing challenges like the credit assignment problem. This capability is particularly useful in tasks requiring extended reasoning, such as construction tasks in Minecraft.
%
% LLMs also enhance sample efficiency \cite{LLMmarlsurvey,ijcai2022p508}. By generating high-fidelity simulations, they can provide additional training data, reducing the need for real-world learning. Moreover, in sparse-reward environments, LLMs can create reward signals to guide policy optimization.
In optimizing reinforcement learning's sample efficiency \cite{ijcai2022p508,schwarzer2021dataefficient}, the generative capabilities of LLMs can provide agents with additional virtual samples through high-fidelity environment simulations \cite{LLMmarlsurvey,ijcai2022p508}. This not only reduces the cost of real-world learning but also offers high-quality trajectories that serve as valuable references for policy optimization. Furthermore, in sparse reward environments, LLMs can accelerate policy learning by automatically designing reward signals.

\subsection{Application Expansion: Driving Intelligent Collaboration in Complex Scenarios}
% \cite{} . 
The potential of LLMs-enhanced MARL in practical applications is enormous, especially in scenarios that require complex collaboration and real-time decision-making \cite{LLMmarlsurvey,Huinstructed,huang2021gymmurts,sukhbaatar2016learning}. For example, in the field of autonomous driving \cite{liu2024diversedriving,LuMARLautoDriving,JayawardanaMBLautodrive}, the integration of LLMs with MARL can simultaneously process sensor data and natural language information (such as traffic regulations, passenger instructions, etc. \cite{ChuMARLonTrafficSingal}), thereby enhancing the safety and accuracy of decision-making \cite{WangMARLautoDriving,Wang10295775}. In the field of collaborative robots, LLMs can help multiple robots form a more intuitive communication mechanism, achieving highly complex task division and dynamic adjustment. In addition, in multi-objective optimization tasks such as smart grids and intelligent healthcare, LLMs can provide domain knowledge and optimization suggestions to assist reinforcement learning, design more practical reward functions, and thus improve the overall efficiency of the system.
In dynamic and complex environments such as disaster relief \cite{qazzaz2024rescueUAV}, LLMs can dynamically allocate roles and responsibilities according to task requirements, helping multi-agent systems quickly adapt to changing environments and highly complex task divisions \cite{mezghani2023think,Liinteractive,grigsby2024amago}. This capability provides a solid technical support for a wide range of applications.
% LLM-enhanced MARL has great potential in practical applications, especially in scenarios requiring complex collaboration and real-time decision-making. For instance, in autonomous driving, LLM-enhanced MARL can process both sensor data and natural language inputs, such as traffic laws and passenger commands, improving decision accuracy and safety. In collaborative robotics, LLMs enable agents to establish intuitive communication protocols for complex task allocation and dynamic adjustments. Similarly, in fields like smart grids and healthcare, LLMs can provide domain-specific knowledge and reward design to optimize multi-objective tasks.

% In dynamic environments like disaster rescue, LLMs can assist multi-agent systems by dynamically allocating roles and adapting to changing conditions. This enhances the system's efficiency and responsiveness in high-stakes scenarios.

\subsection{Human Society Coordination: Balancing Technology and Ethics}
% \cite{} . 
The integration of LLMs into MARL opens new avenues for advancing multi-agent systems, while also highlighting exciting research directions in improving technical efficiency and addressing ethical considerations. For instance, enhancing the robustness of LLMs in unfamiliar environments offers the opportunity to develop strategies for minimizing biases and hallucinations, thereby improving decision accuracy. Furthermore, the computational complexity and resource demands of LLMs present a chance to innovate in optimizing inference efficiency and scalability. This is especially relevant in dynamic multi-agent environments where real-time responsiveness is critical.

From an ethical perspective, the incorporation of LLMs calls for advancements in ensuring data privacy, safeguarding against adversarial attacks, and establishing clear accountability frameworks for AI-driven decisions. Sensitive domains such as healthcare and disaster response could particularly benefit from focused research on protecting sensitive information and enhancing system resilience. Additionally, improving the transparency and explainability of LLMs-driven decisions is another promising area for exploration, as it would increase trust and user confidence in multi-agent systems.

By addressing these areas, future research can maximize the potential of LLMs-enhanced MARL systems, ensuring they are both technically effective and ethically sound in diverse, real-world applications.

% \subsection{}
% \cite{} . 
Overall, the combination of LLMs and MARL brings new momentum to research and applications in multi-agent systems. By enhancing collaboration through natural language understanding and leveraging large-scale knowledge, these systems can achieve greater efficiency and robustness in complex scenarios. However, fully unlocking their potential requires further exploration in theoretical methods, technological development, and ethical practices. With systematic advancements in these areas, LLMs-enhanced MARL can become the foundation for next-generation intelligent decision-making systems, transforming fields like autonomous driving, collaborative robotics, and healthcare, while shaping the future of AI research.

\section{Conclusion} 
\label{conclu}
Multi-agent cooperative decision-making has demonstrated remarkable potential in addressing complex tasks through intelligent collaboration and adaptability. In this survey, we systematically review the evolution of multi-agent systems, highlighting the shift from traditional methods, such as rule-based and game-theory approaches, to advanced paradigms like MARL and LLMs. We differentiate these methods by examining their unique capabilities, challenges, and applications in diverse environments, paying particular attention to dynamic and uncertain settings.
In addition, we explore the critical role of simulation environments as a bridge between theoretical advancements and real-world implementation, emphasizing their influence on agent interaction, learning, and decision-making. Practical applications of multi-agent systems in domains such as autonomous driving, disaster response, and robotics further underscore their transformative potential.
By summarizing advanced multi-agent decision-making methodologies, datasets, benchmarks, and future research directions, this survey aims to provide a comprehensive resource for researchers and practitioners. % state-of-the-art
We hope it inspires future studies to address existing challenges, such as improving inter-agent communication and adaptability, while leveraging the innovative potential of DRL and LLMs-based approaches to advance multi-agent cooperative decision-making.

% \textcolor{red}{TODO.}

\section*{Acknowledgment}
% This survey was primarily conducted by Weiqiang Jin during his research at Xi'an Jiaotong University and The University of Hong Kong. 
The corresponding authors of this survey are B. Zhao and G.Yang from Xi`an Jiaotong University and Imperial College London. 
% Guang Yang was supported in part by the ERC IMI (101005122), the H2020 (952172), the MRC (MC/PC/21013), the Royal Society (IEC$\backslash$NSFC$\backslash$211235), the NVIDIA Academic Hardware Grant Program, the SABER project supported by Boehringer Ingelheim Ltd, NIHR Imperial Biomedical Research Centre (RDA01), Wellcome Leap Dynamic Resilience, UKRI guarantee funding for Horizon Europe MSCA Postdoctoral Fellowships (EP/Z002206/1), and the UKRI Future Leaders Fellowship (MR/V023799/1).
Guang Yang was supported in part by the ERC IMI (101005122), the H2020 (952172), the MRC (MC/PC/21013), the Royal Society (IEC$\backslash$NSFC$\backslash$211235), the NVIDIA Academic Hardware Grant Program, the SABER project supported by Boehringer Ingelheim Ltd, NIHR Imperial Biomedical Research Centre (RDA01), Wellcome Leap Dynamic Resilience, UKRI guarantee funding for Horizon Europe MSCA Postdoctoral Fellowships (EP/Z002206/1), and the UKRI Future Leaders Fellowship (MR/V023799/1).
The authors would like to thank the editors and anonymous reviewers, who significantly enhanced the quality of the survey. 

\section*{Declaration of Competing Interest}
The authors declare that they have no known competing financial interests or personal relationships that could have appeared to influence the work reported in this paper.

\section*{Data Availability}
No data was used for the research described in the article.

\section*{Declaration of Generative AI and AI-assisted Technologies in the Writing Process}
During the preparation of this work, the authors utilized generative AI and AI-assisted technologies for proofreading and enhancing readability and language clarity in certain sections. 
% The use of these technologies was conducted with human oversight and control. 
The authors have carefully reviewed these contents to ensure accuracy and completeness, acknowledging that AI can generate authoritative-sounding output that may be incorrect, incomplete, or biased.
\appendix
\section{Technological Comparisons between Single-Agent and Multi-Agent (Under Reinforce- \\ ment Learning)} % Markov Decision Processes (MDP) vs. Partially Observable Markov Decision Processes (POMDP) in Reinforcement Learning
\label{app_MDPPOMDP}
Here, we discuss a series of technological comparisons of both DRL-based single-agent and MARL-based multi-agent research.

In solving these single-agent sequential decision-making problems, Markov Decision Processes (MDP) is a powerful mathematical modeling framework, especially in uncertain environments. Since the decision-making process of an agent can inherently be modeled as a sequence of decisions, the single-agent decision-making process can be formulated as an typical MDP, similar to a Markov chain.

In contrast to single-agent DRL systems, multi-agent systems under the MARL techniques involve multiple agents interacting within a shared environment. POMDP is a powerful mathematical modeling framework. It is an extension of the MDP framework that is particularly well-suited for modeling decision-making in environments where the agent does not have full visibility of the entire state space. POMDPs extend MDPs to environments where the agent cannot fully observe the underlying state. Instead, the agent maintains a belief state, which is a probability distribution over the possible states.

\begin{figure*}[!t]
\centering
\includegraphics[width=0.95\linewidth]{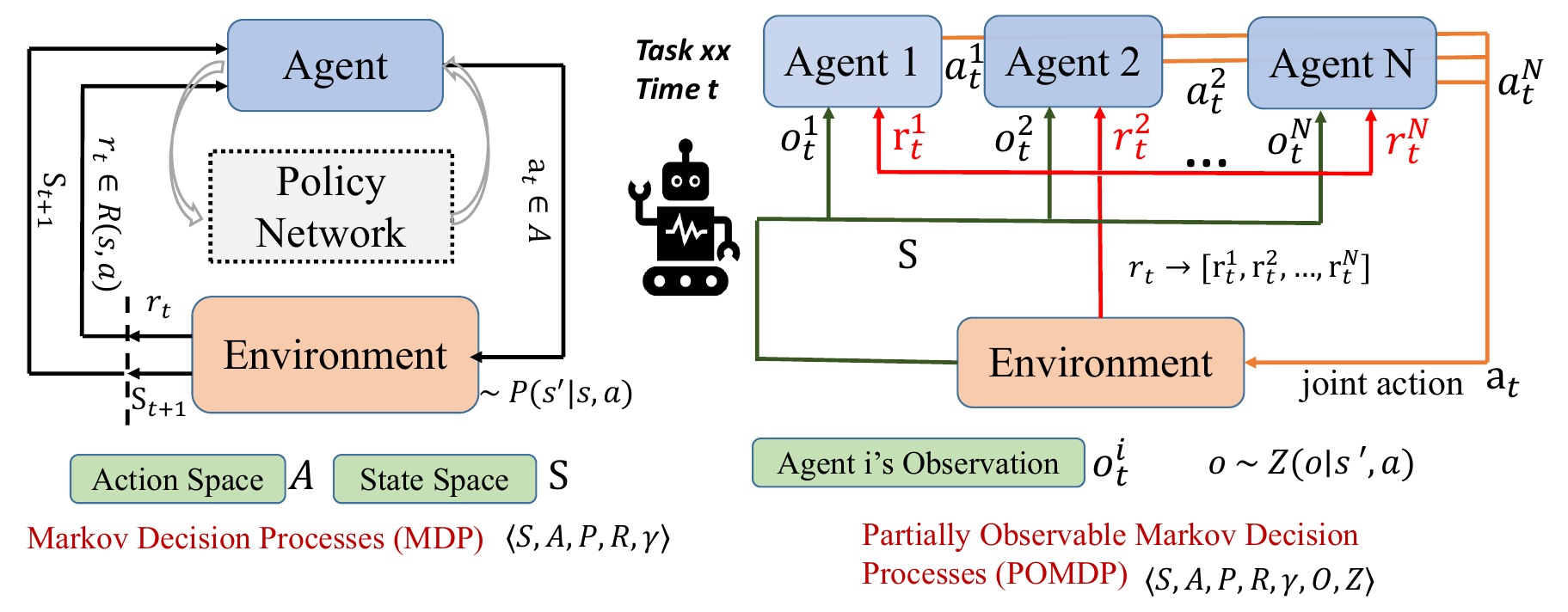}
\caption{The \textit{Markov Decision Process} modeling for the single-agent reinforcement learning paradigm (left) and the \textit{Partially Observable Markov Decision Process} modeling for the multi-agent reinforcement learning paradigm (right).}
\label{fig_MDPPOMDP}
\end{figure*}

% TODO 补全关于MDP和POMDP的这个图的介绍和理论展示
% Figure \ref{fig_MDPPOMDP} \cite{} .
Figure \ref{fig_MDPPOMDP} provides a comparative illustration of \textit{Markov Decision Processes (MDP)} and \textit{Partially Observable Markov Decision Processes (POMDP)}, which correspond to \textit{single-agent} and \textit{multi-agent} reinforcement learning paradigms, respectively.

The left side of Figure \ref{fig_MDPPOMDP} depicts an \textit{MDP}, which models \textit{single-agent decision-making} in a fully observable environment. The agent selects an action \( a \) from the \textit{action space} \( A \) based on the current \textit{state} \( s \) from the \textit{state space} \( S \). The environment transitions to a new state \( s' \) following the transition probability function \( P(s' \mid s, a) \), and the agent receives a reward \( r \). The objective is to optimize a policy \( \pi^* \) that maximizes the cumulative reward. Since the entire state is observable, the decision-making process is relatively straightforward.

On the right side, the \textit{POMDP} framework extends MDPs to \textit{multi-agent settings} where agents operate under \textit{partial observability}. Each agent \( i \) receives only a \textit{partial observation} \( o^i \) rather than the full state \( S \). The agents take individual actions \( a^i \), forming a joint action \( a_t \), which influences state transitions and results in individual rewards \( r^i \). The observations are generated according to the observation function \( Z(o \mid s', a) \), requiring each agent to infer the missing state information and maintain a \textit{belief state} for effective decision-making.

In summary, \textit{MDPs} are well-suited for \textit{single-agent systems}, where the environment is static and fully observable, allowing the agent to make optimal decisions based on complete knowledge of the state. On the other hand, \textit{POMDPs} are crucial for \textit{multi-agent reinforcement learning} scenarios, where multiple agents interact dynamically in an uncertain environment with limited information. This setting introduces challenges such as coordination, competition, and reward interdependencies, making decision-making significantly more complex.

% 在这里画 Markov 的对比图

%%       \cite{<label>} ==> [1]

% %%
% Example citation, See \cite{lamport94}.

% %% If you have bib database file and want bibtex to generate the
% %% bibitems, please use
% %%
% %%  \bibliographystyle{elsarticle-num} 
% %%  \bibliography{<your bibdatabase>}

% %% else use the following coding to input the bibitems directly in the
% %% TeX file.

% %% Refer following link for more details about bibliography and citations.
% %% https://en.wikibooks.org/wiki/LaTeX/Bibliography_Management


% %% For numbered reference style

%   \textit{\LaTeX: a document preparation system},
%   Addison Wesley, Massachusetts,

% \end{thebibliography}
% ###############################################
\bibliographystyle{elsarticle-num} 
\bibliography{reference}

\end{document}